\documentstyle[12pt,aps,eqsecnum]{revtex}

\newcommand{\sign}{\mathop{\rm sign}\nolimits}
\newcommand{\dive}{\mathop{\rm div}\nolimits}
\newcommand{\rot}{\mathop{{\mbox{\boldmath $\rm rot$}}}\nolimits}
\newcommand{\Real}{\mathop{\rm Re}\nolimits}

\def\ab{{\mbox{\boldmath $a$}}}
\def\bb{{\mbox{\boldmath $b$}}}
\def\eb{{\mbox{\boldmath $e$}}}
\def\fb{{\mbox{\boldmath $f$}}}
\def\gb{{\mbox{\boldmath $g$}}}
\def\kb{{\mbox{\boldmath $k$}}}
\def\nb{{\mbox{\boldmath $n$}}}
\def\nablab{{\mbox{\boldmath $\nabla$}}}
\def\rb{{\mbox{\boldmath $r$}}}
\def\vb{{\mbox{\boldmath $v$}}}
\def\ub{{\mbox{\boldmath $u$}}}
\def\Bb{{\mbox{\boldmath $B$}}}
\def\Db{{\mbox{\boldmath $D$}}}
\def\Fb{{\mbox{\boldmath $F$}}}
\def\Ib{{\mbox{\boldmath $I$}}}
\def\Kb{{\mbox{\boldmath $K$}}}
\def\Pb{{\mbox{\boldmath $P$}}}
\def\Rb{{\mbox{\boldmath $R$}}}
\def\Sb{{\mbox{\boldmath $S$}}}
\def\Tb{{\mbox{\boldmath $T$}}}
\def\Ub{{\mbox{\boldmath $U$}}}
\def\Vb{{\mbox{\boldmath $V$}}}
\def\Wb{{\mbox{\boldmath $W$}}}
\def\Omb{{\mbox{\boldmath $\Omega$}}}

\def\lambdab{{\mbox{\boldmath $\lambda$}}}
\def\mub{{\mbox{\boldmath $\mu$}}}
\def\xib{{\mbox{\boldmath $\xi$}}}

\def\Fbc{{\mbox{\boldmath $\cal{F}$}}}
\def\Tbc{{\mbox{\boldmath $\cal{T}$}}}

\begin{document}

\title{Velocity and Pressure Fields Induced by Spheres in an Unbounded Fluid}
\medskip
\author{A~S.~Usenko\footnote{E-mail address:
usenko@bitp.kiev.ua}} \medskip
\address{Bogolyubov Institute for Theoretical Physics, Kiev-143, Ukraine
03143} \medskip
\date{\today}

\maketitle
\bigskip

\begin{abstract}

We propose a procedure for the determination of the time-dependent velocity and
pressure fields of an unbounded incompressible viscous fluid in an external
force field induced by an arbitrary number of spheres moving and rotating in it
as well as the forces and torques exerted by the fluid on the particles. Within
the completely linearized scheme, we express the velocity and pressure fields
of the fluid in terms of induced surface force densities and derive the
explicit form for all quantities contained in these relations not imposing any
additional restrictions on the size of particles, distances between them, and
the frequency range.  We show the incorrectness of similar results obtained
earlier by several authors because these results are expressed in terms of
nonexistent inverse tensors.  We explain the reasons leading to this and
propose a procedure for the elimination of divergent quantities. In the
stationary case, using the proposed procedure, we obtained the translational,
rotational, and coupled friction and mobility tensors for a system containing
an arbitrary number of spheres up to, respectively, the second, forth, and
third orders in the dimensionless parameter equal to the ratio of a typical
radius of a sphere to a typical distance between two spheres. In various
particular cases, the results obtained in the present paper agree with the
well-known results derived by other methods.

\end{abstract}

\bigskip



\newpage

\section{Introduction}  \label{Introduction}

Investigation of hydrodynamic interactions between spheres immersed in an
incompressible viscous fluid is of considerable interest for a wide class of
problems of physics of suspensions and colloidal crystals.  Due to the
long-range nature of the hydrodynamic interactions, their account is
essentially important for the study of these systems, for which many-particle
interactions should be taken into account.  As a rule, for the study of these
interactions, the linearized Navier--Stokes equation for the fluid is used
\cite{ref.Happel}. Usually, the problem is reduced to the determination of the
forces and torques exerted by the fluid on particles moving and rotating in it
with given translational and rotational velocities.  According to the classical
approach used for one particle
\cite{ref.Lamb,ref.Milne,ref.Batch,ref.Landau,ref.Loitsyan}, to find these
forces and torques, first, it is necessary to determine the velocity and
pressure fields of the fluid induced by particles moving and rotating in it.
However, in the case of several particles, this problem is extremely
complicated.  The corresponding exact relations for the forces and torques
exerted on particles due to hydrodynamic interactions between them were derived
only for the particular stationary case of two spheres moving along the line
passing through their centers with equal \cite{ref.Stimson} or different
velocities \cite{ref.Maude} and perpendicular to this line and rotating along
the axis perpendicular to the direction of motion of the spheres and the line
connecting their centers \cite{ref.Wakiya,ref.Davis}.

For this reason, several methods aimed at the determination of approximate
solutions have been developed.  Among these methods, there is the well-known
classical method of reflections first proposed by Smoluchowski for analysis of
the forces exerted by the fluid on  $n$  spheres moving in it with constant
velocities.  Later, this method was used for the solution of many stationary
problems of hydrodynamic interactions of particles (mainly, two particles) in an
incompressible viscous fluid (the detailed review of results is given in
\cite{ref.Happel}) including also the cases of permeable spheres
\cite{ref.Jones1,ref.Jones2,ref.Jones3} and mixed slip-stick boundary conditions
at the surfaces of the spheres
\cite{ref.Feld1,ref.Feld2,ref.Schmitz1,ref.Schmitz2,ref.Schmitz3}.  The
corresponding results are presented in the form of power series in the
dimensionless parameter  $\sigma$  equal to the ratio of a typical radius of a
sphere to a typical distance between two spheres calculated to a certain order
of this parameter.

Since the method of reflections and its further modifications are based on the
solution of the corresponding boundary-value problems, their direct application
to the solution of nonstationary many-particle problems seems to be rather
problematical.  Moreover, even in the stationary case, the procedure of
determination of the velocity field of the fluid induced by particles becomes
essentially more complicated with increase in the number of particles.  For
this reason, the methods of reflections are usually used for two spheres in the
stationary case.  In \cite{ref.MazurBed}, Mazur and Bedeaux proposed a new
method (called the method of induced forces) for the determination of the force
exerted by the fluid on a single sphere moving with time-dependent velocity in
the case where the fluid moves with nonstationary and nonhomogeneous velocity.
Later, this method was developed for the determination of the friction
\cite{ref.Mazur} and mobility tensors \cite{ref.MazurSaarl} in the case of the
stationary fluid containing an arbitrary number of particles, furthermore, the
results for the mobility tensors were generalized to the nonstationary case
\cite{ref.Saarl}.  At finite frequencies, the expressions for the mobility
tensors are obtained up to the third order in two dimensionless parameters,
namely, the parameter  $\sigma$,  which is typical of the stationary case, and
the parameter  $\kappa$,  which is proportional to the ratio of a typical
radius of a sphere to the penetration depth of transverse waves.  The method of
induced forces essentially differs from the method of reflections because it is
not based on the necessity of the knowledge of the explicit form for the fluid
velocity induced by particles.

Other methods for the solution of problems of hydrodynamic interactions between
particles in the fluid were proposed in
\cite{ref.Freed1,ref.Freed2,ref.Pien1,ref.Pien2,ref.Clercx1,ref.Clercx2,ref.Hofman}.
Despite the fact that the solution of nonstationary many-particle problems in
\cite{ref.Pien1,ref.Pien2} is also based on the introduction of unknown induced
forces, this approach essentially differs from the method of induced forces
used in \cite{ref.Mazur,ref.MazurSaarl,ref.Saarl}.  Indeed, within the
framework of this approach, the forces and torques exerted by the fluid on
particles as well as the velocity field of the fluid induced by these particles
are expressed in terms of the introduced induced surface forces, the explicit
form for which was determined up to the third order of two dimensionless
parameters mentioned above, while the method of induced forces in
\cite{ref.Mazur,ref.MazurSaarl,ref.Saarl} is developed in such a way that the
mobility and friction tensors (not the velocity field of the fluid that cannot
be found with the use of this method) are found without determination of the
explicit form of the introduced induced surface forces.  However, the final
results presented in \cite{ref.Pien1,ref.Pien2} are expressed in terms of
multiindex tensors, the explicit expressions for which are not given and that
are only defined as tensors inverse to certain rather complicated multiindex
tensors. For this reason, the question of the agreement between the results
obtained in \cite{ref.Pien1,ref.Pien2} and \cite{ref.Saarl} is open.
Furthermore, this problem remains unsolved even for the stationary case.  Even
in the simplest case of a single sphere rotating with constant angular velocity
in a fluid, the fluid velocity followed from the general relations given in
\cite{ref.Pien1,ref.Pien2} is also expressed in terms of the inverse tensor
and, hence, cannot be reduced to the classical result for the fluid velocity
induced by a rotating sphere \cite{ref.Happel,ref.Landau} without the knowledge
of the explicit form for the inverse tensor.

The aim of the present paper is to develop a method for the solution of various
problems (both stationary and nonstationary) of hydrodynamic interactions
between any number of particles immersed in an unbounded incompressible viscous
fluid in an arbitrary force field such that, as opposed to
\cite{ref.Pien1,ref.Pien2}, the results obtained by this method (the velocity
and pressure fields of the fluid induced by these particles, the forces and
torques exerted by the fluid on the particles, various mobility tensors, etc.)
coincide with the corresponding results obtained in particular cases by other
methods as well as to explain the reasons that do not enable one to represent
the results given in \cite{ref.Pien1,ref.Pien2,ref.Yosh} in the conventional
form.

In Sec.~2, we reduce the problem of  $n$  spheres in an unbounded incompressible
viscous fluid in an arbitrary force field to an equivalent problem for this fluid
without particles in an efficient force field.  Unlike the generally excepted
procedure of the zeroth extension of the external force
\cite{ref.MazurBed,ref.Mazur,ref.Freed1,ref.Freed2,ref.Pien1,ref.Pien2} and the
fluid pressure \cite{ref.MazurBed,ref.Mazur,ref.MazurSaarl,ref.Saarl} to the
domains occupied by the particles, we use another extensions for these
quantities and discuss possible problems connected with different methods of
extensions of the required quantities.  We give relations for the total forces
and torques exerted by the fluid and force fields on particles.

In Sec.~3, within the framework of the completely linearized scheme (both with
respect to the fluid velocity and the velocities of the particles), we express
the required distributions of the fluid velocity and pressure in terms of the
unknown induced surface forces distributed over the surfaces of the particles.
The results are obtained without imposing any additional restrictions on the
size of particles, distances between them, and frequency range.  All required
quantities contained in the relations for the velocity and pressure fields of
the fluid are given in the explicit form in terms of special functions of
dimensionless parameters.  We derive the system of algebraic linear equations in
the unknown harmonics of the induced surface force densities in Sec.~4.

In Sec.~5, we consider the stationary case. Taking into account the obtained
explicit form of the quantities in the system of algebraic equations, which is
similar to the system given in \cite{ref.Yosh}, we solve this system by the
method of successive approximations using the system of noninteracting
particles as a zero approximation.  In this approximation, we show that the
determinant of the system of equations corresponding to the harmonics with $l =
1$  is equal to zero.  Therefore, the inverse tensor, in terms of which the
relations for the forces and torques exerted by the fluid on the particles
given in \cite{ref.Yosh} for the stationary case as well as analogous results
and the velocity field of the fluid induced by the particles given in
\cite{ref.Pien1,ref.Pien2} for the nonstationary case, does not exist.  We
analyze the reasons for this fact and show that this means that the required
induced surface forces can be determined only up to arbitrary potential
components that have no influence on the fluid velocity.  We formulate the
additional conditions that enable one to uniquely determine the induced surface
force densities.  Using the proposed procedure, we obtained the translational,
rotational, and coupled friction and mobility tensors for a system containing
an arbitrary number of spheres up to the second, forth, and third orders in the
dimensionless parameter  $\sigma$, respectively, which in various particular
cases, agree with the well-known results obtained by other methods.  Within the
framework of the considered approach, we formulate an algorithm for the
determination of the velocity and pressure fields of the fluid as well as the
forces and torques exerted by the fluid on particles corresponding to a given
power of the parameter  $\sigma$.  The details of calculation of several
integrals of the products of three and two Bessel functions necessary for
determination of the hydrodynamic interaction tensors are given in Appendix.


\section{General Relations}  \label{General}

We consider an unbounded incompressible viscous fluid in a certain external
force field.  In the linear approximation, the fluid is described by the
linearized Navier--Stokes equation

\begin{equation}
 \rho \frac{\partial \vb(\rb,t)}{\partial t} + \dive \Pb(\rb,t) =
  \Fb^{ext}(\rb,t)
\end{equation}

\noindent  and the continuity equation

\begin{equation}
 \dive \vb(\rb,t) = 0.
\end{equation}

\noindent  Here,  $\Pb(\rb,t)$  is the stress tensor of the fluid with the
components

\begin{equation}
 P_{ij}(\rb,t) =\delta_{ij} \, p(\rb,t) - \eta\left(\frac{\partial
  \vb_i(\rb,t)}{\partial r_j}
  + \frac{\partial\vb_j(\rb,t)}{\partial r_i}\right),  \qquad  i,j = x,y,z,
\end{equation}

\noindent   $p(\rb,t)$  and  $\vb(\rb,t)$  are the hydrostatic pressure and
velocity fields of the fluid,  $\rho$  and  $\eta$  are its density and
viscosity, respectively,  $\delta_{ij}$  is the Kronecker symbol, and
$\Fb^{ext}(\rb,t)$  is the external force acting on a unit volume of the fluid.

We represent the external force  $\Fb^{ext}(\rb,t)$  as a superposition of
the potential  $\Fb^{(p)ext}(\rb,t)$  and solenoidal  $\Fb^{(sol)ext}(\rb,t)$
components

\begin{eqnarray}
 \Fb^{ext}(\rb,t) &=& \Fb^{(p)ext}(\rb,t) + \Fb^{(sol)ext}(\rb,t),\\
 \Fb^{(p)ext}(\rb,t) &=& -\rho \nablab \varphi(\rb,t),
\end{eqnarray}

\noindent  where  $\varphi(\rb,t)$  is the potential of the external
conservative force.

Assuming that the quantities in Eqs.~(2.1)--(2.3) can be expanded in the
Fourier integral of the form

\begin{equation}
 A(\rb,t) = \frac{1}{(2\pi)^4} \int\limits_{-\infty}^\infty \!\!d\omega
  \int\limits_{-\infty}^\infty \!\!d\kb \, A(\kb,\omega)
  \exp(i(\kb \cdot \rb - \omega t)),
\end{equation}

\noindent  we represent the frequency Fourier transform of the solution of
Eqs.~(2.1)--(2.3) as follows:

\begin{eqnarray}
 \vb(\rb,\omega) &=& \frac{1}{(2\pi)^3 \eta} \int\limits_{-\infty}^\infty \!\!
  d\kb \, \frac{\exp(i\kb \cdot \rb)}{k^2 + \kappa^2} \,
  \Fb^{(sol)ext}(\kb,\omega) + \vb^{inf}(\omega),  \\
 \rho(\rb,\omega) &=& -\frac{i}{(2\pi)^3} \int\limits_{-\infty}^\infty \!\!
  d\kb \, \frac{\exp(i\kb \cdot \rb)}{k^2} \, \kb \cdot \Fb^{ext}(\kb,\omega)
  + \rho^{inf}(\omega) \nonumber \\
  &=& -\rho \varphi(\rb,\omega) + \rho^{inf}(\omega),
\end{eqnarray}

\noindent  where

\begin{equation}
 \Fb^{(sol)ext}(\kb,\omega) = (\Ib -\nb_k \nb_k) \cdot \Fb^{ext}(\kb,\omega)
\end{equation}

\noindent  $\Ib$  is the unit tensor,  $\nb_k = \kb/k$  is the unit vector
directed along the vector  $\kb$,  $\kappa = \sqrt{\omega/(i\nu)} = (1 - i
\sign \omega)/\delta$,  $\delta = \sqrt{2\nu/|\omega|}$  is the depth of
penetration of a plane transverse wave of frequency  $\omega$  created by an
oscillating solid surface into the fluid \cite{ref.Landau},  $\nu = \eta/\rho$
is the kinematic viscosity of the fluid, $\vb^{inf}(\omega) = 2\pi
\delta(\omega)\vb^{inf}$  and $\rho^{inf}(\omega) = 2\pi
\delta(\omega)\rho^{inf}$,  where  $\delta(\omega)$ is the Dirac
delta-function,   $\vb^{inf}$  and  $\rho^{inf}$  are certain constant velocity
and pressure at infinity defined by the condition for the fluid at infinity
where the external force field is absent.  According to relation (2.7), the
fluid velocity is determined only by the solenoidal component of the external
force, which is the natural consequence of the continuity equation (2.2).  The
fluid pressure is determined by the potential component of the external force.

Let, in the considered fluid,  $N$  homogeneous macroscopic spheres of radii
$a_\alpha$  and masses  $m_\alpha$,  where  $\alpha = 1, 2,\ldots,N$,  be
present.  At the time  $t$,  the position of the center of sphere  $\alpha$  is
defined by the radius vector  $\Rb_{\alpha}(t)$  relative to the fixed
Cartesian coordinate system with origin at the point  $O$  (in what follows,
the system  $O$). In parallel with this fixed system, we also introduce  $N$
local moving Cartesian coordinate systems with origins  $O_\alpha$  at the
centers of spheres coinciding with their centers of mass (in what follows,
systems $O_\alpha$) so that sphere  $\alpha$  does not move relative to system
$O_\alpha$.  Therefore, the radius vector  $\rb$  of any point of the space can
be represented in the form  $\rb = \Rb_\alpha + \rb_\alpha$,  where
$\rb_\alpha$  is the radius vector of this point relative to the local system
$O_\alpha$.  The motion of the spheres is described by the equations

\begin{eqnarray}
 m_\alpha \frac{d\Ub_\alpha(t)}{dt}  &=& \Fb_\alpha^{tot}(t),  \\
 I_\alpha \frac{d\Omb_\alpha(t)}{dt} &=& \Tb_\alpha^{tot}(t),
\end{eqnarray}

\noindent  where

\begin{eqnarray}
 \Ub_\alpha(t) &=& \frac{d\Rb_\alpha(t)}{dt}  \nonumber
\end{eqnarray}

\noindent  and  $\Omb_\alpha (t)$  are, respectively, the translational and
angular velocities of sphere  $\alpha$,  $I_\alpha = (2/5) m_\alpha a_\alpha$
is its moment of inertia,  $\Fb_\alpha^{tot}(t)$  is the total force and
$\Tb_\alpha^{tot}(t)$  is the total torque (here and in what follows, all
torques corresponding to particle  $\alpha$  are considered relative to its
center) acting on sphere  $\alpha$

\begin{eqnarray}
 \Fb_\alpha^{tot}(t) &=& \Fb_\alpha^{ext}(t) + \Fb_\alpha^{f}(t),\\
 \Tb_\alpha^{tot}(t) &=& \Tb_\alpha^{ext}(t) + \Tb_\alpha^{f}(t),
\end{eqnarray}

\noindent  where  $\Fb_\alpha^{ext}(t)$  and  $\Tb_\alpha^{ext}(t)$  are the
force and torque acting on sphere  $\alpha$  due to the external force field
and $\Fb_\alpha^{f}(t)$  and  $\Tb_\alpha^{f}(t)$  are the force and torque
exerted by the fluid on sphere  $\alpha$

\begin{eqnarray}
 \Fb_\alpha^{f}(t) &=& -\int\limits_{S_\alpha(t)} \! \Pb(\rb,t)\cdot \nb_\alpha
  \, dS_\alpha,  \\
 \Tb_\alpha^{f}(t) &=& -\int\limits_{S_\alpha(t)} \! \Bigl((\rb - \Rb_\alpha(t))
  \times \Pb(\rb,t) \Bigr) \cdot \nb_\alpha \, dS_\alpha,
\end{eqnarray}

\noindent  $S_\alpha(t)$  is the surface of sphere  $\alpha$  at the time
$t$  and  $\nb_\alpha$  is the outward unit vector to this surface.

The fluid is described by the same equations (2.1)--(2.3) as in the case of the
absence of particles but defined only in the domain outside the spheres
$r_\alpha \geq a_\alpha$, \quad  $\alpha = 1, 2\ldots,N$,  where  $r_\alpha
\equiv |\rb_\alpha| = |\rb - \Rb_\alpha (t)|$.

The problem is to determine the velocity and pressure fields of the fluid
induced by moving particles in it as well as the forces and torques exerted by
the fluid on the particles.  For the solution of this problem, we use the method
of induced forces proposed in \cite{ref.MazurBed} and developed in
\cite{ref.Mazur,ref.MazurSaarl,ref.Saarl}.  We extend the quantities
$p(\rb,t)$,  $\vb(\rb,t)$,  and  $\Fb^{ext}(\rb,t)$  defined in Eqs.~(2.1) and
(2.2) for  $r_\alpha \geq a_\alpha$  to the domains  $r_\alpha < a_\alpha$
occupied by the spheres as follows:

\begin{eqnarray}
 \vb(\rb,t) &=& \Ub_\alpha(\rb,t) = \Ub_\alpha(t)
  + \left(\Omb_\alpha(t)\times\rb_\alpha\right),  \quad r_\alpha < a_\alpha,  \\
 p(\rb,t) &=& -\rho\varphi(\rb,t) + p^{inf} - \rho\left(\rb \cdot \frac{\partial
  \Ub_\alpha(t)}{\partial t}\right),  \quad r_\alpha < a_\alpha.
 \end{eqnarray}

We extend the external force field  $\Fb^{ext}(\rb,t)$  given for  $r_\alpha
\geq a_\alpha$  to the domains  $r_\alpha < a_\alpha$  so that in the entire
space, it is described by the same analytic expression as in the domain
$r_\alpha \geq a_\alpha$.  (For short, we call this extension an analytic
extension.)  The nonzero extension of the fluid pressure defined by (2.17) and
the analytic extension of the external force field differ from the usually
accepted zeroth extension of both the fluid pressure
\cite{ref.MazurBed,ref.Mazur,ref.MazurSaarl,ref.Saarl} and the external force
field \cite{ref.Freed1,ref.Freed2,ref.Pien1,ref.Pien2} to the domains $r_\alpha
< a_\alpha$.  It is worth noting that the considered problem can be solved for
various extensions of the quantities  $p(\rb,t)$  and $\Fb^{ext}(\rb,t)$  to
the domains  $r_\alpha < a_\alpha$  including the zeroth extension
\cite{ref.MazurBed,ref.Mazur,ref.MazurSaarl,ref.Saarl,ref.Freed1,ref.Freed2,%
ref.Pien1,ref.Pien2} and the extensions used in the present paper.  However,
the complexity of calculations and the possibility of interpretation of certain
results essentially depend on the specific method of extension.

If the fluid pressure is extended to the domains  $r_\alpha < a_\alpha$
according to relation (2.17), then the function  $p(\rb,t)$  defined in the
entire space has a discontinuity at the surfaces  $r_\alpha = a_\alpha$.  For
this reason, in what follows, we write the fluid pressure at the point $\rb =
\Rb_\alpha + \ab_\alpha$  as $p(\Rb_\alpha + \ab_\alpha + 0,t)$.

Note that for the analytic extension of the external force field
$\Fb^{ext}(\rb,t)$ to the domains  $r_\alpha < a_\alpha$,  the first two terms
in (2.17) represent the pressure of the unbounded fluid without particles in
the external force field $\Fb^{ext}(\rb,t)$.  In the case where the fluid
velocity satisfies the stick boundary conditions at the surfaces of the spheres
\cite{ref.Happel,ref.Batch,ref.Landau} (this case is considered in the present
paper), condition (2.16) ensures the continuity of the function  $\vb(\rb,t)$
given in the entire space at  $r_\alpha = a_\alpha$.

In \cite{ref.Freed2,ref.Pien1,ref.Pien2}, the external force field
$\Fb^{ext}(\rb,t)$  is extended to the domains  $r_\alpha < a_\alpha$  by zero
and the stress tensor  $\Pb(\rb,t)$  in these domains is defined as follows:

\begin{equation}
 \dive \Pb(\rb,t) = -\rho \frac{\partial \Ub_\alpha(\rb,t)}{\partial t},
  \quad r_\alpha < a_\alpha.
\end{equation}

We note that in the case where the fluid velocity  $\vb(\rb,t)$  is extended to
the domains  $r_\alpha < a_\alpha$  according to relation (2.16) (in fact, this
is used in \cite{ref.Freed2,ref.Pien1,ref.Pien2}), the stress tensor
$\Pb(\rb,t)$  for  $r_\alpha < a_\alpha$  must have another analytic
representation than (2.3).  Otherwise, if the stress tensor  $\Pb(\rb,t)$  in
the domains  $r_\alpha < a_\alpha$  is defined by relation (2.3), where the
quantity  $\vb(\rb,t)$  is defined by relation (2.16) in these domains, then for
$r_\alpha < a_\alpha$,  we have

\begin{equation}
 \Pb(\rb,t) = p(\rb,t)\Ib,
\end{equation}

\noindent  where the quantity  $p(\rb,t)$  for  $r_\alpha < a_\alpha$  depends
on the method of extension of the fluid pressure  $p(\rb,t)$  defined for
$r_\alpha \geq a_\alpha$  to  $r_\alpha < a_\alpha$.  However, for any
extension of the fluid pressure  $p(\rb,t)$  to the domains  $r_\alpha <
a_\alpha$,  according to (2.19), the left-hand side of Eq.~(2.18) contains the
potential vector  $\nablab p(\rb,t)$,  while the right-hand side of this
equations contains the solenoidal vector  -$\rho \frac{\partial}{\partial t}
\left(\left( \Omb_\alpha(t)\times \rb \right)\right)$  because

\begin{equation}
 \left( \Omb_\alpha(t) \times \rb\right) = \frac{1}{3}
  \rot \Bigl(\left( \Omb_\alpha(t) \times \rb\right) \times \rb \Bigr).
\end{equation}

\noindent  This means that the extension of the fluid velocity
$\vb(\rb,t)$  to the domains  $r_\alpha < a_\alpha$  according to (2.16)
eliminates the possibility of extension (2.18) (used in
\cite{ref.Freed2,ref.Pien1,ref.Pien2}) for the stress tensor  $\Pb(\rb,t)$
defined by relations (2.3) in the entire space to these domains for the case of
the time-dependent angular velocity  $\Omb_\alpha(t)$.

The quantities  $\vb(\rb,t)$  and  $p(\rb,t)$  extended to  $r_\alpha <
a_\alpha$  according to (2.16) and (2.17) and  $\Fb^{ext}(\rb,t)$
analytically extended to these domains are given in the entire space and
satisfy the continuity equation (2.2) and the equation

\begin{equation}
 \rho \frac{\partial \vb(\rb,t)}{\partial t} + \dive \Pb(\rb,t) =
  \Fb^{ext}(\rb,t) + \Fb^{ind}(\rb,t)
\end{equation}

\noindent  given in the entire space.  Equation (2.21) differs from Eq.~(2.1)
by the additional term (the induced force density)  $\Fb^{ind}(\rb,t)$  that
appears due to the extension of the linearized Navier--Stokes equation given
for  $r_\alpha \geq a_\alpha$  to the domains  $r_\alpha < a_\alpha$.  For the
analytic extension of the external force field, the quantity $\Fb^{ext}(\rb,t)$
in Eqs.~(2.1) and (2.21) is described by the same analytic expressions, while,
for the zeroth extension of the external force field to $r_\alpha <a_\alpha$
\cite{ref.MazurBed,ref.Freed1,ref.Freed2,ref.Pien1,ref.Pien2}, the analytic
expressions describing the quantity  $\Fb^{ext}(\rb,t)$  in the domains
$r_\alpha \geq a_\alpha$  and  $r_\alpha < a_\alpha$  are different. Therefore,
in the latter case, the procedure where the quantity $\Fb^{ind}(\rb,t)$ is
formally considered to be equal to zero in Eq.~(2.21) is not equivalent to the
removal of the particles from the fluid because Eq.~(2.1) given in the entire
space and Eq.~(2.21) with $\Fb^{ind}(\rb,t) = 0$  describe the unbounded fluid
in various force fields, namely, in  $\Fb^{ext}(\rb,t)$ and in
$\sum\limits_{\alpha= 1}^{N} \Theta(r_\alpha - a_\alpha) \Fb^{ext}(\rb,t)$,
where  $\Theta(x)$  is the Heaviside function, respectively.

The induced force density   $\Fb^{ind}(\rb,t)$  is presented in the form

\begin{equation}
 \Fb^{ind}(\rb,t) = \sum\limits_{\alpha = 1}^{N} \Fb_\alpha^{ind}(\rb,t),
\end{equation}

\noindent  where  $\Fb_\alpha^{ind}(\rb,t)$  is the induced force density for
particle  $\alpha$.  This quantity has the volume  $\Fb_\alpha^{(V)ind}(\rb,t)$
and surface  $\Fb_\alpha^{(S)ind}(\rb,t)$  components given, respectively,
inside the volume  $V_\alpha = (4/3)\pi a_\alpha^3$  occupied by the particle
$\alpha$  and on its surface  $S_{\alpha}(t)$  \cite{ref.Saarl}

\begin{equation}
 \Fb_\alpha^{ind}(\rb,t) = \Fb_\alpha^{(V)ind}(\rb,t) +
  \Fb_\alpha^{(S)ind}(\rb,t).
\end{equation}

\noindent  Here,

\begin{eqnarray}
 \Fb_\alpha^{(V)ind}(\rb,t) &=& \Theta(a_\alpha - r_\alpha)
  \left\{\rho\frac{\partial}{\partial t}\left(\Omb_\alpha(t) \times
  \rb_\alpha\right) - \Fb^{(sol)ext}(\rb,t)\right \}, \\
 \Fb_\alpha^{(S)ind}(\rb,t) &=& \int \!\! d\Omega_\alpha \,
  \delta(\rb - \Rb_\alpha(t) - \ab_\alpha) \fb_\alpha(\ab_\alpha,t),
\end{eqnarray}

\noindent  where  $\ab_\alpha \equiv (a_\alpha, \theta_\alpha, \varphi_\alpha)$
is the vector directed from the center of sphere  $\alpha$  to a point on its
surface characterized by the polar  $\theta_\alpha$  and azimuth
$\varphi_\alpha$  angles in the local spherical coordinate system  $O_\alpha$, \quad
$d\Omega_\alpha = \sin{\theta_\alpha} \, d\theta_\alpha d\varphi_\alpha$  is
the solid angle, and  $\fb_\alpha(\ab_\alpha,t)$  is a certain unknown density
of the induced surface force distributed over the surface  $S_{\alpha}(t)$.

Unlike \cite{ref.MazurSaarl}, the volume component
$\Fb_{\alpha}^{(V)ind}(\rb,t)$  of the induced force does not contain the
translational velocity $\Ub_\alpha(t)$  of particle  $\alpha$, which is caused
by the different extensions of the fluid pressure to  $r_\alpha < a_\alpha$
used in the present paper and in \cite{ref.MazurSaarl}. In addition, owing to
extension (2.17) for the fluid pressure to  $r_\alpha < a_\alpha$,
$\Fb_\alpha^{(V)ind}(\rb,t)$  is defined only by the solenoidal component of
the analytically extended external force field.  Thus, in the particular case
of conservative external forces, $\Fb_\alpha^{(V)ind}(\rb,t)$ is independent of
external forces.

It is easy to verify that the representation of the induced force density in
the form (2.22)--(2.25) as well as relations (2.16) and (2.17) and the analytic
extension of  $\Fb^{ext}(\rb,t)$  to  $r_\alpha < a_\alpha$  insure the
validity of Eq.~(2.21) in the entire space.  In this case,

\begin{equation}
 \dive \Pb(\rb,t) = -\rho \frac{d\Ub_\alpha(t)}{dt} + \Fb^{(p)ext}(\rb,t)
  + \Fb_\alpha^{(S)ind}(\rb,t),  \quad r_\alpha \leq a_\alpha.
\end{equation}

Using relation (2.26), we can represent force (2.14) and torque (2.15) exerted
by the fluid on sphere  $\alpha$  as follows:

\begin{eqnarray}
 \Fb_\alpha^{f}(t) &=& \Fb_\alpha (t) + \Fb_\alpha^{in}(t)
  - \tilde{\Fb}_\alpha^{(p)ext}(t),  \\
 \Tb_\alpha^{f}(t) &=& \Tb_\alpha (t) - \tilde{\Tb}_\alpha^{(p)ext}(t),
\end{eqnarray}

\noindent  where  $\Fb_\alpha (t)$  and  $\Tb_\alpha(t)$  are, respectively,
the force and the torque exerted on the sphere  $\alpha$  due to the induced
force density distributed over its surface

\begin{eqnarray}
 \Fb_\alpha (t) &=& -\int \!\! d\rb \, \Fb_\alpha^{(S)ind}(\rb,t)
  = -\int \!\! d\Omega_\alpha \, \fb_\alpha(\ab_\alpha,t),  \\
 \Tb_\alpha (t) &=& -\int \!\! d\rb \, \left(\rb_\alpha \times
  \Fb_\alpha^{(S)ind}(\rb,t)\right) = -\int \!\! d\Omega_\alpha \,
  \Bigl ( \ab_\alpha \times \fb_\alpha(\ab_\alpha,t) \Bigr ),
\end{eqnarray}

\noindent  $\tilde{\Fb}_\alpha^{(p)ext}(t)$  and
$\tilde{\Tb}_\alpha^{(p)ext}(t)$  are, respectively, the force and the torque
acting on the fluid sphere occupying the volume  $V_\alpha$  instead of
spherical particle  $\alpha$  due to the potential component
$\Fb^{(p)ext}(\rb,t)$  of the external force field analytically extended to
this domain

\begin{eqnarray}
 \tilde{\Fb}_\alpha^{(p)ext}(t) &=& \int\limits_{V_\alpha} \!\! d\rb \,
  \Fb^{(p)ext}(\rb,t),  \\
 \tilde{\Tb}_\alpha^{(p)ext}(t) &=& \int\limits_{V_\alpha} \!\! d\rb \,
  \left(\rb_\alpha \times \Fb^{(p)ext}(\rb,t)\right),
\end{eqnarray}

\noindent  and

\begin{equation}
 \Fb_\alpha^{in}(t) = \tilde{m}_\alpha \frac{d\Ub_\alpha (t)}{dt}
\end{equation}

\noindent  is the inertial force that is necessary to be applied to a fluid
sphere of volume  $V_\alpha$  in order that it move with the acceleration
$d\Ub_\alpha(t)/dt$  \cite{ref.MazurBed},  where  $\tilde{m}_\alpha = \rho
V_\alpha$  is the mass of the fluid displaced by particle  $\alpha$.

In the particular case of the time-independent homogeneous gravitational force
field

\begin{equation}
 \varphi(\rb,t) \equiv \varphi(\rb) = -\gb \rb,
  \qquad  \Fb^{(sol)ext}(\rb,t) = 0,
\end{equation}

\noindent  we have

\begin{equation}
 \begin{array}{lrllrl}
  \Fb_\alpha^{ext}(t) &\equiv& \Fb_\alpha^{ext} = m_\alpha \gb,  \qquad
   & \tilde{\Fb}_\alpha^{(p)ext}(t) &=& -\Fb_\alpha^A,  \nonumber  \\
  \Tb_\alpha^{ext}(t) &=& 0,  \qquad
   & \tilde{\Tb}_\alpha^{(p)ext}(t) &=& 0,
 \end{array}
\end{equation}

\noindent  where  $\Fb_\alpha^A = -\tilde{m}_\alpha \gb$  is the Archimedes
force acting on particle  $\alpha$  and  $\gb$  is the acceleration of gravity.
Therefore, in the considered case,  $\Fb_\alpha^{ext}(t) -
\tilde{\Fb}_\alpha^{ext}(t) = \left(m_\alpha - \tilde{m}_\alpha\right) \gb$  in
relation (2.12) is the gravity force of particle  $\alpha$  corrected for the
buoyancy force.

In view of relations (2.29) and (2.30), the forces and torques exerted by the
fluid on particles are completely determined by the surface induced force
densities. If it is necessary to determine only the forces  $\Fb_\alpha(t)$ and
the torques  $\Tb_\alpha(t)$  in the linear approximation with respect to the
velocities of the fluid and the spheres, it is convenient to use the procedure
proposed in \cite{ref.Mazur,ref.MazurSaarl,ref.Saarl}, which does not require
the explicit form of  $\Fb^{(S)ind}(\rb,t)$.  On the basis of this procedure,
for an arbitrary number of spheres in a fluid, one obtains the mobility tensors
of particles that move and rotate in the fluid both in the stationary case
\cite{ref.Mazur,ref.MazurSaarl} and with regard for the time dependence
\cite{ref.Saarl}.  In a more general case where the velocity field of the fluid
should be determined, the problem becomes essentially more complicated because
the explicit form of the induced surface force densities must be obtained.  In
the nonstationary case, this problem is considered in
\cite{ref.Pien1,ref.Pien2} where the zeroth extension of the external forces to
the domains occupied by spheres is used and the stress tensor in these domains
is defined by relation (2.18).


\section{Representation of the Required Quantities in Terms of Harmonics of
Induced Surface Force Densities}  \label{Representation}

To determine the velocity and pressure fields of the fluid in the presence of
$N$  spheres in it, we note the same structure of Eqs.~(2.1) and (2.21).
Therefore, we can use solution (2.7), (2.8) of Eqs.~(2.1)--(2.3) with the
substitution $\Fb^{ext}(\kb,\omega) + \Fb^{ind}(\kb,\omega)$  for
$\Fb^{ext}(\kb,\omega)$.

We consider the problem linearized both with respect to the velocity of the
fluid and the velocities of the spheres.  Within the framework of this
approximation, we neglect the time dependence of the positions of centers of
the spheres defined by  $\Rb_\alpha (t)$  and their surfaces  $S_\alpha (t)$.
This means that the spheres do not displace considerably for considered time
intervals (for details, see \cite{ref.Saarl}).  This enables us to represent
$\Fb_\alpha^{(S)ind} (\kb,\omega)$  and  $\Fb_\alpha^{(V)ind} (\kb,\omega)$  as
follows:

\begin{eqnarray}
 \Fb_\alpha^{(S)ind} (\kb,\omega) &=& \int \!\! d\Omega_\alpha \, \exp{(-i\kb
  \cdot \left(\Rb_\alpha + \ab_\alpha \right))} \,
  \fb_\alpha(\ab_\alpha,\omega),  \\
 \Fb_\alpha^{(V)ind} (\kb,\omega) &=& -\left\{i \frac{2}{3} \xi_{\alpha}
  b_\alpha^2 \frac{j_2(ka_\alpha)}{k^2} \Bigl (\Omb_\alpha(\omega)\times\kb\Bigr )
  + \tilde{\Fb}_\alpha^{(sol)ext} (\kb,\omega) \right\} \exp{(-i\kb \cdot \Rb_\alpha)},
\end{eqnarray}

\noindent  where  $\xi_\alpha = 6 \pi \eta a_\alpha$  is the Stokes friction
coefficient for a sphere of radius  $a$  uniformly moving in the fluid,
$b_\alpha = \kappa a_\alpha$  is the dimensionless parameter that characterizes
the ratio of the radius  $a$  of sphere  $\alpha$  to the length  $\delta$
[$|b_\alpha | = \sqrt{2}\,a_\alpha/\delta)$],  $j_n(x)$  is the spherical
Bessel function of order  $n$,  and

\begin{equation}
 \tilde{\Fb}_\alpha^{(sol)ext} (\kb,\omega) = \int \!\! d\rb_\alpha \,
  \exp{(-i\kb \cdot \rb_\alpha)} \, \Theta(a_\alpha - r_\alpha) \,
  \Fb^{(sol)ext} (\Rb_\alpha + \rb_\alpha,\omega).
\end{equation}

Note that the quantity  $\tilde{\Fb}_\alpha^{(sol)ext} (\kb,\omega)$  defined
by relation (3.3) differs from the Fourier transform  $\Fb_\alpha^{(sol)ext}
(\kb,\omega)$  of the solenoidal component of the external force field because
the integral in (3.3) is taken over the finite space domain  $r_\alpha \le
a_\alpha$,  while for  $\Fb_\alpha^{(sol)ext} (\kb,\omega)$,  the Fourier
integral is taken over the entire space.

By using relations (2.7), (2.8), and (3.1)--(3.3), we can represent the
required Fourier transforms for the quantities  $\vb(\rb,\omega)$  and
$p(\rb,\omega)$ defined in the entire space as follows:

\begin{eqnarray}
 \vb(\rb,\omega) &=& \vb^{(0)}(\rb,\omega) + \vb^{ind}(\rb,\omega),  \\
 p(\rb,\omega)   &=&  p^{(0)}(\rb,\omega)  +  p^{ind}(\rb,\omega).
\end{eqnarray}

\noindent  Here, the quantities  $p^{(0)}(\rb,\omega)$  and
$\vb^{(0)}(\rb,\omega)$, where

\begin{eqnarray}
 \vb^{(0)}(\rb,\omega) &=& \vb^{(0)sol}(\rb,\omega) + \vb^{inf}(\omega),  \\
 \vb^{(0)sol}(\rb,\omega) &=& \frac{1}{(2 \pi)^3 \eta}
  \int\limits_{-\infty}^\infty \!\! d\kb \, \frac{\exp{(i\kb \cdot \rb)}}
  {k^2 + \kappa^2} \, \Fb^{(sol)ext}(\kb,\omega),
\end{eqnarray}

\noindent  defined at any point of the space, for  $r_\alpha \ge a_\alpha$,
are, respectively, the pressure and the velocity of the unbounded fluid in the
absence of particles defined by relations (2.8) and (3.6)--(3.7). The
quantities  $\vb^{ind}(\rb,\omega)$  and $p^{ind}(\rb,\omega)$  can be
represented in the form

\begin{eqnarray}
 \vb^{ind}(\rb,\omega) &=& \sum\limits_{\beta = 1}^N
  \vb_\beta^{ind}(\rb,\omega),  \\
 p^{ind}(\rb,\omega)   &=& \sum\limits_{\beta = 1}^N
  p_\beta^{ind}(\rb,\omega),
\end{eqnarray}

\noindent  where

\begin{eqnarray}
 \vb_\beta^{ind}(\rb,\omega) &=& \vb_\beta^{(V)ind}(\rb,\omega)
  + \vb_\beta^{(S)ind}(\rb,\omega),  \\
 p_\beta^{ind}(\rb,\omega)  &=&  p_\beta^{(V)ind}(\rb,\omega)
  + p_\beta^{(S)ind}(\rb,\omega).
\end{eqnarray}

\noindent  For  $r_\alpha > a_\alpha$,  $\vb^{ind}(\rb,\omega)$  and
$p^{ind}(\rb,\omega)$ are, respectively, the fluid velocity and pressure
induced by all particles.  Here,

\begin{eqnarray}
 \vb_\beta^{(S)ind}(\rb,\omega) &=& \frac{1}{(2 \pi)^3}
  \int\limits_{-\infty}^\infty \!\! d\kb \,
  \exp{(i\kb \cdot \left(\rb-\Rb_\beta\right))}
  \int \!\! d\Omega_\beta \, \exp{(-i\kb\ab_\beta)} \,
  \Sb(\kb,\omega) \cdot \fb_\beta(\ab_\beta,\omega), \nonumber  \\
  && \\
 p_\beta^{(S)ind}(\rb,\omega)  &=& -\frac{i}{(2 \pi)^3}
  \int\limits_{-\infty}^\infty \!\! d\kb \,
  \frac{\exp{(i\kb \cdot \left(\rb-\Rb_\beta\right))}}{k^2}
  \int \!\! d\Omega_\beta \, \exp{(-i\kb\ab_\beta)} \,
  \kb \cdot \fb_\beta(\ab_\beta,\omega).
\end{eqnarray}

\noindent  Relations (3.12) and (3.13) are valid for any  $\rb$.  For $r_\alpha
> a_\alpha$,  the quantities  $\vb_\beta^{(S)ind}(\rb,\omega)$  and
$p_\beta^{(S)ind}(\rb,\omega)$  can be interpreted, respectively, as the fluid
velocity and pressure at the point  $\rb$  generated by the induced surface
force $\Fb_\beta^{(S)ind}(\rb,\omega)$ distributed over the surface of sphere
$\beta$.  The quantity

\begin{equation}
 \Sb(\kb,\omega) = \frac{1}{\eta \left(k^2 + \kappa^2 \right)}
 \left(\Ib - \nb_k \nb_k \right)
\end{equation}

\noindent  is the dynamic Oseen tensor.

The quantities  $\vb_\beta^{(V)ind}(\rb,\omega)$  and
$p_\beta^{(V)ind}(\rb,\omega)$  defined in the entire space, in the domain
$r_\alpha > a_\alpha$,  are, respectively, the fluid velocity and pressure at
the point $\rb$ generated by the induced volume force
$\Fb_\beta^{(V)ind}(\rb,\omega)$ given in the volume $V_\beta$ occupied by
particle  $\beta$,  moreover,

\begin{equation}
 p_\beta^{(V)ind}(\rb,\omega) = \frac{i}{(2 \pi)^3}
  \int\limits_{-\infty}^\infty \!\! d\kb \,
  \frac{\exp{(i\kb \cdot \left(\rb-\Rb_\beta\right))}}{k^2} \,
  \kb \cdot \tilde{\Fb}_\beta^{(sol)ext}(\kb,\omega).
\end{equation}

\noindent  Thus, the contribution to the fluid pressure due to the volume
induced forces is nonzero only if the considered system is in a nonconcervative
force field.

It is convenient to represent the quantity  $\vb_\beta^{(V)ind}(\rb,\omega)$
as the sum of two terms

\begin{equation}
 \vb_\beta^{(V)ind}(\rb,\omega) = \vb_\beta^{(r)ind}(\rb,\omega)
 + \vb_\beta^{(sol)ind}(\rb,\omega),
\end{equation}

\noindent  where the first term is caused by the rotation of particle  $\beta$

\begin{equation}
 \vb_\beta^{(r)ind}(\rb,\omega) = -i \xi_\beta b_\beta^2 \frac{2}{3\eta}
 \frac{1}{(2\pi)^3} \int\limits_{-\infty}^\infty \!\! d\kb \,
 \frac{\exp{(i\kb \cdot \left(\rb-\Rb_\beta\right))}}{k^2 +\kappa^2}
 \frac{j_2(\kappa a_\beta)}{k^2} \Bigl (\Omb_\beta(\omega) \times \kb \Bigr )
\end{equation}

\noindent  and the second is caused by the solenoidal component of the external
force filed

\begin{equation}
 \vb_\beta^{(sol)ind}(\rb,\omega) = -\frac{1}{(2\pi)^3 \eta}
 \int\limits_{-\infty}^\infty \!\! d\kb
 \frac{\exp{(i\kb \cdot \left(\rb-\Rb_\beta\right))}}{k^2 +\kappa^2}
 \left(\Ib - \nb_k \nb_k \right) \cdot \tilde{\Fb}_\beta^{(sol)ext}(\kb,\omega).
 \end{equation}

Note that for the zeroth extension of the external forces to the domains
$r_\alpha < a_\alpha$  used in
\cite{ref.Mazur,ref.Freed1,ref.Freed2,ref.Pien1,ref.Pien2}, the interpretation
of  $\vb^{(0)}(\rb,\omega)$  as the nonperturbed velocity of the fluid, i.e.,
the fluid velocity in the absence of the spheres, is not a quite correct
because, in this case,  $\vb^{(0)}(\rb,\omega)$  is the velocity of the fluid
not containing particles but in the external force field  $\sum\limits_{\alpha
= 1}^N \Theta(r_\alpha - a_\alpha) \Fb^{ext}(\rb,t)$  instead of
$\Fb^{ext}(\rb,t)$ given in the entire space.  The same remarks are also valid
for the interpretation of  $p^{(0)}(\rb,\omega)$.

We expand the induced surface force densities  $\fb_\beta(\ab_\beta,\omega)$  in
the series in the spherical harmonics

\begin{equation}
 \fb_\beta(\ab_\beta,\omega) = \frac{1}{2 \sqrt{\pi}} \sum\limits_{lm}
  \fb_{\beta, lm}(a_\beta,\omega) Y_{lm}(\theta_\beta,\varphi_\beta),
\end{equation}

\noindent  where, for short, the symbol  $\sum\limits_{lm}$  means
$\sum\limits_{l = 0}^\infty \sum\limits_{m = -l}^l$  and the arguments
$a_\beta$ in  $\fb_{\beta, lm}(a_\beta,\omega)$  is omitted in what follows,
and the spherical harmonics $Y_{lm}(\theta,\varphi)$  are defined as follows
\cite{ref.Nikiforov}:

\begin{eqnarray}
 Y_{lm}(\theta,\varphi) &=& \frac{\exp{(im\varphi)}}{\sqrt{2\pi}}
  \Theta_{lm}(\cos\theta), \qquad  l = 0, 1, \ldots, \qquad  -l \leq m \leq l,
  \nonumber \\
 \Theta_{lm}(x) &=& (-1)^{\frac{m - |m|}{2}} \sqrt{\frac{2l + 1}{2}
  \frac{(l - |m|)!}{(l + |m|)!}}\, P_l^{|m|}(x),
\end{eqnarray}

\noindent  and  $P_l^m(x)$  is the associated Legendre polynomial.

By using expansion (3.19) and relations (2.29) and (2.30), we can represent the
Fourier transforms of the force  $\Fb_\alpha(\omega)$  and the torque
$\Tb_\alpha(\omega)$  exerted by the fluid on particle  $\alpha$  in terms of
Fourier harmonics of the induced surface force density  $\fb_{\alpha,
lm}(\omega)$  as follows:

\begin{eqnarray}
 \Fb_\alpha (\omega) &=& -\fb_{\alpha, 00}(\omega),  \\
 \Tb_\alpha (\omega) &=& -\frac{a_\alpha}{\sqrt{3}} \sum\limits_{m = -1}^1
  \left( \eb_m \times \fb_{\alpha, 1m}(\omega) \right),
\end{eqnarray}

\noindent  where

\begin{equation}
 \eb_0 \equiv \eb_z,  \qquad  \eb_{\pm 1} = \frac{1}{\sqrt{2}}
 \left(i\eb_y \pm \eb_x \right),
\end{equation}

\noindent  i.e., up to the factor  $(-1)^m,$  where  $ m = 0, \pm1,$  are the
cyclic covariant unit vectors \cite{ref.Varshalovich}, and  $\eb_x,$  $\eb_y,$
and $\eb_z$  are the Cartesian unit vectors.

Thus, the problem of determination of the forces and torques exerted by the
fluid on particles is reduced to the determination of only the zeroth and first
(with respect to  $l$) harmonics of the induced surface force densities
\cite{ref.Pien1,ref.Pien2,ref.Yosh}.

For any point of the space, its position defined by the radius vector  $\rb$
relative to the fixed system  $O$  can be represented in the form $\rb =
\Rb_\alpha + \rb_\alpha$,  where  $\Rb_\alpha$  is the radius vector that
defines the position of the center of certain sphere  $\alpha$  and
$\rb_\alpha$  is the radius vector directed from this center to the point at
hand.   Since there are  $N$  particles, there exist  $N$  different
representations for  $\rb$.  Among these representations, in the case where the
point of observation belongs to the domain occupied by the fluid, it is
convenient to take  $\rb_\alpha$  corresponding to the minimum difference
$r_\beta - a_\beta$,  where  $\beta = 1, 2, \ldots, N$,  which means the
minimum distance from the point at hand to the surface of sphere  $\alpha$.  If
the point of observation lies inside the domain occupied by sphere $\alpha$,
then $\rb_\alpha$ is the radius vector directed from the center of this sphere
to this point.

In this case, for  $r_\alpha > a_\alpha$,  the quantities
$\vb_\beta^{(S)ind}(\Rb_\alpha+\rb_\alpha,\omega)$,
$p_\beta^{(S)ind}(\Rb_\alpha+\rb_\alpha,\omega)$  and
$\vb_\beta^{(V)ind}(\Rb_\alpha+\rb_\alpha,\omega)$,
$p_\beta^{(V)ind}(\Rb_\alpha+\rb_\alpha,\omega)$,  where  $\beta = 1,
2,\ldots,N,$  are the velocities and the pressures of the fluid at the point
$\rb = \Rb_\alpha +\rb_\alpha$  induced, respectively, by the surface and
volume forces distributed over the surface and inside the volume of particle
$\beta$  with the center defined by the vector  $\Rb_\beta$.  The particular
case  $\beta = \alpha$  corresponds to the velocities and the pressures of the
fluid generated at this point, respectively, by the surface and volume forces
distributed over the surface and inside the volume of the particle that is the
most closely situated to this point.  This enables us to separate contributions
to the induced velocity and pressure of the fluid caused by the closest
particle and more distant ones.

At any point  $\rb$  of the space, we can represent the quantities defined by
relations (3.12), (3.13), (3.15), (3.17), and (3.18) as expansions in the
spherical harmonics

\begin{eqnarray}
 \vb_\beta^{(S)ind}(\Rb_\alpha+\rb_\alpha,\omega) &=& 2\sqrt{\pi} \,
  \sum\limits_{lm} \vb_{\beta, lm}^{(S)ind}(\Rb_\alpha,r_\alpha,\omega)
  Y_{lm}(\theta_\alpha,\varphi_\alpha), \\
 p_\beta^{(S)ind}(\Rb_\alpha+\rb_\alpha,\omega) &=& 2\sqrt{\pi} \,
  \sum\limits_{lm} p_{\beta, lm}^{(S)ind}(\Rb_\alpha,r_\alpha,\omega)
  Y_{lm}(\theta_\alpha,\varphi_\alpha), \\
 \vb_\beta^{(r)ind}(\Rb_\alpha+\rb_\alpha,\omega) &=& 2\sqrt{\pi} \,
  \sum\limits_{lm} \vb_{\beta, lm}^{(r)ind}(\Rb_\alpha,r_\alpha,\omega)
  Y_{lm}(\theta_\alpha,\varphi_\alpha), \\
 \vb_\beta^{(sol)ind}(\Rb_\alpha+\rb_\alpha,\omega) &=& 2\sqrt{\pi} \,
  \sum\limits_{lm} \vb_{\beta, lm}^{(sol)ind}(\Rb_\alpha,r_\alpha,\omega)
  Y_{lm}(\theta_\alpha,\varphi_\alpha), \\
 p_\beta^{(V)ind}(\Rb_\alpha+\rb_\alpha,\omega) &=& 2\sqrt{\pi} \,
  \sum\limits_{lm} p_{\beta, lm}^{(V)ind}(\Rb_\alpha,r_\alpha,\omega)
  Y_{lm}(\theta_\alpha,\varphi_\alpha),
\end{eqnarray}

\noindent  where  $\rb_\alpha \equiv (r_\alpha,\theta_\alpha,\varphi_\alpha)$
is the radius vector from the center of the particle closets to the considered
point if  $r_\alpha \ge a_\alpha$  or the radius vector from the center of the
particle to the point of observation lying inside this particle if $r_\alpha <
a_\alpha$ and $\theta_\alpha$ and $\varphi_\alpha$ are, respectively, the polar
and azimuth angles of this vector in the local spherical coordinate system
$O_\alpha$.  The expansion coefficients in series (3.24)--(3.28) of the
corresponding quantities in the spherical harmonics
$Y_{lm}(\theta_\alpha,\varphi_\alpha)$  can be presented in the following form:
For  $\beta  \neq  \alpha$,  we have

\begin{eqnarray}
 \vb_{\beta, l_1 m_1}^{(S)ind}(\Rb_\alpha,r_\alpha,\omega) &=&
  \sum\limits_{l_2 m_2}
  \Tb_{l_1 m_1}^{l_2 m_2}(r_\alpha,a_\beta,\Rb_{\alpha\beta},\omega)
  \cdot \fb_{\beta,l_2 m_2}(\omega),  \\
 p_{\beta, l_1 m_1}^{(S)ind}(\Rb_\alpha,r_\alpha,\omega) &=&
  \sum\limits_{l_2 m_2}
  \Db_{l_1 m_1}^{l_2 m_2}(r_\alpha,a_\beta,\Rb_{\alpha\beta})
  \cdot \fb_{\beta,l_2 m_2}(\omega),  \\
 \vb_{\beta, l_1 m_1}^{(r)ind}(\Rb_\alpha,r_\alpha,\omega) &=&
  \frac{2}{3} \xi_\beta b_\beta^2 \sum\limits_{l_2 m_2} (-1)^{l_2}
  P_{l_1 2,l_2}(r_\alpha,a_\beta,R_{\alpha\beta},\omega)
  \left(\Omb_\beta(\omega) \times
  \Wb_{l_1 m_1,00}^{l_2 m_2} \right) \nonumber \\
  & \times & Y_{l_2 m_2}^*(\Theta_{\alpha\beta},\Phi_{\alpha\beta}),  \\
 \vb_{\beta, l_1 m_1}^{(sol)ind}(\Rb_\alpha,r_\alpha,\omega) &=&
  -2 \sqrt{\pi} \int \!\! d\rb_\beta \Theta(a_\beta - r_\beta)
  \sum\limits_{l_2 m_2}
  \Tb_{l_1 m_1}^{l_2 m_2}(r_\alpha,r_\beta,\Rb_{\alpha\beta},\omega) \nonumber\\
  & \cdot &\Fb^{(sol)ext}(\Rb_\beta + \rb_\beta,\omega)
  Y_{l_2 m_2}^*(\theta_{\beta},\varphi_{\beta}),  \\
 p_{\beta, l_1 m_1}^{(V)ind}(\Rb_\alpha,r_\alpha,\omega) &=&
  2 \sqrt{\pi} \int \!\! d\rb_\beta \Theta(a_\beta - r_\beta)
  \sum\limits_{l_2 m_2}
  \Db_{l_1 m_1}^{l_2 m_2}(r_\alpha,r_\beta,\Rb_{\alpha\beta}) \nonumber \\
  & \cdot & \Fb^{(sol)ext}(\Rb_\beta + \rb_\beta,\omega)
  Y_{l_2 m_2}^*(\theta_{\beta},\varphi_{\beta}),
\end{eqnarray}

\noindent  where  $\Rb_{\alpha\beta} = \Rb_\alpha - \Rb_\beta \equiv
(R_{\alpha\beta},\Theta_{\alpha\beta},\Phi_{\alpha\beta})$  is the vector
between the centers of spheres  $\alpha$  and  $\beta$  pointing from sphere
$\beta$  to sphere  $\alpha$  in the spherical coordinate system  $O$,

\begin{eqnarray}
 \Tb_{l_1 m_1}^{l_2 m_2}(r_\alpha,r_\beta,\Rb_{\alpha\beta},\omega) &=&
  \sum\limits_{lm}
  \Tb_{l_1 m_1,lm}^{l_2 m_2}(r_\alpha,r_\beta,R_{\alpha\beta},\omega)
  Y_{lm}(\Theta_{\alpha\beta},\Phi_{\alpha\beta}),  \\
 \Tb_{l_1 m_1,lm}^{l_2 m_2} (r_\alpha,r_\beta,R_{\alpha\beta},\omega) &=&
  F_{l_1 l_2,l}(r_\alpha,r_\beta,R_{\alpha\beta},\omega)
  \Kb_{l_1 m_1,lm}^{l_2 m_2}, \\
 F_{l_1 l_2,l}(r_\alpha,r_\beta,R_{\alpha\beta},\omega) &=&
  \frac{2}{\pi \eta} \int\limits_0^\infty \!\! dk\, \frac{k^2}{k^2 + \kappa^2}
  \, j_{l_1}(kr_\alpha) j_{l_2}(kr_\beta) j_l(kR_{\alpha\beta}),
\end{eqnarray}

\begin{equation}
 \Kb_{l_1 m_1, lm}^{l_2 m_2} =
  i^{l_1 - l_2 + l} \int \!\! d\Omega_k \, (\Ib - \nb_k \nb_k)\,
  Y_{l_1 m_1}^*(\theta_k,\varphi_k) Y_{l_2 m_2}(\theta_k,\varphi_k)
  Y_{lm}^*(\theta_k,\varphi_k),
\end{equation}

\begin{eqnarray}
 \Db_{l_1 m_1}^{l_2 m_2}(r_\alpha,r_\beta,\Rb_{\alpha\beta}) &=&
  \sum\limits_{lm} \Db_{l_1 m_1,lm}^{l_2 m_2}(r_\alpha,r_\beta,R_{\alpha\beta})
  Y_{lm}(\Theta_{\alpha\beta},\Phi_{\alpha\beta}),  \\
 \Db_{l_1 m_1,lm}^{l_2 m_2} (r_\alpha,r_\beta,R_{\alpha\beta}) &=&
  C_{l_1 l_2,l}(r_\alpha,r_\beta,R_{\alpha\beta}) \Wb_{l_1 m_1,lm}^{l_2 m_2}, \\
 C_{l_1 l_2,l}(r_\alpha,r_\beta,R_{\alpha\beta}) &=& \frac{2}{\pi}
  \int\limits_0^\infty \!\! dk\,
  k \, j_{l_1}(kr_\alpha) j_{l_2}(kr_\beta) j_l(kR_{\alpha\beta}), \\
 \Wb_{l_1 m_1,lm}^{l_2 m_2} &=&
  i^{l_1 - l_2 + l - 1} \int \!\! d\Omega_k \, \nb_k \,
  Y_{l_1 m_1}^*(\theta_k,\varphi_k) Y_{l_2 m_2}(\theta_k,\varphi_k)
  Y_{lm}^*(\theta_k,\varphi_k),  \\
 P_{l_1 l_2,l}(r_\alpha,a_\beta,R_{\alpha\beta},\omega) &=&
  \frac{2}{\pi \eta} \int\limits_0^\infty \!\! dk\, \frac{k}{k^2 + \kappa^2}
  \, j_{l_1}(kr_\alpha) j_{l_2}(ka_\beta) j_l(kR_{\alpha\beta}).
\end{eqnarray}

The quantities  $\Kb_{l_1 m_1, lm}^{l_2 m_2}$  and  $\Wb_{l_1 m_1, lm}^{l_2
m_2}$  defined by relations (3.37) and (3.41) can be represented in the explicit
form in terms of the Wigner 3j-symbols \cite{ref.Varshalovich,ref.Davydov}.
(For the tensor  $\Kb_{l_1 m_1, lm}^{l_2 m_2}$,  this representation is given in
\cite{ref.Yosh}.)  The corresponding relations have the form

\begin{eqnarray}
 \Kb_{l_1 m_1, lm}^{l_2 m_2} &=& \frac{i^{l_1 - l_2 + l}}{3\sqrt{\pi}}
  \sqrt{(2l_1 +1)(2l_2 +1)(2l + 1)} \,(-1)^{m_1 + m}
  \left \{
   \left ( \begin{array}{rrr}
   l_1  &  l_2  &  l  \\
   0  &  0  &  0  \\
   \end{array} \right)
  \left ( \begin{array}{rrr}
   l_1  &  l_2  &  l  \\
  -m_1  &  m_2  &  m  \\
  \end{array} \right)
 \Ib \right. \nonumber  \\
 &-& (-1)^m \frac{1}{2} \sqrt{\frac{3}{2}} \sum\limits_{k = -2}^2
 (-1)^k \Kb_k \sum\limits_{j = j_{min}}^1 (2L + 1)
  \left ( \begin{array}{rrr}
  2  &  l  &  L  \\
  0  &  0  &  0  \\
  \end{array} \right)
  \left ( \begin{array}{rrr}
  l_1  &  l_2  &  L  \\
  0    &  0    &  0  \\
  \end{array} \right)  \nonumber  \\
 & \times & \left.
  \left ( \begin{array}{rrl}
  2  &   l  &  L  \\
  k  &  -m  &  m - k  \\
  \end{array} \right)
  \left ( \begin{array}{rrl}
   l_1   &  l_2  &  L  \\
  -m_1   &  m_2  &  k - m  \\
  \end{array} \right)
 \right \},  \qquad \qquad  l_1,l_2,l \geq 0,
\end{eqnarray}

\noindent  where

\begin{eqnarray}
 L = l + 2j,  \qquad \qquad
  j_{min} = \left \{
  \begin{array}{rll}
   1,  \quad  &  \mbox{if}  &  \quad l =    0  \\
   0,  \quad  &  \mbox{if}  &  \quad l =    1  \\
  -1,  \quad  &  \mbox{if}  &  \quad l \geq 2, \\
  \end{array}  \right.  \nonumber
\end{eqnarray}

\begin{eqnarray}
 \Kb_0 &=& \sqrt{\frac{2}{3}} \left (-\eb_x \eb_x - \eb_y \eb_y + 2\eb_z\eb_z
  \right) = \sqrt{\frac{2}{3}} \left (\eb_1 \eb_{-1} + \eb_{-1} \eb_1
  + 2 \eb_0 \eb_0 \right),  \nonumber  \\
 \Kb_1 &=& \eb_x \eb_z + \eb_z \eb_x -i\left (\eb_y \eb_z + \eb_z \eb_y \right)
  = -\sqrt{2} \left (\eb_{-1} \eb_0 + \eb_0 \eb_{-1} \right), \nonumber  \\
 \Kb_{-1} &=& -\Kb_1^* = -\sqrt{2} \left (\eb_1 \eb_0 + \eb_0 \eb_1 \right), \\
  \nonumber
 \Kb_2 &=& \eb_x \eb_x - \eb_y \eb_y - i\left (\eb_x \eb_y + \eb_y \eb_x \right)
   = 2 \eb_{-1} \eb_{-1},  \nonumber  \\
 \Kb_{-2} &=& \Kb_2^* = 2 \eb_1 \eb_1,  \nonumber
\end{eqnarray}

\noindent  and

\begin{eqnarray}
 \Wb_{l_1 m_1, lm}^{l_2 m_2} &=& \frac{i^{l_1 - l_2 + l - 1}}{2\sqrt{\pi}}
  \sqrt{(2l_1 +1)(2l_2 +1)(2l + 1)} \,(-1)^{m - m_2}
  \sum\limits_{k = -1}^1 \eb_k^* \sum\limits_{j = j_{min}}^1 (2L + 1)
  \nonumber  \\
   & \times &\left ( \begin{array}{rrr}
   1  &  l  &  L  \\
   0  &  0  &  0  \\
   \end{array} \right)
   \left ( \begin{array}{rrr}
   l_1  &  l_2  &  L  \\
   0    &  0    &  0  \\
   \end{array} \right)
   \left ( \begin{array}{rrl}
   1  &   l  &  L  \\
   k  &  -m  &  m - k  \\
   \end{array} \right)
   \left ( \begin{array}{rrl}
    l_1  &  l_2  &  L  \\
   -m_1  &  m_2  &  k - m  \\
   \end{array} \right), \nonumber \\
   & & \qquad \qquad \qquad \qquad \qquad \qquad \qquad \qquad \qquad
   \qquad \qquad  l_1,l_2,l \geq 0.
\end{eqnarray}

\noindent  where

\begin{eqnarray}
 L = l + 2j -1,  \qquad \qquad
  j_{min} = \left \{
  \begin{array}{rll}
   1,  \quad  &  \mbox{if}  &  l =    0  \\
   0,  \quad  &  \mbox{if}  &  l \geq 1.  \\
  \end{array}  \right.  \nonumber
\end{eqnarray}

Relation (3.43) coincides with relation (28) given in  \cite{ref.Yosh} up to
the factor  $(-1)^{(m_1 - |m_1| + m_2 - |m_2| + m - |m|)/2}$,  which is caused
by the different definition of the spherical harmonics $Y_{lm}(\theta,\varphi)$
for  $m < 0$  in the present paper and in \cite{ref.Yosh}.

In view of the properties of 3j-symbols, we have

\begin{equation}
 \Kb_{l_1 m_1, lm}^{l_2 m_2} \neq 0 \qquad  \mbox{only for}  \qquad
 l = l_1 + l_2 - 2p,  \qquad  l \geq 0,
\end{equation}

\noindent  where  $p = -1,0,1,\ldots,p_{max},$  $p_{max} = \min
\Bigl(\left[(l_1 + l_2)/2\right], 1 + \min (l_1, l_2) \Bigr)$, $\left[a\right]$
is the integer part of  $a$,  and  $\min (a,b)$ means the smallest quantity of
$a$ and  $b$, and

\begin{equation}
 \Wb_{l_1 m_1, lm}^{l_2 m_2} \neq 0 \qquad  \mbox{only for}  \qquad
 l = l_1 + l_2 - 2p + 1,  \qquad  l \geq 0,
\end{equation}

\noindent  where  $p = 0,1,\ldots,\tilde{p}_{max},$  $\tilde{p}_{max} = \min
\Bigl(\left[(l_1 + l_2 + 1)/2\right], 1 + \min (l_1, l_2) \Bigr)$.

According to (3.46) and (3.47), the sums over  $l$  in relations (3.34) and
(3.38) contain only terms with  $l$  satisfying these conditions.  Therefore,
it is necessary to investigate the quantities
$F_{l_1 l_2,l}(r_\alpha,r_\beta,R_{\alpha\beta},\omega)$  and
$C_{l_1 l_2,l}(r_\alpha,r_\beta,R_{\alpha\beta})$  only for the values of
$l$  given by (3.46) and (3.47), respectively.  According to relations (3.29),
(3.30) and (3.32), (3.33), the corresponding quantities in them are defined by
general relations (3.34)--(3.41) for  $r_\beta = a_\beta$  and  $0 \leq r_\beta
\leq a_\beta$,  respectively.

In the particular case  $l = 0$,  the general relations (3.43) and (3.45) for
the quantities  $\Kb_{l_1 m_1, lm}^{l_2 m_2}$  and  $\Wb_{l_1 m_1, lm}^{l_2
m_2}$ are simplified to the form

\begin{eqnarray}
 \Kb_{l_1 m_1, 00}^{l_2 m_2} &=&
  \frac{i^{l_1 - l_2}}{2 \sqrt{\pi}} \int \!\! d\Omega_k \, (\Ib - \nb_k
  \nb_k)\,
  Y_{l_1 m_1}^*(\theta_k,\varphi_k) Y_{l_2 m_2}(\theta_k,\varphi_k) \nonumber \\
  &=& \frac{1}{3\sqrt{\pi}} \left \{ \delta_{l_1, l_2} \, \delta_{m_1, m_2} \, \Ib
  - \sqrt{\frac{3 (2l_1 + 1)(2l_2 + 2)}{8}} \, i^{l_1 - l_2} (-1)^{m_1}
   \left ( \begin{array}{rrr}
   l_1  &  l_2  &  2  \\
    0  &    0  &  0  \\
   \end{array} \right)
  \right.  \nonumber  \\
  &\times & \left. \sum\limits_{k = -2}^2 \delta_{k, m_1 - m_2}
   \left ( \begin{array}{rrr}
    l_1  &  l_2  &  2  \\
   -m_1  &  m_2  &  k  \\
   \end{array} \right)
  \Kb_k \right \}
\end{eqnarray}

\noindent  and

\begin{eqnarray}
 \Wb_{l_1 m_1, 00}^{l_2 m_2} &=&
  \frac{i^{l_1 - l_2 - 1}}{2 \sqrt{\pi}} \int \!\! d\Omega_k \, \nb_k \,
  Y_{l_1 m_1}^*(\theta_k,\varphi_k) Y_{l_2 m_2}(\theta_k,\varphi_k) \nonumber \\
  &=& \frac{i^{l_1 - l_2 - 1}}{2\sqrt{\pi}}
  \sqrt{(2l_1 + 1)(2l_2 + 2)} (-1)^{m_1}
   \left ( \begin{array}{rrr}
   l_1  &  l_2  &  1  \\
    0  &    0  &  0   \\
   \end{array} \right) \nonumber  \\
  &\times& \sum\limits_{k = -1}^1 \delta_{k, m_1 - m_2}
   \left ( \begin{array}{rrr}
    l_1  &  l_2  &  1  \\
   -m_1  &  m_2  &  k  \\
   \end{array} \right)
   \eb_k^*.
\end{eqnarray}

Taking the properties of 3j-symbols into account, we obtain that

\begin{equation}
 \Kb_{l_1 m_1, 00}^{l_2 m_2} \neq 0 \quad  \mbox{only for}  \quad
 l_2 = l_1 + 2p \geq 0, \quad \mbox{where} \quad
  p = \left \{
  \begin{array}{rrl}
   0, 1,     \quad  &  \mbox{if}  &  l_1 =    0,1  \\
   0, \pm 1, \quad  &  \mbox{if}  &  l_1 \geq 2,
  \end{array}  \right.
\end{equation}

\noindent  and

\begin{equation}
 \Wb_{l_1 m_1, 00}^{l_2 m_2} \neq 0 \quad  \mbox{only for}  \quad
 l_2 = l_1 + 2p +1 \geq 0,  \quad \mbox{where} \quad
  p = \left \{
  \begin{array}{rrl}
   0,     \quad  &  \mbox{if}  &  l_1 =    0,1  \\
   0, -1, \quad  &  \mbox{if}  &  l_1 \geq 1.
  \end{array}  \right.
\end{equation}

According to (3.51), the sum over  $l_2$  in relation (3.31) contains only terms
for  $l_2 = l_1 \pm 1 \geq 0$.  This means that the quantity
$P_{l_1 2,l_2}(r_\alpha,a_\beta,R_{\alpha\beta},\omega)$  contained in this
relation must be determined only for these values of  $l_2$.

In the domains  $r_\alpha \leq R_{\alpha\beta} - r_\beta$  and  $r_\alpha \geq
R_{\alpha\beta} + r_\beta$  (far from both particles), where, according to
(3.32) and (3.33),  $r_\beta \leq a_\beta$,  the integrals in (3.36) and (3.40)
that define the quantities
$F_{l_1l_2,l}(r_\alpha,r_\beta,R_{\alpha\beta},\omega)$  and
$C_{l_1l_2,l}(r_\alpha,r_\beta,R_{\alpha\beta})$  can be represented in the
explicit form for the values of  $l$  given by conditions (3.46) and (3.47),
respectively, (for details of calculation, see Appendix).  For $r_\alpha \leq
R_{\alpha\beta} - r_\beta$,  where  $0 \leq r_\beta \leq a_\beta$,  we have

\begin{eqnarray}
 F_{l_1 l_2,l}(r_\alpha,r_\beta,R_{\alpha\beta},\omega) &=&
  (-1)^p \frac{2\kappa}{\pi\eta} \left\{ \tilde{j}_{l_1}(x_\alpha)
  \tilde{j}_{l_2}(x_\beta) \tilde{h}_l(y_{\alpha\beta})
  - \delta_{l, l_1 + l_2 + 2}
  \frac{\pi^{3/2}}{2}
  \frac{\Gamma\left(l_1 + l_2 + \frac{5}{2} \right)}
  {\Gamma\left(l_1 + \frac{3}{2} \right)\Gamma\left(l_2 + \frac{3}{2} \right)}
  \right. \nonumber \\
  &\times &  \left. \frac{r_\alpha^{l_1} r_\beta^{l_2}}
  {y_{\alpha\beta}^3 R_{\alpha\beta}^{l_1 + l_2}} \right \},
  \quad  l = l_1 + l_2 -2p \geq 0,  \quad  p = -1, 0, 1,\ldots, p_{max},  \\
 C_{l_1 l_2,l}(r_\alpha,r_\beta,R_{\alpha\beta}) &=&
  \delta_{l, l_1 + l_2 + 1} \frac{\sqrt{\pi}}{2}
  \frac{\Gamma\left(l_1 + l_2 + \frac{3}{2} \right)}
  {\Gamma\left(l_1 + \frac{3}{2} \right)\Gamma\left(l_2 + \frac{3}{2} \right)}
  \frac{r_\alpha^{l_1} r_\beta^{l_2}}{R_{\alpha\beta}^{l_1 + l_2 +2}},
  \nonumber \\
  && \qquad \qquad \qquad l = l_1 + l_2 - 2p + 1 \geq 0,
  \qquad  p = 0, 1,\ldots, \tilde{p}_{max},
\end{eqnarray}

\noindent  where  $\tilde{j}_l(x) = \sqrt{\pi/(2x)} I_{l +\frac{1}{2}}(x)$  and
$\tilde{h}_l(x) = \sqrt{\pi/(2x)} K_{l +\frac{1}{2}}(x)$  are the modified
spherical Bessel functions of the first and third kind, respectively,
\cite{ref.Abram},  $\Gamma(z)$  is the gamma function,  $x_\alpha = \kappa
r_\alpha,$  $x_\beta = \kappa r_\beta,$  and  $y_{\alpha\beta} = \kappa
R_{\alpha\beta}$.  The dimensionless parameter  $y_{\alpha\beta} = \kappa
R_{\alpha\beta}$  [$|y_{\alpha\beta}| = \sqrt{2}\,R_{\alpha\beta}/\delta$]
characterizes the ratio of the distance between particles  $\alpha$  and
$\beta$  to the depth of penetration  $\delta$  of a plane transverse wave of
frequency $\omega$  into the fluid.  In a certain frequency range, the quantity
$|y_{\alpha\beta}|$  can be both smaller (for short distances between
particles) and greater (for space-apart particles) than one.

Note that relations (3.52) and (3.53) for $F_{l_1
l_2,l}(r_\alpha,r_\beta,R_{\alpha\beta},\omega)$  and $C_{l_1
l_2,l}(r_\alpha,r_\beta,R_{\alpha\beta})$  defining the induced fluid velocity
and pressure were obtained without imposing any additional restrictions on the
size of particles, distances between them, and the frequency range.

In the particular case of equal spheres ($a_\alpha \equiv a, \quad \alpha =
1,2,\ldots,N$), if the observation point lies at the surface of a sphere
($r_\alpha = a$) and  $r_\beta = a$,  relation (3.52) for  $l = l_1 + l_2 -2p,$
where  $p = 0, 1,\ldots,p_{max}$  coincides with the corresponding relation
(4.11) in \cite{ref.Pien2} up to the factor  $(-1)^p$.  For  $l = l_1 + l_2 +
2$,  relation (3.52) agrees with (4.11) in \cite{ref.Pien2} up to the factor
$(l_1 + l_2 + 5/2)$  in the first and second terms in (4.11) and the omitted
factor  $1/\kappa!$  (in terms of the notation used in \cite{ref.Pien2}) in the
third term presented in (4.11) in \cite{ref.Pien2} as the infinite series.

For  $r_\alpha \geq R_{\alpha\beta} + r_\beta$,  where $0 \leq r_\beta \leq
a_\beta$,  we have

\begin{eqnarray}
 F_{l_1 l_2,l}(r_\alpha,r_\beta,R_{\alpha\beta},\omega) &=&
  (-1)^{l_{2} - p} \, \frac{2\kappa}{\pi\eta} \Biggl \{ \tilde{h}_{l_1}(x_\alpha)
  \tilde{j}_{l_2}(x_\beta) \tilde{j}_l(y_{\alpha\beta})
  - \delta_{l, l_1 - l_2 - 2}
  \frac{\pi^{3/2}}{2}  \nonumber \\
  &\times & \frac{\Gamma\left(l_1 + \frac{1}{2} \right)}
  {\Gamma\left(l_2 + \frac{3}{2} \right)
  \Gamma\left(l_1 - l_2 - \frac{1}{2} \right)}
  \frac{r_\beta^{l_2} R_{\alpha\beta}^l}
  {x_\alpha^3 r_\alpha^{l_1 - 2}} \Biggr \}, \nonumber \\
  && \qquad \qquad \qquad l = l_1 + l_2 - 2p \geq 0,
  \qquad  p = -1, 0, 1,\ldots, p_{max},
\end{eqnarray}

\begin{eqnarray}
 C_{l_1 l_2,l}(r_\alpha,r_\beta,R_{\alpha\beta}) &=&
  \frac{\sqrt{\pi}}{2} \frac{\Gamma\left(l_1 + \frac{1}{2} \right)}
  {\Gamma\left(l_2 + \frac{3}{2} \right)}
  \frac{r_\beta^{l_2} R_{\alpha\beta}^{l_1 - l_2 - 1}}
  {r_\alpha^{l_1 + 1}} \left \{ \delta_{l, l_1 - l_2 - 1} \,
  \frac{1}{\Gamma\left(l_1 - l_2 + \frac{1}{2} \right)}
  +  \delta_{l, l_1 - l_2 - 3} \right. \nonumber  \\
  &\times & \left. \frac{2}{\Gamma\left(l_1 - l_2 - \frac{3}{2} \right)}
  \left [ \frac{1}{2l_1 - 1} \left (\frac{r_\alpha}{R_{\alpha\beta}} \right )^2
  - \frac{1}{2l_2 + 3}
  \left (\frac{r_\beta}{R_{\alpha\beta}} \right )^2
  - \frac{1}{2l + 3} \right ] \right \}
  \nonumber \\
  && \qquad \qquad \qquad l = l_1 + l_2 - 2p + 1 \geq 0,
  \qquad  p = 0, 1,\ldots, \tilde{p}_{max}.
\end{eqnarray}

Note that the quantities $F_{l_1
l_2,l}(r_\alpha,r_\beta,R_{\alpha\beta},\omega)$  and $C_{l_1
l_2,l}(r_\alpha,r_\beta,R_{\alpha\beta})$  defined by (3.54) and (3.55) for
$r_\alpha \geq R_{\alpha\beta} + r_\beta$  can also be represented in the form
(3.52) and (3.53), which are valid for  $r_\alpha \leq R_{\alpha\beta} -
r_\beta$,  performing the changes  $r_\alpha \leftrightarrow R_{\alpha\beta}$
and  $l_1 \leftrightarrow l$  and putting, respectively, $l_1 = l + l_2 - 2s$,
where  $s = -1,0,1,\ldots,s_{max}$  and $s_{max} = \min\Bigl(\left[(l +
l_2)/2\right], 1 + \min (l, l_2) \Bigr)$,  and  $l_1 = l + l_2 + 1 - 2s$, where
$s = 0,1,\ldots,\tilde{s}_{max}$  and $\tilde{s}_{max} = \min\Bigl(\left[(l +
l_2 + 1)/2\right], 1 + \min (l, l_2) \Bigr)$.

The dimensionless parameter  $\sigma_{\beta\alpha} = a_\beta/R_{\alpha\beta}$
(as well as  $\sigma_{\alpha\beta} = a_\alpha/R_{\alpha\beta}$), which is the
ratio of the radius of particles to the distance between them, is always
smaller than one (for spheres of equal radii, it cannot be greater than $1/2$).
In diluted suspensions,  $\sigma_{\beta\alpha} \ll 1$.

For  $l = l_1 \pm 1 \geq 0$,  the quantity
$P_{l_12,l}(r_\alpha,a_\beta,R_{\alpha\beta},\omega)$  defined by relation
(3.42) with  $l_2 = 2$  can be represented as follows (see Appendix):

\begin{equation}
 P_{l_1 2,l}(r_\alpha,a_\beta,R_{\alpha\beta},\omega) = (-1)^p
  \frac{2}{\pi \eta} \, \tilde{j}_{l_1}(x_\alpha)
  \tilde{j}_2(b_\beta) \tilde{h}_l(y_{\alpha\beta})
\end{equation}

\noindent  for  $r_\alpha \leq R_{\alpha\beta} - a_\beta$  and

\begin{equation}
 P_{l_1 2,l}(r_\alpha,a_\beta,R_{\alpha\beta},\omega) = (-1)^{p + 1}
  \frac{2}{\pi \eta} \, \tilde{h}_{l_1}(x_\alpha)
  \tilde{j}_2(b_\beta) \tilde{j}_l(y_{\alpha\beta})
\end{equation}

\noindent  for  $r_\alpha \geq R_{\alpha\beta} + a_\beta$.  Here,  $b_\beta =
\kappa a_\beta$.

In the particular case $\beta = \alpha$, we have

\begin{eqnarray}
 \vb_{\alpha, l_1 m_1}^{(S)ind}(\Rb_\alpha,r_\alpha,\omega) &=&
  \sum\limits_{l_2 m_2}
  \Tb_{l_1 m_1}^{l_2 m_2}(r_\alpha,a_\alpha,\omega) \cdot
  \fb_{\alpha,l_2 m_2}(\omega),  \\
 p_{\alpha, l_1 m_1}^{(S)ind}(\Rb_\alpha,r_\alpha,\omega) &=&
  \sum\limits_{l_2 m_2}
  \Db_{l_1 m_1}^{l_2 m_2}(r_\alpha,a_\alpha) \cdot
  \fb_{\alpha,l_2 m_2}(\omega),  \\
 \vb_{\alpha, l_1 m_1}^{(r)ind}(\Rb_\alpha,r_\alpha,\omega) &=&
  \delta_{l_1, 1} \, \frac{\xi_\alpha b_\alpha^2}{6\pi \sqrt{3}}
  P_{12}(r_\alpha,a_\alpha,\omega)
  \Bigl(\Omb_\alpha(\omega) \times \eb_{m_1}^* \Bigr),  \\
 \vb_{\alpha, l_1 m_1}^{(sol)ind}(\Rb_\alpha,r_\alpha,\omega) &=&
  -\int \!\! d\rb_\alpha^{\,\prime} \Theta(a_\alpha - r_\alpha^{\,\prime})
  \sum\limits_{l_2 m_2}
  \Tb_{l_1 m_1}^{l_2 m_2}(r_\alpha,r_\alpha^{\,\prime},\omega) \nonumber\\
  & \cdot &\Fb^{(sol)ext}(\Rb_\alpha + \rb_\alpha^{\,\prime},\omega)
  Y_{l_2 m_2}^*(\theta_\alpha^{\,\prime},\varphi_\alpha^{\,\prime}),  \\
 p_{\alpha, l_1 m_1}^{(V)ind}(\Rb_\alpha,r_\alpha,\omega) &=&
  \int \!\! d\rb_\alpha^{\,\prime} \Theta(a_\alpha - r_\alpha^{\,\prime})
  \sum\limits_{l_2 m_2}
  \Db_{l_1 m_1}^{l_2 m_2}(r_\alpha,r_\alpha^{\,\prime}) \nonumber \\
  & \cdot & \Fb^{(sol)ext}(\Rb_\alpha + \rb_\alpha^{\,\prime},\omega)
  Y_{l_2 m_2}^*(\theta_\alpha^{\,\prime},\varphi_\alpha^{\,\prime}),
\end{eqnarray}

\noindent  where

\begin{eqnarray}
 \Tb_{l_1 m_1}^{l_2 m_2}(r_\alpha,r_\alpha^{\,\prime},\omega) &=&
  \frac{1}{2 \sqrt{\pi}} \, F_{l_1 l_2}(r_\alpha,r_\alpha^{\,\prime},\omega)
  \Kb_{l_1 m_1,00}^{l_2 m_2}, \\
 F_{l_1 l_2}(r_\alpha,r_\alpha^{\,\prime},\omega) &=&
  \frac{2}{\pi \eta} \int\limits_0^\infty \!\! dk\, \frac{k^2}{k^2 + \kappa^2}
  \, j_{l_1}(kr_\alpha) j_{l_2}(k r_\alpha^{\,\prime}), \\
 \Db_{l_1 m_1}^{l_2 m_2}(r_\alpha,r_\alpha^{\,\prime}) &=&
  \frac{1}{2 \sqrt{\pi}} \, C_{l_1 l_2}(r_\alpha,r_\alpha^{\,\prime})
  \Wb_{l_1 m_1,00}^{l_2 m_2}, \\
 C_{l_1 l_2}(r_\alpha,r_\alpha^{\,\prime}) &=&
  \frac{2}{\pi} \int\limits_0^\infty \!\! dk\,
  k \, j_{l_1}(kr_\alpha) j_{l_2}(k r_\alpha^{\,\prime}), \\
 P_{l_1 l_2}(r_\alpha,a_\alpha,\omega) &=&
  \frac{2}{\pi \eta} \int\limits_0^\infty \!\! dk\, \frac{k}{k^2 + \kappa^2}
  \, j_{l_1}(kr_\alpha) j_{l_2}(ka_\alpha).
\end{eqnarray}

Taking into account that according to (3.60) only one harmonic of $\vb_{\alpha,
l_1 m_1}^{(r)ind}(\Rb_\alpha,r_\alpha,\omega)$  is not equal to zero, we can
represent this component as follows:

\begin{equation}
 \vb_\alpha^{(r)ind}(\Rb_\alpha + \rb_\alpha,\omega) =
  \frac{\xi_\alpha b_\alpha^2}{6\pi} \, P_{1,2}(r_\alpha,a_\alpha,\omega)
  \Bigl( \Omb_\alpha(\omega) \times \nb_\alpha \Bigr),
\end{equation}

\noindent  where  $\nb_\alpha = \rb_\alpha/r_\alpha$.

According to (3.50) and (3.51), the quantities  $\Kb_{l_1 m_1, 00}^{l_2 m_2}$
and  $\Wb_{l_1 m_1, 00}^{l_2 m_2}$  are nonzero only for at most three and two
values of  $l_2$,  respectively.  Therefore, the infinite sums
$\sum\limits_{l_2 = 0}^\infty$  in relations (3.58), (3.61) and (3.59), (3.62)
are replaced by sums containing, respectively, at most three and two terms
corresponding to different values of  $l_2$.  This means that the quantities
$F_{l_1 l_2}(r_\alpha,r_\alpha^{\,\prime},\omega)$  and
$C_{l_1 l_2}(r_\alpha,r_\alpha^{\,\prime})$  should be determined only for these
values of  $l_2$.

Finally, for  $r_\alpha \geq a_\alpha$  and  $0 \leq r_\alpha^{\,\prime} \leq
a_\alpha$,  we can represent these quantities in the form (for details of
calculation, see Appendix):

\begin{eqnarray}
 F_{l_1 l_2}(r_\alpha,r_\alpha^{\,\prime},\omega) =
  (-1)^p \frac{2\kappa}{\pi \eta} \left \{
  \tilde{h}_{l_1}(x_\alpha) \tilde{j}_{l_2}(x_\alpha^{\,\prime})
  - \delta_{l_2, l_1 -2} \left (l_1 - \frac{1}{2} \right) \frac{\pi}{x_\alpha^3}
  \left( \frac{r_\alpha^{\,\prime}}{r_\alpha} \right)^{l_1 -2} \right \},
  \nonumber \\
   \qquad \qquad \qquad \qquad l_2 = l_1 + 2p \geq 0,  \quad p = 0, \pm 1, \\
 C_{l_1 l_2}(r_\alpha,r_\alpha^{\,\prime}) =
  \delta_{l_2, l_1 - 1} \frac{1}{r_\alpha^2}
  \left( \frac{r_\alpha^{\,\prime}}{r_\alpha} \right)^{l_1 - 1}
  + \delta_{r_\alpha, a} \, \delta_{r_\alpha^{\,\prime}, a} \, \frac{1}{2 a_\alpha^2}
  \left( \delta_{l_2, l_1 + 1} - \delta_{l_2, l_1 - 1} \right),
  \,\, l_2 = l_1 \pm 1 \geq 0,
\end{eqnarray}

\noindent  where  $x_\alpha^{\,\prime} = \kappa r_\alpha^{\,\prime}$.

For the quantity  $P_{1,2}(r_\alpha,a_\alpha,\omega)$,  in the domain $r_\alpha
\geq a_\alpha$,  we obtain

\begin{equation}
 P_{1,2}(r_\alpha,a_\alpha,\omega) =
  \frac{2}{\pi \eta} \, \tilde{h}_1(x_\alpha) \tilde{j}_2(b_\alpha).
\end{equation}

In the particular case  $r_\alpha = a_\alpha$  (the observation point lies at
the surface of sphere  $\alpha$) and  $r_\alpha^{\,\prime} = a_\alpha$,
relations (3.69)--(3.71) are reduced to the form (see Appendix)

\begin{eqnarray}
 F_{l_1 l_2}(a_\alpha,a_\alpha,\omega) &=&
  (-1)^p \frac{2\kappa}{\pi \eta} \,
  \tilde{j}_{l_{max})}(b_\alpha) \tilde{h}_{l_{min}}(b_\alpha),
  \quad  l_2 = l_1 + 2p \geq 0,  \quad  p = 0, \pm 1,  \\
 C_{l_1 l_2}(a_\alpha,a_\alpha) &=&  \frac{1}{2 a_\alpha^2}
  \left( \delta_{l_2, l_1 + 1} + \delta_{l_2, l_1 - 1} \right),
  \quad \qquad  l_2 = l_1 \pm 1 \geq 0,  \\
 P_{1,2}(a_\alpha,a_\alpha,\omega) &=&
  \frac{2}{\pi \eta} \, \tilde{h}_1(b_\alpha) \tilde{j}_2(b_\alpha),
\end{eqnarray}

\noindent  where  $l_{max} = \max(l_1, l_2)$  and  $l_{min} = \min(l_1, l_2)$.

For  $l_2 = l_1$  and  $l_2 = l_1 + 2$,  expression (3.72) coincides with the
corresponding result obtained in \cite{ref.Pien2}.  However, the quantity
$\Tb_{l_1 m_1}^{l_2m_2}(a_\alpha,a_\alpha,\omega)$  defined by relation (3.63)
for $r_\alpha = a_\alpha$  and  $r_\alpha^{\,\prime} = a_\alpha$  is not equal
to zero for three values of  $l_2$  [$l_2 = l_1, \, l_1 + 2,$  and  $l_1 - 2$
(for $l_1 \geq 2)$], while, according to \cite{ref.Pien2}, $\Tb_{l_1
m_1}^{l_2m_2}(a_\alpha,a_\alpha,\omega) \neq 0$  only for  $l_2 = l_1$ and $l_2
= l_1 + 2$.

Relations (3.4), (3.6)--(3.8), (3.10), (3.12), (3.24), (3.26), (3.27), (3.29),
(3.31), (3.32), (3.58), (3.60) and (3.61) completely determine the function
$\vb (\rb,\omega)$ in the entire space provided that the induced surface force
densities are known.  Analogously, relations (2.8), (3.5), (3.9), (3.11),
(3.13), (3.15), (3.25), (3.28), (3.30), (3.33), (3.59), and (3.62) uniquely
reproduce the original function  $p(\rb,\omega)$  at any point of the space
with the exception of the points of the surfaces of the spheres ($r_\alpha =
a_\alpha, \quad \alpha = 1, 2,\ldots, N$) where the obtained quantity is equal
to the half-sum of the original function given at  $\rb = \Rb_\alpha +
\ab_\alpha + 0$ and $\rb = \Rb_\alpha + \ab_\alpha - 0$  because the original
function defined as the fluid pressure for  $r_\alpha \ge a_\alpha$ and (2.17)
for  $r_\alpha < a_\alpha$ is discontinuous at the surfaces  $r_\alpha =
a_\alpha$.  With regard for (2.8), (2.17), and (3.5), the fluid pressure at the
surfaces of the spheres can be represented as follows:

\begin{equation}
 p(\Rb_\alpha + \ab_\alpha + 0,\omega) = p^{(0)}(\Rb_\alpha + \ab_\alpha,\omega)
  + p^{ind}(\Rb_\alpha + \ab_\alpha + 0,\omega),
\end{equation}

\noindent  where

\begin{equation}
 p^{ind}(\Rb_\alpha + \ab_\alpha + 0,\omega) = 2 p^{ind}(\Rb_\alpha + \ab_\alpha,\omega)
  - i\omega \rho \left( \Rb_\alpha + \ab_\alpha \right) \cdot
  \Ub_\alpha (\omega)
\end{equation}

\noindent  and the quantities  $p^{(0)}(\Rb_\alpha + \ab_\alpha,\omega)$  and
$p^{ind}(\Rb_\alpha + \ab_\alpha,\omega)$  are defined by the corresponding
above-derived expressions for  $p^{(0)}(\rb,\omega)$  and $p^{ind}(\rb,\omega)$
given at $\rb = \Rb_\alpha + \ab_\alpha$.


\section{System of Equations for Harmonics of Induced Surface Force Densities}
 \label{Boundary Conditions}

In the previous section, the required distributions of the velocity and
pressure fields of the fluid as well as the forces and torques exerted by the
fluid on the particles were expressed in terms of harmonics of the induced
surface force densities.  To determine these harmonics, we use the stick
boundary conditions for the fluid velocity at the surfaces of the particles
\cite{ref.Happel,ref.Batch,ref.Landau}

\begin{equation}
 \vb(\rb,t) = \Ub_\alpha(t) + \Bigl(\Omb_\alpha(t)\times\rb_\alpha\Bigr)
  \biggr |_{\rb=\Rb_\alpha+\ab_\alpha},  \qquad \alpha = 1,2,\ldots,N.
\end{equation}

Passing in relations (4.1) to the Fourier transform with respect to the
frequency and using representation (3.4) for the fluid velocity, we obtain

\begin{equation}
 \Vb_\alpha(\ab_\alpha,\omega) = \vb^{ind}(\Rb_\alpha + \ab_\alpha,\omega),
  \qquad \alpha = 1,2,\ldots,N,
\end{equation}

\noindent  where

\begin{equation}
 \Vb_\alpha(\ab_\alpha,\omega) = \Ub_\alpha(\omega)
  + \Bigl(\Omb_\alpha(\omega)\times\ab_\alpha\Bigr)
  - \vb^{(0)}(\Rb_\alpha + \ab_\alpha,\omega)
\end{equation}

\noindent  is the velocity of the point  $\rb_\alpha = \Rb_\alpha +\ab_\alpha$
of the surface of sphere  $\alpha$  relative to the fluid velocity
$\vb^{(0)}(\Rb_\alpha + \ab_\alpha,\omega)$   at this point in the absence of
particles in the fluid.

For each  $\alpha$,  we expand the quantities in relations (4.2) in the
spherical harmonics  $Y_{lm}(\theta,\varphi)$  analogously to expansion (3.19)
for the induced surface force densities.  Note that representations of the
induced velocity and pressure of the fluid in the form (3.24)--(3.28) are, in
fact, the expansions of these quantities in the spherical harmonics
$Y_{lm}(\theta,\varphi)$  at the surface of the sphere of an arbitrary radius
$r_\alpha$  with the center at the point  $O_\alpha$.  Therefore, putting
$\rb_\alpha = \ab_\alpha$  in relations (3.24), (3.26), and (3.27), we
immediately obtain the required expansion for the quantity
$\vb^{ind}(\Rb_\alpha + \ab_\alpha,\omega)$ at the surface of particle
$\alpha$ in the spherical harmonics $Y_{lm}(\theta,\varphi)$.  As a result, we
obtain the following system of algebraic equations in the unknown quantities
$\fb_{\beta,lm}(\omega)$:

\begin{equation}
 \sum\limits_{\beta = 1}^N \sum\limits_{l_2 m_2}
 \Tb_{\alpha, l_1 m_1}^{\beta, l_2 m_2}(\omega)
  \cdot \fb_{\beta, l_2 m_2}(\omega) = \Vb_{\alpha, l_1 m_1}(\omega)
  - \sum\limits_{\beta = 1}^N \left \{\vb_{\alpha, l_1 m_1}^{\beta (r)}(\omega)
  + \vb_{\alpha, l_1 m_1}^{\beta (sol)}(\omega)\right \},
\end{equation}

\noindent  where

\begin{eqnarray}
 \Vb_{\alpha, l_1 m_1}(\omega) &=& \delta_{l_1, 0}\, \delta_{m_1, 0}
  \left\{\Ub_\alpha(\omega) -\vb^{inf}(\omega)\right \}
  + \delta_{l_1, 1}\, \frac{a_\alpha}{\sqrt{\pi}}\,
  \Bigl(\Omb_\alpha(\omega)\times\ab_\alpha\Bigr)
  - \vb_{\alpha, l_1 m_1}^{(0)sol}(\omega),\\
 \vb_{\alpha, lm}^{(0)sol}(\omega) &=& \frac{1}{2\sqrt{\pi}} \int \!\!
  d\Omega_\alpha \,\vb^{(0)sol}(\Rb_\alpha +\ab_\alpha,\omega)
  Y_{lm}^*(\theta_\alpha,\varphi_\alpha),
\end{eqnarray}

\noindent  and, to simplify the representation of relations, we introduce the
following notation:

\[
\begin{array}{lrllrl}
  \Tb_{\alpha, l_1 m_1}^{\beta, l_2m_2}(\omega) & \equiv &
   \Tb_{l_1  m_1}^{l_2m_2}(a_\alpha,a_\beta,R_{\alpha\beta},\omega),  \qquad
   & \Tb_{\alpha, l_1 m_1}^{\alpha, l_2m_2}(\omega) & \equiv &
   \Tb_{l_1  m_1}^{l_2m_2}(a_\alpha,a_\alpha,\omega),  \\
  \vb_{\alpha,l_1 m_1}^{\beta (r)}(\omega) & \equiv &
   \vb_{\beta,l_1 m_1}^{(r)ind}(\Rb_\alpha,a_\alpha,\omega), \qquad
   & \vb_{\alpha,l_1 m_1}^{\alpha (r)}(\omega) & \equiv &
   \vb_{\alpha,l_1 m_1}^{(r)ind}(\Rb_\alpha,a_\alpha,\omega), \\
  \vb_{\alpha,l_1 m_1}^{\beta (sol)}(\omega) & \equiv &
   \vb_{\beta,l_1 m_1}^{(sol)ind}(\Rb_\alpha,a_\alpha,\omega), \qquad
   & \vb_{\alpha,l_1 m_1}^{\alpha (sol)}(\omega) & \equiv &
   \vb_{\alpha,l_1 m_1}^{(sol)ind}(\Rb_\alpha,a_\alpha,\omega), \qquad
\end{array}
\]

\noindent  where the quantities
$\Tb_{l_1m_1}^{l_2m_2}(a_\alpha,a_\beta,R_{\alpha\beta},\omega)$,  \quad
$\Tb_{l_1m_1}^{l_2m_2}(a_\alpha,a_\alpha,\omega)$,  \quad
$\vb_{\beta,l_1m_1}^{(r)ind}(\Rb_\alpha,a_\alpha,\omega)$,  \quad
$\vb_{\alpha,l_1 m_1}^{(r)ind}(\Rb_\alpha,a_\alpha,\omega)$,  \quad
$\vb_{\beta,l_1 m_1}^{(sol)ind}(\Rb_\alpha,a_\alpha,\omega)$,  \quad and \quad
$\vb_{\alpha,l_1 m_1}^{(sol)ind}(\Rb_\alpha,a_\alpha,\omega)$,  \quad are
defined, respectively, by relations (3.34), (3.63), (3.31), (3.60), (3.32), and
(3.61) for $r_\alpha = a_\alpha$.

In the absence of the sum on the right-hand side of Eqs.~(4.4) (this sum is
absent in the case of the stationary problem and conservative external fields),
the tensor  $\Tb_{\alpha,l_1m_1}^{\beta,l_2 m_2}(\omega)$  defines harmonic
$(l_1 m_1)$  of the expansion of the fluid velocity at the surface of particle
$\alpha$  induced by harmonic  $(l_2 m_2)$  of the expansion of the induced
surface force density  $\fb_\beta (\ab_\beta,\omega)$  distributed over the
surface of particle  $\beta$  (another particle if  $\beta \neq \alpha$  or the
same if  $\beta = \alpha$).  For the stationary case  $(\omega = 0)$,  these
quantities were introduced in \cite{ref.Yosh} and called hydrodynamic
interaction tensors.  Their generalization to the nonstationary case was
performed in \cite{ref.Pien1,ref.Pien2}.  In what follows, we use this
terminology.  Equations (4.4) agree with the corresponding equations (3.7) in
\cite{ref.Pien2} up to the terms  $\vb_{\alpha,l_1 m_1}^{\beta (r)ind}(\omega)$
and  $\vb_{\alpha,l_1 m_1}^{\beta (sol)ind}(\omega)$,  which is caused by
extension (2.18) for the stress tensor mentioned above and the zero extension
of the external forces to the domains occupied by the particles used in
\cite{ref.Pien2}.


\section{Stationary Case}  \label{Stationary}

In the previous sections, we reduced the problem of determination of the
velocity and pressure fields of the unbounded viscous fluid induced by an
arbitrary number of spheres immersed in it to the solution of the infinite
system of linear algebraic equations (4.4) in the harmonics of induced surface
force densities  $\fb_{\beta,lm}(\omega)$.  In this paper, we consider the
important particular case corresponding to the stationary mode, i.e., all
quantities are time independent.  It is easy to perform the passage to this
case using the general relations obtained above, setting $\omega = 0$ in them,
and assuming that all quantities are independent of $\omega$ [for example,
$\vb^{ind}(\rb,\omega = 0) \rightarrow \vb^{ind}(\rb)$,
$\Tb_{\alpha,l_1m_1}^{\beta,l_2m_2}(\omega = 0) \rightarrow
\Tb_{\alpha,l_1m_1}^{\beta,l_2m_2}$,  etc., i.e., simply omitting the argument
$\omega$  (or  $t$)].  In view of relations (3.31) and (3.60), we get

\begin{equation}
 \begin{array}{lrllrll}
  \vb_{\beta, l_1 m_1}^{(r)ind}(\Rb_\alpha,\rb_\alpha) &=& 0, \qquad
   &\vb_{\alpha, l_1 m_1}^{\beta (r)ind} &=& 0, \qquad & \beta = 1, 2,\ldots,N,
   \nonumber \\
  \vb_\beta^{(r)ind}(\rb) &=& 0,  \qquad
   &\vb_\beta^{(V)ind}(\rb) &=& \vb_\beta^{(sol)ind}(\rb), \qquad
   & \beta = 1, 2,\ldots,N.
 \end{array}
\end{equation}

Thus, in the stationary case, the fluid velocity caused by the induced volume
forces is determined only by the solenoidal component of the external force
field.  This is quite natural because, according to (2.24), in the stationary
mode,

\begin{equation}
 \Fb_\alpha^{(V)ind}(\rb) = -\Theta(a_\alpha - r_\alpha) \Fb^{(sol)ext}(\rb).
\end{equation}

Passing in relations (3.52), (3.54), (3.69), and (3.72) to the limit as $\omega
\rightarrow 0$,  we obtain the corresponding relations for the quantities
defining the fluid velocity in the stationary mode.

For  $\beta \neq \alpha$,  \, $\lim\limits_{\omega \to 0} F_{l_1
l_2,l}(r_\alpha,r_\beta,R_{\alpha\beta},\omega)$,  where $0 \leq r_\beta \leq
a_\beta$  and  $l = l_1 + l_2 - 2p$, $\, p = -1, 0, 1,\ldots,p_{max}$, gives
that \, $F_{l_1 l_2,l}(r_\alpha,r_\beta,R_{\alpha\beta}) \neq 0$  only if  $l =
l_1 + l_2 \, \, (p = 0)$  and  $l = l_1 + l_2 + 2 \, \, (p = -1)$

\begin{eqnarray}
 F_{l_1 l_2,l_1+l_2}(r_\alpha,r_\beta,R_{\alpha\beta}) &=& \frac{\sqrt{\pi}}
  {4\eta R_{\alpha\beta}} \frac{\Gamma\left(l_1+l_2+\frac{1}{2}\right)}
  {\Gamma\left(l_1+\frac{3}{2}\right) \Gamma\left(l_2+\frac{3}{2}\right)}
  \frac{r_\alpha^{l_1} r_\beta^{l_2}}{R_{\alpha\beta}^{l_1 + l_2}}, \\
 F_{l_1 l_2,l_1+l_2+2}(r_\alpha,r_\beta,R_{\alpha\beta}) &=&
  F_{l_1 l_2,l_1+l_2}(r_\alpha,r_\beta,R_{\alpha\beta})
  \left(l_1+l_2+\frac{1}{2}\right)  \nonumber \\
  & \times & \left\{1-\left(l_1+l_2+\frac{3}{2}\right)
  \left[\frac{1}{l_1+\frac{3}{2}}\left(\frac{r_\alpha}{R_{\alpha\beta}}\right)^2
  + \frac{1}{l_2+\frac{3}{2}}\left(\frac{r_\beta}{R_{\alpha\beta}}\right)^2
  \right] \right\}
\end{eqnarray}

\noindent  in the domain  $r_\alpha \leq R_{\alpha\beta} - r_\beta$,  and only
if  $l = l_1 - l_2 \,\, (l_1 \geq l_2, \, p = l_2)$  and  $l = l_1 - l_2 - 2
\,\, ({l_1 \geq l_2 + 2,} \, p = l_2 + 1)$

\begin{eqnarray}
 F_{l_1 l_2,l_1-l_2}(r_\alpha,r_\beta,R_{\alpha\beta}) &=& \frac{\sqrt{\pi}}
  {4\eta r_\alpha} \frac{\Gamma\left(l_1+\frac{1}{2}\right)}
  {\Gamma\left(l_2+\frac{3}{2}\right) \Gamma\left(l_1-l_2+\frac{3}{2}\right)}
  \frac{r_\beta^{l_2} R_{\alpha\beta}^{l_1 - l_2}}{r_\alpha^{l_1}},
  \qquad l_1 \geq l_2, \\
 F_{l_1 l_2,l_1-l_2-2}(r_\alpha,r_\beta,R_{\alpha\beta}) &=& \frac{\sqrt{\pi}}
  {4\eta r_\alpha} \frac{\Gamma\left(l_1-\frac{1}{2}\right)}
  {\Gamma\left(l_2+\frac{3}{2}\right) \Gamma\left(l_1-l_2-\frac{1}{2}\right)}
  \frac{r_\beta^{l_2} R_{\alpha\beta}^{l_1 - l_2 - 2}}{r_\alpha^{l_1-2}}
  \left\{1 - \left(l_1-\frac{1}{2}\right) \right. \nonumber \\
  & \times & \left. \left[\frac{1}{l_1-l_2-\frac{1}{2}}
  \left(\frac{R_{\alpha\beta}}{r_\alpha}\right)^2
  + \frac{1}{l_2+\frac{3}{2}}\left(\frac{r_\beta}{r_\alpha}\right)^2\right]
  \right\}, \qquad  l_1 \geq l_2+2
\end{eqnarray}

\noindent  in the domain  $r_\alpha \geq R_{\alpha\beta} + r_\beta$.

Thus, in the stationary case, the sum over  $l$  in relation (3.34) contains
only two terms with  $l = l_1 + l_2$  and  $l = l_1 + l_2 + 2$  for  $r_\alpha
\leq R_{\alpha\beta} - r_\beta$  and with  $l = l_1 - l_2 \geq 0$  and   $l =
l_1 - l_2 - 2 \geq 0$  for  $r_\alpha \geq R_{\alpha\beta} + r_\beta$.

For  $\beta = \alpha$,  for  $r_\alpha \geq a_\alpha$,  we have

\begin{equation}
 F_{l_1 l_2}(r_\alpha, a_\alpha) = \delta_{l_2, l_1} \, F_{l_1 l_1}(r_\alpha, a_\alpha)
  + \delta_{l_2, l_1 - 2} \, F_{l_1 l_2}(r_\alpha, a_\alpha),
  \quad l_2 = l_1 + 2n \geq 0,  \, n = 0, \pm 1,
\end{equation}

\noindent  where

\begin{eqnarray}
 F_{l_1 l_1}(r_\alpha, a_\alpha) &=& \frac{1}{(2l_1 + 1)\eta r_\alpha}
  \left( \frac{a_\alpha}{r_\alpha} \right)^{l_1},  \\
 F_{l_1, l_1 - 2}(r_\alpha, a_\alpha) &=& \frac{1}{2\eta r_\alpha}
  \left( \frac{a_\alpha}{r_\alpha} \right)^{l_1 - 2}
  \left[ 1 - \left( \frac{a_\alpha}{r_\alpha} \right)^2 \right],  \qquad  l_1 \geq 2.
\end{eqnarray}

According to (5.7)--(5.9),

\begin{equation}
 F_{l_1 l_2}(a_\alpha, a_\alpha) = \delta_{l_2, l_1} F_{l_1}(a_\alpha),
  \qquad  l_2 = l_1 + 2n \geq 0,  \quad  n = 0, \pm 1,
\end{equation}

\noindent  where

\begin{equation}
 F_{l_1}(a_\alpha) \equiv F_{l_1 l_1}(a_\alpha, a_\alpha)
  = \frac{1}{(2l_1 + 1)\eta a_\alpha} \, .
\end{equation}

Therefore, the tensor

\begin{equation}
 \Tb_{\alpha,l_1 m_1}^{\alpha, l_2 m_2} = \delta_{l_2, l_1} \,
  \Tb_{\alpha,l_1 m_1}^{\alpha, l_1 m_2}
\end{equation}

\noindent  is diagonal with respect to the indices  $l_1$  and  $l_2$
\cite{ref.Yosh}.  Here,

\begin{equation}
 \Tb_{\alpha,l_1 m_1}^{\alpha, l_1 m_2} = \frac{3 \sqrt{\pi}}{(2l_ 1 + 1)
  \xi_\alpha} \Kb_{l_1 m_1, 00}^{l_1 m_1} \, .
\end{equation}

We represent the system of equations (4.4) in the form

\begin{equation}
 \sum\limits_{m_2 = -l_1}^{l_1} \Tb_{\alpha, l_1 m_1}^{\alpha, l_1 m_2} \cdot
  \fb_{\alpha, l_1 m_2} - \Vb_{\alpha, l_1 m_1}
  + \vb_{\alpha, l_1 m_1}^{\alpha (sol)}
  = - \sum\limits_{\beta \neq \alpha} \left \{ \sum\limits_{l_2 m_2}
  \Tb_{\alpha, l_1 m_1}^{\beta, l_2 m_2} \cdot \fb_{\beta, l_2 m_2}
  + \vb_{\alpha, l_1 m_1}^{\beta (sol)} \right \}
\end{equation}

\noindent  separating terms corresponding to the hydrodynamic interaction
between different particles on the right-hand side of the equations.  In the
absence of the external force field, system (5.14) coincides with the
corresponding system derived in \cite{ref.Yosh}.  Taking into account the
obtained explicit form for the quantities  $\Tb_{\alpha, l_1 m_1}^{\alpha, l_1
m_2}$  and  $\Tb_{\alpha, l_1 m_1}^{\beta, l_2 m_2}$,  we get for any $l_1, l_2
\geq 0$  \cite{ref.Yosh}

\begin{eqnarray}
 a_\alpha \Tb_{\alpha,l_1 m_1}^{\alpha, l_2 m_2} & \sim & \delta_{l_2, l_1} \,
  \sigma_{\alpha\beta}^0,  \nonumber  \\
 a_\alpha \Tb_{\alpha,l_1 m_1}^{\beta, l_2 m_2} & \sim& \sigma_{\alpha\beta}^{l_1 + 1}
  \sigma_{\beta\alpha}^{l_2} \, .
\end{eqnarray}

\noindent  This enables us to seek a solution of system (5.14) by the method of
successive approximations.  As the zeroth iteration, we use system (5.14) with
the zeroth right-hand side, which corresponds to the absence of interaction
between the particles (the induced surface force density on the surface of
particle  $\alpha$  is determined by the characteristics of the fluid and this
particle and independent of the characteristics of the rest particles).  The
solution of system (5.14) in the form of a series, each term of which, in fact,
determines the contribution of a separate iteration to the total solution is
given in \cite{ref.Yosh}.  Moreover, the final result is expressed in terms of
the tensor  $\Tb_{\alpha,l_1 m_1}^{\beta, l_2 m_2}$  and the inverse tensor
$\tilde{\Tb}_{\alpha,l_1 m_1}^{\alpha, l_1 m_2}$  defined by the condition

\begin{equation}
 \sum\limits_{m_3 = -l_1}^{l_1} \tilde{\Tb}_{\alpha, l_1 m_1}^{\alpha, l_1 m_3}
  \cdot \Tb_{\alpha, l_1 m_3}^{\alpha, l_1 m_2} = \delta_{m_1, m_2} \Ib.
\end{equation}

\noindent  Taking relation (5.13) into account, the inverse tensor
$\tilde{\Tb}_{\alpha,l_1 m_1}^{\alpha, l_1 m_2}$  can be represented in the
form

\begin{equation}
 \tilde{\Tb}_{\alpha,l_1 m_1}^{\alpha, l_1 m_2}
  = \frac{(2 l_1 + 1) \xi_\alpha}{3\sqrt{\pi}}
  \tilde{\Kb}_{l_1 m_1, 00}^{l_1 m_2}
\end{equation}

\noindent  where  $\tilde{\Kb}_{l_1 m_1, 00}^{l_2 m_2}$  is the tensor inverse
to the tensor  $\Kb_{l_1 m_1, 00}^{l_2 m_2}$  defined by the condition

\begin{equation}
 \sum\limits_{m_3 = -l_1}^{l_1} \tilde{\Kb}_{l_1 m_1, 00}^{l_1 m_3} \cdot
  \Kb_{l_1 m_3,00}^{l_1 m_2} = \delta_{m_1, m_2} \Ib.
\end{equation}

However, the explicit form for the tensor $\tilde{\Tb}_{\alpha,l_1
m_1}^{\alpha, l_1 m_2}$  is not given in \cite{ref.Yosh} except for  $l_1 = 0$.
Furthermore, it is stated that this tensor always exists due to the uniqueness
of the solution of the Stokes equation.  Being the central point for the
determination of the solution in \cite{ref.Yosh}, the derivation of the forces
and torques exerted by the fluid on particles is based on this statement.  As a
result, these forces and torques are also expressed in terms of the inverse
tensor $\tilde{\Tb}_{\alpha,l_1 m_1}^{\alpha, l_1 m_2}$.  For this reason, in
the present paper, we dwell on the derivation of a similar solution in detail.
To this end, we represent  $\fb_{\alpha, l_ 1 m_1}$  in the form

\begin{equation}
 \fb_{\alpha, l_1 m_1} = \sum\limits_{n = 0}^\infty \fb_{\alpha, l_1 m_1}^{(n)},
\end{equation}

\noindent  where  $\fb_{\alpha, l_1 m_1}^{(n)}$  is a solution of system (5.14)
corresponding the  {\it n\/}th iteration.

In a similar form, we represent the force  $\Fb_\alpha$  and the torque
$\Tb_\alpha$  exerted by the fluid on particle  $\alpha$  (as well as the
induced velocity and pressure of the fluid)

\begin{eqnarray}
 \Fb_\alpha &=& \sum\limits_{n = 0}^\infty \Fb_\alpha^{(n)},  \\
 \Tb_\alpha &=& \sum\limits_{n = 0}^\infty \Tb_\alpha^{(n)},
\end{eqnarray}

\noindent  where  $\Fb_\alpha^{(n)}$  and  $\Tb_\alpha^{(n)}$  are the force
and the torque corresponding the  {\it n\/}th iteration

\begin{eqnarray}
 \Fb_\alpha^{(n)} &=& -\fb_{\alpha, 00}^{(n)} \, ,  \\
 \Tb_\alpha^{(n)} &=& -\frac{a_\alpha}{\sqrt{3}} \sum\limits_{m = -1}^1
  \left( \eb_m \times \fb_{\alpha, 1m}^{(n)} \right).
\end{eqnarray}


\subsection{$n = 0$.  Noninteracting Particles}  \label{n = 0}

In this approximation (zero iteration), the infinite system of equations (5.14)
is reduced to the collection of independent systems of equations for each
particle  $\alpha$  and each  $l_1$

\begin{equation}
 \sum\limits_{m_2 = -l_1}^{l_1} \Tb_{\alpha, l_1 m_1}^{\alpha, l_1 m_2} \cdot
  \fb_{\alpha, l_1 m_2}^{(0)} - \Vb_{\alpha, l_1 m_1}
  + \vb_{\alpha, l_1 m_1}^{\alpha (sol)} = 0.
\end{equation}

For  $l_1 = 0$,

\begin{equation}
 \Kb_{00,00}^{00} = \frac{1}{3 \sqrt{\pi}}, \qquad
 \Tb_{\alpha, 00}^{00} = \xi_\alpha^{-1} \Ib,
\end{equation}

\noindent  and, according to (5.18) and (5.16), we have

\begin{equation}
 \tilde{\Kb}_{00,00}^{00} = 3 \sqrt{\pi}, \qquad
 \tilde{\Tb}_{\alpha, 00}^{\alpha,00} = \xi_\alpha \Ib.
\end{equation}

We represent the solution  $\fb_{\alpha, 00}^{(0)}$  of system (5.24)  with
$l_1 = 0$  in the form

\begin{equation}
 \fb_{\alpha, 00}^{(0)} = \fb_{\alpha, 00}^{(t,0)} + \fb_{\alpha, 00}^{(ext,0)},
\end{equation}

\noindent  where the harmonics  $\fb_{\alpha, 00}^{(t,0)}$  and  $\fb_{\alpha,
00}^{(ext,0)}$  associated, respectively, with the relative translational motion
of particle  $\alpha$  with the velocity

\begin{equation}
 \Ub_\alpha^{(t)} = \Ub_\alpha - \vb^{inf}
\end{equation}

\noindent  and the external force field are determined as follows:

\begin{eqnarray}
 \fb_{\alpha, 00}^{(t,0)} &=& \xi_\alpha \Ub_\alpha^{(t)},  \\
 \fb_{\alpha, 00}^{(ext,0)} &=& - \xi_\alpha
 \overline{\vb^{(0)sol}(\rb)}^{{\,}S_\alpha} + \tilde{\Fb}_\alpha^{(sol)ext}.
\end{eqnarray}

\noindent  Here,  $\overline{\vb^{(0)sol}(\rb)}^{{\,}S_\alpha} = \vb_{\alpha,
00}^{(0)sol}$  is the fluid velocity induced by the solenoidal component of the
external force field in the absence of particles averaged over the surface of
particle  $\alpha$  and

\begin{equation}
 \tilde{\Fb}_\alpha^{(sol)ext} = \int\limits_{V_\alpha} \!\! d\rb \,
  \Fb^{(sol)ext}(\rb)
\end{equation}

\noindent  is the force acting by the solenoidal component
$\Fb^{(sol)ext}(\rb)$  of the external force field  $\Fb^{ext}(\rb)$  on the
fluid occupying the volume  $V_\alpha$.

Analogously to (5.27), in what follows, for any {\it n\/}th iteration, we
represent the quantity  $\fb_{\alpha, lm}^{(n)}$  as the superposition of the
components

\begin{equation}
 \fb_{\alpha, lm}^{(n)} = \fb_{\alpha, lm}^{(t,n)} + \fb_{\alpha, lm}^{(r,n)}
  + \fb_{\alpha, lm}^{(ext,n)}, \qquad  n = 0, 1, 2,\ldots
\end{equation}

\noindent  associated, respectively, with the translational motion of particle
$\alpha$, its rotation, and the external force field (as is shown in what
follows, in the particular case  $l = 1$,  $\fb_{\alpha, 1m}^{(n)}$  may be
represented as a superposition of four components).  To separate the
contributions caused by the translational motion of particles, their rotation,
and the external force field to the induced velocity and pressure of the fluid
as well as to the forces and torques exerted by the fluid on particles, we also
use analogous representations for these quantities.  According to (5.27) and
(5.32), for zero iteration, we have

\begin{equation}
 \fb_{\alpha, 00}^{(r,0)} = 0 \, .
\end{equation}

Now we consider the case  $l_1 = 1$.  According to (5.24), we have

\begin{equation}
 \sum\limits_{m_2 = -1}^1 \Kb_{1 m_1,00}^{1 m_2} \cdot \fb_{\alpha,1m_2}^{(0)}
  = \frac{\xi_\alpha}{\sqrt{\pi}} \left \{ \frac{a_\alpha}{\sqrt{3}}
  \left(\Omb_\alpha \times \eb_{m_1}^* \right)
  - \ub_{\alpha, 1 m_1}^{sol} \right \},
\end{equation}

\noindent  where

\begin{equation}
 \ub_{\alpha, 1m}^{sol} =  \vb_{\alpha, 1m}^{(0)sol}
  + \vb_{\alpha, 1m}^{\alpha (sol)} \, .
\end{equation}

Using relation (3.48) for  $\Kb_{l_1 m_1,00}^{l_2 m_2}$  and setting  $l_2 = l_1
= 1$  in it, we obtain

\begin{equation}
 \det \Kb_{1 m_1,00}^{1 m_2} = 0 \, .
\end{equation}

Therefore, the tensors  $\Kb_{1 m_1,00}^{1 m_2}$  and  $\Tb_{\alpha,
1m_1}^{\alpha,1 m_2}$  are degenerate and, hence, the inverse tensors
$\tilde{\Kb}_{1m_1,00}^{1m_2}$  and  $\tilde{\Tb}_{\alpha, 1 m_1}^{\alpha,1
m_2}$  do not exist or they should be defined in another way \cite {ref.Freed1}
than by relations (5.18) and (5.16). Thus, even for the case of noninteracting
particles, it is difficult to interpret the relation for  $\fb_{\alpha, 1m}$
obtained in \cite{ref.Yosh} and expressed in terms of the nonexistent (in the
ordinary sense) inverse tensor  $\tilde{\Tb}_{\alpha, 1 m_1}^{\alpha,1 m_2}$.
Furthermore, with regard for the interaction between particles, not only
harmonics with  $l_1 = 1$  but all harmonics  $\fb_{\alpha, l_1 m_1}$  are
expressed in terms of  $\tilde{\Tb}_{\alpha, 1 m_1}^{\alpha,1 m_2}$  [relation
(38) in \cite{ref.Yosh}].  To understand the reason leading to the
impossibility of the existence of $\tilde{\Tb}_{\alpha, 1 m_1}^{\alpha,1 m_2}$,
we investigate the system of equations (5.34) in detail.  To this end, first,
we decompose all vectors in (5.34) into the independent unit vectors $\eb_0$
and $\eb_{\pm 1}$  defined by relations (3.23) taking into account that any
vector $\ab \equiv (a_x, a_y, a_z)$ may be represented in terms of these unit
vectors in the form $\ab \equiv (a_{+1}, a_{-1},a_0)$,  i.e.,

\begin{equation}
 \ab = a_0 \eb_0 - \left( a_{-1} \eb_1 + a_{+1} \eb_{-1} \right),
\end{equation}

\noindent  where  $a_{\pm 1} = (ia_y \pm a_x)/\sqrt{2}$.

Going from $\Omb_\alpha \equiv (\Omega_{\alpha x},\Omega_{\alpha
y},\Omega_{\alpha z})$, $\fb_{\alpha,1m}^{(0)} \equiv
(f_{\alpha,1m,x}^{(0)},f_{\alpha,1m,y}^{(0)}, f_{\alpha,1m,z}^{(0)})$,  and
$\ub_{\alpha,1m}^{sol} \equiv (u_{\alpha,1m,x}^{sol},
u_{\alpha,1m,y}^{sol},u_{\alpha,1m,z}^{sol})$  to $\Omb_\alpha \equiv
(\Omega_{\alpha,+1},\Omega_{\alpha,-1},\Omega_{\alpha,0})$,
$\fb_{\alpha,m}^{(0)} \equiv (f_{\alpha,m+1}^{(0)},f_{\alpha,m-1}^{(0)},
f_{\alpha,m0}^{(0)})$,  and  $\ub_{\alpha,m}^{sol} \equiv
(u_{\alpha,m+1}^{sol}, u_{\alpha,m-1}^{sol},u_{\alpha,m0}^{sol})$,  we reduce
the system of nine equations (5.34) to three independent systems

\begin{eqnarray}
 && \left \{
  \begin{array}{lrl}
   2f_{\alpha,00}^{(0)} - f_{\alpha,+1+1}^{(0)} - f_{\alpha,-1-1}^{(0)}
    &=& 10\xi_\alpha u_{\alpha,00}^{sol}  \\
   f_{\alpha,00}^{(0)} - 3f_{\alpha,+1+1}^{(0)} + 2f_{\alpha,-1-1}^{(0)}
    &=& -i \varepsilon_\alpha \Omega_{\alpha,0}
    + 10\xi_\alpha u_{\alpha,+1+1}^{sol}  \\
   f_{\alpha,00}^{(0)} + 2f_{\alpha,+1+1}^{(0)} - 3f_{\alpha,-1-1}^{(0)}
    &=& i \varepsilon_\alpha \Omega_{\alpha,0}
    + 10\xi_\alpha u_{\alpha,-1-1}^{sol},  \\
  \end{array}
   \right. \\
   && \nonumber \\
 && \left \{
  \begin{array}{lrl}
   4f_{\alpha,0-1}^{(0)} + f_{\alpha,+10}^{(0)}
    &=& i \varepsilon_\alpha \Omega_{\alpha,-1}
    - 10\xi_\alpha u_{\alpha,0-1}^{sol}  \\
   f_{\alpha,0-1}^{(0)} + 4f_{\alpha,+10}^{(0)}
    &=& i \varepsilon_\alpha \Omega_{\alpha,-1}
    - 10\xi_\alpha u_{\alpha,+10}^{sol} \, ,  \\
  \end{array}
  \right.  \\
   && \nonumber  \\
 && \left \{
  \begin{array}{lrl}
   4f_{\alpha,0+1}^{(0)} + f_{\alpha,-10}^{(0)}
    &=& -i \varepsilon_\alpha \Omega_{\alpha,+1}
    - 10\xi_\alpha u_{\alpha,0+1}^{sol}  \\
   f_{\alpha,0+1}^{(0)} + 4f_{\alpha,-10}^{(0)}
    &=& -i \varepsilon_\alpha \Omega_{\alpha,+1}
    - 10\xi_\alpha u_{\alpha,-10}^{sol} \, ,
  \end{array}
 \right.
\end{eqnarray}

\noindent  where  $\varepsilon_\alpha = 10 a_\alpha \xi_\alpha/\sqrt{3}$, \,
and

\begin{eqnarray}
 f_{\alpha,+1-1}^{(0)} &=& -\frac{10}{3} \, \xi_\alpha  u_{\alpha,+1-1}^{sol},
  \nonumber  \\
 f_{\alpha,-1+1}^{(0)} &=& -\frac{10}{3} \, \xi_\alpha  u_{\alpha,-1+1}^{sol}.
\end{eqnarray}

The solutions of systems (5.39) and (5.40) have the form

\begin{eqnarray}
 f_{\alpha,0+1}^{(0)} &=& -i \frac{\varepsilon_\alpha}{5} \Omega_{\alpha,+1} +
  \frac{2}{3} \, \xi_\alpha \left( u_{\alpha,-10}^{sol} - 4u_{\alpha,0+1}^{sol} \right),
  \nonumber  \\
 f_{\alpha,0-1}^{(0)} &=& i \frac{\varepsilon_\alpha}{5} \Omega_{\alpha,-1} +
  \frac{2}{3} \, \xi_\alpha \left( u_{\alpha,+10}^{sol} - 4u_{\alpha,0-1}^{sol} \right),
  \nonumber  \\
 f_{\alpha,+10}^{(0)} &=& i \frac{\varepsilon_\alpha}{5} \Omega_{\alpha,-1} +
  \frac{2}{3} \, \xi_\alpha \left( u_{\alpha,0-1}^{sol} - 4u_{\alpha,+10}^{sol} \right),
  \nonumber  \\
 f_{\alpha,-10}^{(0)} &=& -i \frac{\varepsilon_\alpha}{5} \Omega_{\alpha,+1} +
  \frac{2}{3} \, \xi_\alpha \left( u_{\alpha,0+1}^{sol} - 4u_{\alpha,-10}^{sol} \right).
\end{eqnarray}

The determinant of system (5.38) is equal to zero.  Indeed, it is easy to see
that the left-hand side of the first equation of system (5.38) is equal to the
sum of the left-hand sides of the second and third equations of this system.
For the consistency of system (5.38), the right-hand sides of the equations of
this system must satisfy the following condition:

\begin{equation}
 \sum\limits_{m = -1}^1 \eb_m \cdot \ub_{\alpha, 1m}^{sol} = 0 \, .
\end{equation}

Using relations (3.61) and (4.6), we verify the validity of this condition
because

\begin{eqnarray}
 \sum\limits_{m = -1}^1 \eb_m \cdot \vb_{\alpha, 1m}^{(0)sol} = 0, \\
 \sum\limits_{m = -1}^1 \eb_m \cdot \vb_{\alpha, 1m}^{\alpha(sol)} = 0 \, .
\end{eqnarray}

Furthermore, relations (5.44) and (5.45) are also true in the general
nonstationary case where  $\omega \neq 0$  including any values of  $r_\alpha$
and not only for  $r_\alpha = a_\alpha$.

Therefore, system (5.38) and, hence the original system (5.34), has an infinite
number of solutions instead of a single solution as it is stated in
\cite{ref.Yosh}.

This result is quite natural.  Indeed, the problem of determination of
$\fb_{\alpha, lm}$  with the use of boundary conditions (4.1) means that we try
to represent the unknown induced surface force densities $\fb_\alpha
(\ab_\alpha)$  in terms of the known fluid velocity at the surfaces of the
particles.  According to the continuity equation (2.2) for the incompressible
fluid, the fluid velocity is a solenoidal vector.  Thus, within the framework
of this approach, the required quantity  $\fb_\alpha (\ab_\alpha)$  can be
determined only up to an arbitrary potential vector.  The potential component
of the induced surface force must make no contribution to the fluid velocity.
Therefore, instead of the first equation in system (5.38), as an additional
equation, we can use the condition of the absence of the potential component of
the induced surface force density  $\fb_\alpha (\ab_\alpha)$.

In the general case where the interaction between the particles is taken into
account, the conditions of the absence of the potential components of the
induced surface forces can be written in the form

\begin{equation}
 Y_\alpha = 0,  \qquad \qquad  \alpha = 1,2, \ldots,N,
\end{equation}

\noindent  where

\begin{equation}
 Y_\alpha = \sum\limits_{m = -1}^1 \eb_m \cdot \fb_{\alpha, 1m}
  = f_{\alpha, 00} + f_{\alpha, +1+1} + f_{\alpha, -1-1},
  \quad  \alpha = 1,2, \ldots,N.
\end{equation}

In the approximation of noninteracting particles considered in this section,
these conditions have the form

\begin{equation}
 Y_\alpha^{(0)} \equiv \sum\limits_{m = -1}^1 \eb_m \cdot \fb_{\alpha, 1m}^{(0)}
  = f_{\alpha, 00}^{(0)} + f_{\alpha, +1+1}^{(0)} + f_{\alpha, -1-1}^{(0)} = 0,
  \quad  \alpha = 1,2, \ldots,N,
\end{equation}

\noindent  where the superscript 0 in  $Y_\alpha^{(0)}$,  just as in
$\fb_{\alpha, lm}^{(0)}$  and other quantities, stands for zero approximation.

In the case where both the solenoidal and potential components of the induced
surface forces should be determined, a certain additional equation linearly
independent of the second and third equations of system (5.38) should be
formulated.

To derive additional equations with regard for the interaction between the
particles in the fluid, we consider relation (3.5) at  $\rb = \Rb_\alpha$  and
equate it to relation (2.17) given at  $\rb = \Rb_\alpha$. As result, we get
the relations

\begin{equation}
 Y_\alpha = 4\pi \sqrt{3} \, a_\alpha^2 \sum\limits_{\beta \neq \alpha}
  \left \{ p_\beta^{(V)ind}(\Rb_\alpha)+ p_\beta^{(S)ind}(\Rb_\alpha)
  \right \}, \quad  \alpha = 1,2, \ldots,N.
\end{equation}

In the derivation of Eqs.~(5.49), we used the explicit form for the quantities
$\Db_{l_1 m_1}^{l_2 m_2}(0,r_\alpha^{\,\prime})$,  where  $r_\alpha^{\,\prime}
\le a_\alpha$,  and obtained that $p_\alpha^{(V)ind}(\Rb_\alpha) = 0$.

Putting  $r_\alpha = \varepsilon$,  where  $\varepsilon \to +0$,  in relations
(3.25), (3.28), (3.38), and (3.40), we can represent the quantity
$p_\beta^{(S)ind}(\Rb_\alpha)$  in the form (for details of calculation of the
required quantities for  $r_\alpha = 0$,  see Appendix)

\begin{eqnarray}
 p_\beta^{(S)ind}(\Rb_\alpha) &=& \frac{1}{4\pi R_{\alpha\beta}^2} \biggl \{
  \left ( \nb_{\alpha\beta} \cdot \fb_{\beta,00} \right )
  + 4 \pi \sum\limits_{l_2 = 1}^\infty \sum\limits_{m_2 = -l_2}^{l_2}
  \sigma_{\beta\alpha}^{l_2} \sum\limits_{m = -(l_2 + 1)}^{l_2 + 1}
  \Wb_{00, l_2 + 1,m}^{l_2 m_2}\cdot  \fb_{\beta, l_2 m_2} \nonumber \\
   & \times & Y_{l_2 + 1,m}(\Theta_{\alpha\beta},\Phi_{\alpha\beta}) \biggr \},
\end{eqnarray}

\noindent  where  $\nb_{\alpha\beta} = \Rb_{\alpha\beta}/R_{\alpha\beta}$.

Thus, in the general case, Eqs.~(5.49) differ from Eqs.~(5.46) corresponding to
the absence of the potential components of the induced surface forces by the
nonzero right-hand side.  However, in the approximation of noninteracting
particles, equating the right-hand of Eqs.~(5.49) to zero, we obtain
Eqs.~(5.48), which means that the induced surface forces have no potential
components in this approximation.

Using Eq.~(5.49) and the second and third equations of system (5.38), we obtain
the system of three equations with nonzero determinant, the solution of which
has the form

\begin{eqnarray}
 f_{\alpha, -1-1}^{(0)} &=&
  - i \frac{\varepsilon_\alpha}{5} \Omega_{\alpha,0} -\frac{2}{3} \, \xi_\alpha
  \left( 4 u_{\alpha, -1-1}^{sol} + u_{\alpha, +1+1}^{sol} \right ), \nonumber \\
 f_{\alpha, +1+1}^{(0)} &=&
   i \frac{\varepsilon_\alpha}{5} \Omega_{\alpha,0} -\frac{2}{3} \, \xi_\alpha
  \left( 4 u_{\alpha, +1+1}^{sol} + u_{\alpha, -1-1}^{sol} \right ), \nonumber \\
 f_{\alpha, 00}^{(0)} &=&  \frac{10}{3} \, \xi_\alpha
  \left( u_{\alpha, +1+1}^{sol} + u_{\alpha, -1-1}^{sol} \right ).
\end{eqnarray}

Returning in relations (5.41), (5.42), and (5.51) to the quantities
$\fb_{\alpha,1m}^{(0)}$,  we get

\begin{equation}
 \fb_{\alpha,1m}^{(0)} = \fb_{\alpha,1m}^{(r,0)} + \fb_{\alpha,1m}^{(ext,0)},
\end{equation}

\noindent  where

\begin{eqnarray}
 \fb_{\alpha,1m}^{(r,0)} &=& \frac{\sqrt{3}}{2} \frac{\xi_\alpha^R}{a_\alpha}
  \left( \Omb_\alpha \times \eb_m^* \right),  \\
 \fb_{\alpha,1m}^{(ext,0)} &=& -\frac{2}{3} \, \xi_\alpha \left( 4 \ub_{\alpha, 1m}^{sol}
  + \left (\ub_{\alpha, 1m}^{sol} \right)^T \right),
\end{eqnarray}

\noindent  and  $\xi_\alpha^R = 8 \pi \eta a_\alpha^3$  is the Stokes friction
coefficient for a rotating sphere of radius  $a_\alpha$.  The quantities
$\left(\ub_{\alpha, 1m}^{sol} \right)^T$  are defined as follows:  We consider
three vectors  $\bb_m \equiv (b_{m,x},b_{m,y},b_{m,z})$,  where  $m = 0, \pm
1$. According to (5.37), we go to the vectors  $\bb_m \equiv
(b_{m+1},b_{m-1},b_{m0})$ and represent  $\bb_m$  in the form

\begin{equation}
 \bb_{m_1} = \sum\limits_{m_2 = -1}^1 \Bb_{m_1 m_2} \cdot \eb_{m_2},
  \qquad  m_1 = 0, \pm 1,
\end{equation}

\noindent  where

\begin{equation}
 \Bb_{m_1 m_2} =
  \left (
   \begin{array}{lll}
    b_{00}  &  -b_{0-1}  &  -b_{0+1} \\
    b_{+10}  &  -b_{+1-1}  &  -b_{+1+1} \\
    b_{-10} &  -b_{-1-1} &  -b_{-1+1}
   \end{array}
  \right ),
  \qquad  m_1, m_2 = 0, \pm 1.
\end{equation}

\noindent  Then  $\bb_m^T$  is the vector defined as follows

\begin{equation}
 \bb_{m_1}^T = \sum\limits_{m_1 = -1}^1 \Bb_{m_1 m_2}^T \cdot \eb_{m_2},
  \qquad  m_1 = 0, \pm 1,
\end{equation}

\noindent  where  $\Bb_{m_1 m_2}^T$  is the matrix transposed to the matrix
$\Bb_{m_1 m_2}$.

By virtue of (5.52), the translational motion of particles makes no
contribution to the harmonic  $\fb_{\alpha,1m}^{(0)}$  \quad [$\,
\fb_{\alpha,1m}^{(t,0)} = 0\,$].

In view of relation (5.54) for  $\fb_{\alpha,1m}^{(ext,0)}$, a similar
representation may be given for any right-hand side of system (5.34) for which
this system is consistent.  Indeed, we can analogously represent the quantity
$\fb_{\alpha,1m}^{(r,0)}$

\begin{equation}
 \fb_{\alpha,1m}^{(r,0)} = \frac{2a_\alpha \xi_\alpha}{3\sqrt{3}}
  \left\{4\bb_{\alpha,m}^{(r)} + \left( \bb_{\alpha,m}^{(r)} \right)^T \right \},
\end{equation}

\noindent  where the vector  $\bb_{\alpha,m}^{(r)} = \left(\Omb_\alpha \times
\eb_m^* \right)$  can be rewritten in the form (5.55) with the asymmetric
tensor  $\Bb_{\alpha,m_1 m_2}^{(r)}$

\begin{equation}
 \Bb_{\alpha, m_1 m_2}^{(r)} =
  \left (
   \begin{array}{ccc}
    0                    &  -i\Omega_{\alpha,-1}  &  i\Omega_{\alpha,+1}  \\
    i\Omega_{\alpha,-1}   &  0                    &  -i\Omega_{\alpha,0} \\
    -i\Omega_{\alpha,+1}  &  i\Omega_{\alpha,0}   &   0
   \end{array}
  \right ),
    \qquad  m_1, m_2 = 0, \pm 1.
\end{equation}

\noindent  According to (5.59), we have $\left(\Bb_{\alpha,m_1
m_2}^{(r)}\right)^T = -\Bb_{\alpha,m_1 m_2}^{(r)}$, which gives  $\left(
\bb_{\alpha,m}^{(r)}\right)^T = -\bb_{\alpha,m}^{(r)}$.

For  $l_1 \geq 2$,  according to (5.24), we have

\begin{equation}
 \sum\limits_{m_2 = -l_1}^{l_1} \Tb_{\alpha,l_1 m_1}^{\alpha,l_1 m_2} \cdot
  \fb_{\alpha,l_1 m_2}^{(0)} = -\ub_{\alpha, l_1 m_1}^{sol},
\end{equation}

\noindent  where

\begin{equation}
 \ub_{\alpha, lm}^{sol} =  \vb_{\alpha, lm}^{(0)sol}
 + \vb_{\alpha, lm}^{\alpha (sol)}.
\end{equation}

Assuming the existence of the inverse matrix  $\tilde{\Kb}_{l_1 m_1,00}^{l_1
m_2}$ defined by relation (5.18) for  $l_1 \geq 2$,  we obtain

\begin{equation}
 \fb_{\alpha,l_1 m_1}^{(0)} \equiv \fb_{\alpha,l_1 m_1}^{(ext,0)}
  = -\sum\limits_{m_2 = -l_1}^{l_1} \tilde{\Tb}_{\alpha,l_1 m_1}^{\alpha,l_1 m_2}
  \cdot \ub_{\alpha,l_1 m_2}^{sol},
\end{equation}

\noindent  i.e., in the approximation of noninteracting particles, the
harmonics of the induced surface force densities  $\fb_{\alpha,l_1 m_1}^{(0)}$
for  $l_1 \geq 2$  are not equal to zero only if the external force field has
the solenoidal component.

Substituting the obtained solutions (5.27), (5.29), (5.30), (5.52)--(5.54), and
(5.62) into (3.58), (3.59), (3.29), and (3.30) and using (3.34), (3.35), and
(3.43)--(3.51), we obtain the following relations for the  $t$,  $r$, and $ext$
components of the harmonics of the fluid velocity and pressure caused by
translational motion of particles, their rotation, and the external
(solenoidal) force field, respectively, in the approximation of noninteracting
particles:

\begin{eqnarray}
 \vb_{\alpha,l_1 m_1}^{(S,t,0)ind}(\Rb_\alpha,r_\alpha) &=&
  \frac{a_\alpha}{r_\alpha} \left \{ \delta_{l_1,0}\, \delta_{m_1,0} \, \Ib
  + \delta_{l_1,2} \, \frac{1}{4} \sqrt{\frac{3}{10}} \left [ 1 - \left(
  \frac{a_\alpha}{r_\alpha} \right)^2  \right] \Kb_{m_1} \right \}
  \cdot \Ub_\alpha^{(t)},  \\
 \vb_{\alpha,l_1 m_1}^{(S,r,0)ind}(\Rb_\alpha,r_\alpha) &=&
  \delta_{l_1,1} \, \frac{1}{\sqrt{3}} \frac{a_\alpha^3}{r_\alpha^2}
  \left( \Omb_\alpha \times \eb_{m_1}^* \right),  \\
 \vb_{\alpha,l_1 m_1}^{(S,ext,0)ind}(\Rb_\alpha,r_\alpha) &=&
  \sum\limits_{m_2 = -l_1}^{l_1}
  \Tb_{l_1 m_1}^{l_1 m_2}(r_\alpha,a_\alpha) \cdot
  \fb_{\alpha,l_1 m_2}^{(ext,0)}, \\
 p_{\alpha,l_1 m_1}^{(S,t,0)ind}(\Rb_\alpha,r_\alpha) &=&
   \left (1 - \frac{1}{2} \, \delta_{r_\alpha,a_\alpha} \right ) \,\delta_{l_1,1} \,
  \frac{\sqrt{3}}{2} \eta \frac{a_\alpha}{r_\alpha^2}\,
  \eb_{m_1}^* \cdot \Ub_\alpha^{(t)},  \\
 p_{\alpha,l_1 m_1}^{(S,ext,0)ind}(\Rb_\alpha,r_\alpha) &=&
  \sum\limits_{m_2 = -l_1}^{l_1} \Db_{l_1 m_1}^{l_1 m_2}(r_\alpha,a_\alpha)
  \cdot \fb_{\alpha,l_1 m_2}^{(ext,0)},  \\
 \vb_{\beta,l_1 m_1}^{(S,t,0)ind}(\Rb_\alpha,r_\alpha) &=& \xi_\beta
  \Tb_{l_1 m_1}^{00}(r_\alpha,a_\beta,\Rb_{\alpha\beta}) \cdot \Ub_\beta^{(t)},
  \qquad  \beta \neq \alpha, \\
 \vb_{\beta,l_1 m_1}^{(S,r,0)ind}(\Rb_\alpha,r_\alpha) &=&
  \frac{\sqrt{3}}{2} \frac{\xi_\beta^R}{a_\beta} \sum\limits_{m_2 = -1}^1
  \Tb_{l_1 m_1}^{1m_2}(r_\alpha,a_\beta,\Rb_{\alpha\beta}) \cdot
  \left( \Omb_\beta \times \eb_{m_2}^* \right),  \qquad  \beta \neq \alpha,  \\
 \vb_{\beta,l_1 m_1}^{(S,ext,0)ind}(\Rb_\alpha,r_\alpha) &=&
  \sum\limits_{l_2 m_2}
  \Tb_{l_1 m_1}^{l_2 m_2}(r_\alpha,a_\beta,\Rb_{\alpha\beta}) \cdot
  \fb_{\beta,l_2 m_2}^{(ext,0)},  \qquad  \beta \neq \alpha, \\
 p_{\beta,l_1 m_1}^{(S,t,0)ind}(\Rb_\alpha,r_\alpha) &=& \xi_\beta
  \Db_{l_1 m_1}^{00}(r_\alpha,a_\beta,\Rb_{\alpha\beta}) \cdot \Ub_\beta^{(t)},
  \qquad  \beta \neq \alpha, \\
 p_{\beta,l_1 m_1}^{(S,ext,0)ind}(\Rb_\alpha,r_\alpha) &=&
  \sum\limits_{l_2 m_2}
  \Db_{l_1 m_1}^{l_2 m_2}(r_\alpha,a_\beta,\Rb_{\alpha\beta}) \cdot
  \fb_{\beta,l_2 m_2}^{(ext,0)},  \qquad  \beta \neq \alpha, \\
 p_{\beta,l_1 m_1}^{(S,r,0)ind}(\Rb_\alpha,r_\alpha) &=& 0,
  \qquad  \beta = 1,2,\ldots,N.
\end{eqnarray}

According to (5.73), the rotation of the particles has no effect on the fluid
pressure

\begin{equation}
 p_\beta^{(S,r,0)ind}(\rb) = 0,  \qquad  \beta = 1,2,\ldots,N.
\end{equation}

Substituting (5.63), (5.64), and (5.66) into (3.24) and (3.25) and taking
relation (3.76) into account, we obtain the following space distributions for
the corresponding components of the velocity and pressure fields of the fluid:

\begin{eqnarray}
 \vb_\alpha^{(S,t,0)ind}(\Rb_\alpha + \rb_\alpha) &=& \frac{a_\alpha}{r_\alpha}
  \left \{ 1 - \frac{1}{4} \left[ 1 - \left( \frac{a_\alpha}{r_\alpha} \right)^2
  \right] \left( \Ib - 3 \nb_\alpha \nb_\alpha \right)\right \}
  \cdot \Ub_\alpha^{(t)}, \\
 \vb_\alpha^{(S,r,0)ind}(\Rb_\alpha + \rb_\alpha) &=&
  \left( \frac{a_\alpha}{r_\alpha} \right)^3
  \left( \Omb_\alpha \times \rb_\alpha \right),  \\
 p_\alpha^{(S,t,0)ind}(\Rb_\alpha + \rb_\alpha) &=& \frac{3\eta}{2}
  \frac{a_\alpha}{r_\alpha^2} \, \nb_\alpha \cdot \Ub_\alpha^{(t)}.
\end{eqnarray}

Relations (5.75) and (5.77) coincide with the well-known distributions of the
fluid velocity and pressure induced by a sphere moving with the constant
velocity  $\Ub_\alpha$ in the immovable fluid ($\vb^{inf} = 0$) or with the
velocity and pressure fields of the fluid uniformly moving with the velocity
$\vb^{inf}$  relative to an immovable sphere ($\Ub_\alpha = 0$)
\cite{ref.Landau}.  Relation (5.76) describes the known distribution of the
velocity of the fluid induced by a sphere rotating in it with the constant
angular velocity  $\Omb_\alpha$  \cite{ref.Landau}.

Using relations (5.68), (5.69), and (5.71), we obtain the following main
contributions to the fluid velocity and pressure induced by particle $\beta$ in
the vicinity of the surface of particle  $\alpha$ for   $r_\alpha \ll
R_{\alpha\beta}$ (the near zone):

\begin{eqnarray}
 \vb_\beta^{(S,t,0)ind}(\Rb_\alpha + \rb_\alpha) & \approx & \frac{3}{4}
  \sigma_{\beta\alpha} \left( \Ib + \nb_{\alpha\beta}\nb_{\alpha\beta} \right)
  \cdot \Ub_\beta^{(t)},  \\
 \vb_\beta^{(S,r,0)ind}(\Rb_\alpha + \rb_\alpha) & \approx &
  a_\beta \sigma_{\beta\alpha}^2
  \left( \Omb_\beta \times \nb_{\alpha\beta} \right), \\
 p_\beta^{(S,t,0)ind}(\Rb_\alpha + \rb_\alpha) & \approx & \frac{3}{2} \eta
  \frac{\sigma_{\beta\alpha}}{R_{\alpha\beta}} \,
  \nb_{\alpha\beta} \cdot \Ub_\beta^{(t)}.
\end{eqnarray}

However, for the correct description of the fluid velocity and pressure up to
these powers of the parameter  $\sigma$, these quantities should be also found
in the first approximation.

Using relations (5.5), (5.6), and (3.55), we obtain the velocity and pressure
fields of the fluid induced by particle  $\beta \neq \alpha$  in the far zone,
i.e., for  $r_\alpha \gg R_{\alpha\beta}$.  In zero approximation with respect
to the ratio $R_{\alpha\beta}/r_\alpha$,  we obtain the following space
distributions of the fluid velocity and pressure induced by a system of $N$
spheres far from it:

\begin{eqnarray}
 \vb^{(S,t,0)ind}(\Rb_\alpha + \rb_\alpha) & \approx &
  \frac{3}{4r_\alpha} \left( \Ib + \nb_\alpha \nb_\alpha \right) \cdot
  \sum\limits_{\beta = 1}^N a_\beta \Ub_\beta^{(t)},  \\
 \vb^{(S,r,0)ind}(\Rb_\alpha + \rb_\alpha) & \approx &
  \frac{1}{r_\alpha^2} \sum\limits_{\beta = 1}^N a_\beta^3 \,
  \left( \Omb_\beta \times \nb_\alpha \right),  \\
 p^{(S,t,0)ind}(\Rb_\alpha + \rb_\alpha) & \approx & \frac{3\eta}{2}
  \frac{1}{r_\alpha^2}
  \sum\limits_{\beta = 1}^N a_\beta \, \nb_\alpha \cdot \Ub_\beta^{(t)}.
\end{eqnarray}

\noindent  In this approximation,  $r_\alpha$  can be interpreted as the
distance from a certain point inside this system (e.g., its center) to the
point of observation.

Naturally, these relations can be also derived by using the superposition
principle for the velocity and pressure fields of the fluid induced by
noninteracting particles keeping only the main terms in them.

In the case of equal particles  $a_\beta = a$,  relations (5.81)--(5.83) can be
represented as follows:

\begin{eqnarray}
 \vb^{(S,t,0)ind}(\Rb_\alpha + \rb_\alpha) & \approx &
  \frac{3}{4} \frac{a}{r_\alpha} \left( \Ib + \nb_\alpha \nb_\alpha \right)
  \cdot \Ub^{tot},  \\
 \vb^{(S,r,0)ind}(\Rb_\alpha + \rb_\alpha) & \approx &
  \frac{a^3}{r_\alpha^2} \left( \Omb^{tot} \times \nb_\alpha \right),  \\
 p^{(S,t,0)ind}(\Rb_\alpha + \rb_\alpha) & \approx & \frac{3\eta}{2}
  \frac{a}{r_\alpha^2} \, \nb_\alpha \cdot \Ub^{tot},
\end{eqnarray}

\noindent  where

\begin{eqnarray}
 \Ub^{tot}  &=& \sum\limits_{\beta = 1}^N \Ub_\beta^{(t)},  \\
 \Omb^{tot} &=& \sum\limits_{\beta = 1}^N \Omb_\beta.
\end{eqnarray}

Thus, the velocity and pressure fields of the fluid induced by the system of
equal spheres in the far zone, in zero approximation with respect to the ratio
of the typical distance between two spheres to the distance to the point of
observation, can represented as the velocity and pressure fields of the fluid
induced by a single sphere moving in the fluid with the translational velocity
$\Ub^{tot} + \vb^{inf}$  and rotating with the angular velocity $\Omb^{tot}$.
In the particular case of two equal spheres rotating in the opposite directions
with equal angular velocities, their rotation has no effect on the fluid
velocity far from the spheres within the framework of the considered
approximation.

If all spheres move with the same translational velocity  $\Ub_\beta = \Ub_0$
($\Ub_\beta^{(t)} \equiv \Ub^{(t)}= \Ub_0 - \vb^{inf}$), then

\begin{eqnarray}
 \vb^{(S,t,0)ind}(\Rb_\alpha + \rb_\alpha) & \approx &
  \frac{3}{4} \frac{a_{tot}}{r_\alpha} \left(\Ib+\nb_\alpha \nb_\alpha \right)
  \cdot \Ub^{(t)},  \\
 p^{(S,t,0)ind}(\Rb_\alpha + \rb_\alpha) & \approx & \frac{3\eta}{2}
  \frac{a_{tot}}{r_\alpha^2} \, \nb_\alpha \cdot \Ub^{(t)},
\end{eqnarray}

\noindent  where  $a_{tot} = \sum\limits_{\beta = 1}^N a_\beta$.  In this case,
the action of the system of spheres in the far zone is equivalent to the action
of a single sphere of radius  $a_{tot}$.

If all spheres rotate with the same angular velocity  $\Omb_\beta = \Omb$,  then
we obtain the following relation for the fluid velocity:

\begin{equation}
 \vb^{(S,r,0)ind}(\Rb_\alpha + \rb_\alpha) \approx \frac{3}{4\pi}
  \frac{V_{tot}}{r_\alpha^2} \left( \Omb \times \nb_\alpha \right),
\end{equation}

\noindent  which corresponds to the fluid velocity induced by a single sphere
occupying the volume  $V_{tot} = \sum\limits_{\beta = 1}^N V_\beta = (4 \pi/3)
\sum\limits_{\beta = 1}^N  a_\beta^3$  far from it.

According to (5.22), (5.27), (5.29), (5.20), and (5.33), in the approximation of
noninteracting particles, we obtain the following relations for the components
of the force acting by the fluid on particle  $\alpha$  caused by translational
and rotational motions of particles and the external force field:

\begin{eqnarray}
 \Fb_\alpha^{(t,0)}   &=& -\xi_\alpha \Ub_\alpha^{(t)},  \\
 \Fb_\alpha^{(r,0)}   &=& 0,  \\
 \Fb_\alpha^{(ext,0)} &=& \xi_\alpha \overline{\vb^{(0)sol}(\rb)}^{{\,}S_\alpha}
  - \tilde{\Fb}_\alpha^{(sol)ext}.
\end{eqnarray}

Relation (5.92) is the classical Stokes force acting by the fluid on a sphere
moving in it with the constant velocity  $\Ub_\alpha$  if the fluid moves with
constant velocity  $\vb^{inf}$.  Relation (5.93) illustrates the well-known
fact of the absence of a force acting on a sphere due to its rotation.  The
first term in relation (5.94) corresponds to the known Fax\'{e}n relation
\cite{ref.Happel} defined by the classical Stokes law with the velocity equal
to the velocity of inhomogeneous motion of the fluid in the absence of spheres
averaged over the surface of sphere  $\alpha$.

Substituting (5.92)--(5.94) into (2.27) and taking into account the absence of
the inertial force  $\Fb_\alpha^{in}$  defined by (2.33) in the stationary
case, we obtain the following relation for the total force acting on particle
$\alpha$:

\begin{equation}
 \Fb_\alpha^{tot} = \Fb_\alpha^{ext} - \tilde{\Fb}_\alpha^{ext}
  + \Fb_\alpha^{(t,0)} + \xi_\alpha \overline{\vb^{(0)sol}(\rb)}^{{\,}S_\alpha},
\end{equation}

\noindent  where

\begin{equation}
 \tilde{\Fb}_\alpha^{ext} = \tilde{\Fb}_\alpha^{(p)ext} +
  \tilde{\Fb}_\alpha^{(sol)ext} = \int\limits_{V_\alpha} \!\! d\rb \,
  \Fb^{ext}(\rb)
\end{equation}

\noindent  is the force acting by the external force field  $\Fb^{ext}(\rb)$ on
the fluid occupying the volume  $V_\alpha$.  The difference $\Fb_\alpha^{ext} -
\tilde{\Fb}_\alpha^{ext}$  in (5.95) is the force acting on particle  $\alpha$
in an arbitrary force field  $\Fb^{ext}(\rb)$  corrected for the buoyancy force
for this force field.

To determine the torque  $\Tb_\alpha^{(0)}$  exerted by the fluid on sphere
$\alpha$,  we substitute relations (5.52)--(5.54) into (5.23) and perform
certain transformations taking into account the validity of the following
relation:

\begin{equation}
 \sum\limits_{m = -1}^1 \left( \eb_m \times \bb_m^T \right)
  = - \sum\limits_{m = -1}^1 \left( \eb_m \times \bb_m \right)
\end{equation}

\noindent  for any vectors  $b_m$  and  $b_m^T$  defined by relations (5.55)
and (5.57).  As a result, we get

\begin{eqnarray}
 \Tb_\alpha^{(t,0)}   &=& 0,  \\
 \Tb_\alpha^{(r,0)}   &=& -\xi_\alpha^R \Omb_\alpha,  \\
 \Tb_\alpha^{(ext,0)} &=& 2\xi_\alpha
  \overline{\Bigl( \ab_\alpha \times \vb^{(0)sol}(\rb)\Bigr)}^{{\,}S_\alpha}
  - \tilde{\Tb}_\alpha^{(sol)ext}.
\end{eqnarray}

According to (5.98), we have the well-known result of the absence of the torque
acting on a sphere moving with a constant velocity in an unbounded fluid.
Relation (5.99) is the classical result for the torque exerted by the fluid on
a sphere rotating in it with constant angular velocity $\Omb_\alpha$
\cite{ref.Happel}.  The first term in (5.100) corresponds to the torque exerted
by the fluid on sphere  $\alpha$  due to the inhomogeneous motion of the fluid
in the absence of particles caused by the solenoidal component of the external
force field.  For a homogeneous distribution of the fluid velocity, this torque
is equal to zero.  The second term in (5.100) defined as follows:

\begin{equation}
 \tilde{\Tb}_\alpha^{(sol)ext} = \int\limits_{V_\alpha} \!\! d\rb \,
 \left( \rb_\alpha \times \Fb^{(sol)ext}(\rb) \right)
\end{equation}

\noindent  is the torque exerted on the fluid sphere occupying the volume
$V_\alpha$  due to the solenoidal component  $\Fb^{(sol)ext}(\rb)$  of the
external force field  $\Fb^{ext}(\rb)$.

Substituting (5.98)--(5.100) into (2.28), we obtain the following relation for
the total torque exerted on sphere  $\alpha$:

\begin{equation}
 \Tb_\alpha^{tot} = \Tb_\alpha^{ext} - \tilde{\Tb}_\alpha^{ext}
  + \Tb_\alpha^{(r,0)} + 2 \xi_\alpha
  \overline{\Bigl( \ab_\alpha \times \vb^{(0)sol}(\rb)\Bigr)}^{{\,}S_\alpha},
\end{equation}

\noindent  where

\begin{equation}
 \tilde{\Tb}_\alpha^{ext} = \tilde{\Tb}_\alpha^{(p)ext} +
 \tilde{\Tb}_\alpha^{(sol)ext} = \int\limits_{V_\alpha} \!\! d\rb \,
 \left( \rb_\alpha \times \Fb^{ext}(\rb) \right)
\end{equation}

\noindent  is the torque acting due to the external force field
$\Fb^{ext}(\rb)$  on the fluid occupying the volume  $V_\alpha$.  The
difference $\Tb_\alpha^{ext} - \tilde{\Tb}_\alpha^{ext}$  in (5.102) is the
torque acting on sphere $\alpha$  in an arbitrary force field $\Fb^{ext}(\rb)$
corrected for the buoyancy torque for this force field.

Thus, the external force field makes the contribution to the total force and
torque exerted on sphere  $\alpha$  both due to the buoyancy force
$\tilde{\Fb}_\alpha^{ext}$  and torque  $\tilde{\Tb}_\alpha^{ext}$  on the
one hand, and due to the inhomogeneous distribution of the fluid velocity in
this force field in the absence of particles, on the other.


\subsection{$n = 1$.  The First Iteration}  \label{n = 1}

For the first iteration, with regard for (5.24), the system of equations (5.14)
is reduced to the form

\begin{equation}
 \sum\limits_{m_2 = -l_1}^{l_1} \Tb_{\alpha, l_1 m_1}^{\alpha, l_1 m_2}
  \cdot \fb_{\alpha, l_1 m_2}^{(1)} = - \sum\limits_{\beta \neq \alpha}
  \left \{ \sum\limits_{m_2 = -l_2}^{l_2}
  \Tb_{\alpha, l_1 m_1}^{\beta, l_2 m_2} \cdot \fb_{\beta, l_2 m_2}^{(0)}
  + \vb_{\alpha, l_1 m_1}^{\beta (sol)} \right \},
\end{equation}

\noindent  where the quantities  $\fb_{\alpha, l_2 m_2}^{(0)}$  are determined
by relations (5.27), (5.29), (5.30) for  $l_2 = 0$,  (5.52)--(5.54)  for  $l_2
= 1$,  and (5.62) for  $l_2 \geq 2$.

Taking into account relations (3.34), (3.35), (3.37), (5.3), and (5.4), we
represent the solution of system (5.104) for  $l_1 = 0$ in the form (5.32) with
$n = 1$,  where

\begin{eqnarray}
 \fb_{\alpha,00}^{(t,1)} &=& -\xi_\alpha \sum\limits_{\beta \neq \alpha}
  \xi_\beta \Tb_M(\Rb_{\alpha\beta}) \cdot \Ub_\beta^{(t)},  \\
 \fb_{\alpha,00}^{(r,1)} &=& \sum\limits_{\beta \neq \alpha}
  \xi_\beta a_\beta \sigma_{\alpha\beta} \sigma_{\beta\alpha}
  \left( \nb_{\alpha\beta} \times \Omb_\beta \right),  \\
 \fb_{\alpha,00}^{(ext,1)} &=& -\xi_\alpha \sum\limits_{\beta \neq \alpha}
  \biggl \{ \vb_{\alpha,00}^{\beta(sol)} + \Tb_M(\Rb_{\alpha\beta})
  \cdot \left( \tilde{\Fb}_\beta^{(sol)ext} - \xi_\beta
  \overline{\vb^{(0)sol}(\rb)}^{{\,}S_\beta} \right) \nonumber  \\
  &+& \sum\limits_{l_2 = 1}^\infty \sum\limits_{m_2 = -l_2}^{l_2}
  \Tb_{\alpha,00}^{\beta,l_2 m_2} \cdot \fb_{\beta,l_2 m_2}^{(ext,0)} \biggr \},
\end{eqnarray}

\noindent  where

\begin{eqnarray}
 \Tb_M(\Rb_{\alpha\beta}) \equiv
  \Tb_{\alpha,00}^{\beta,00}(a_\alpha,a_\beta,\Rb_{\alpha\beta})
  &=& \frac{1}{8\pi\eta R_{\alpha\beta}} \biggl \{
  \left( \Ib + \nb_{\alpha\beta}\nb_{\alpha\beta} \right)  \nonumber \\
  &+& \left( \sigma_{\alpha\beta}^2 + \sigma_{\beta\alpha}^2 \right)
  \left( \frac{1}{3} \Ib - \nb_{\alpha\beta}\nb_{\alpha\beta} \right) \biggr \}
\end{eqnarray}

\noindent  is the modified Oseen tensor \cite{ref.Yosh}.

According to (5.105)--(5.107), in this approximation, $\fb_{\alpha,00}^{(r,1)}
\sim \sigma^2$,  $\fb_{\alpha,00}^{(t,1)}$  contains terms proportional to
$\sigma$  and  $\sigma^3$,  and $\fb_{\alpha,00}^{(ext,1)}$  has terms $\sim
1/R_{\alpha\beta}^n$,  where $n = 1, 2,\ldots$,  furthermore,

\begin{equation}
 \fb_{\alpha,00}^{(ext,1)} \approx \frac{3}{4} \xi_\alpha
  \sum\limits_{\beta \neq \alpha} \xi_\beta \sigma_{\beta\alpha}
  \left( \Ib + \nb_{\alpha\beta}\nb_{\alpha\beta} \right) \cdot
  \overline{\vb^{(0)sol}(\rb)}^{{\,}S_\beta}
  + o\left(\frac{1}{R_{\alpha\beta}^2} \right).
\end{equation}

For  $l_ 1 = 1$, system (5.104) has the form

\begin{equation}
 \sum\limits_{m_2 = -1}^1 \Kb_{1m_1,00}^{1m_2} \cdot
  \fb_{\alpha, 1m_2}^{(1)} = - \frac{\xi_\alpha}{\sqrt{\pi}}
  \sum\limits_{\beta \neq \alpha}
  \left \{ \sum\limits_{m_2 = -l_2}^{l_2}
  \Tb_{\alpha, 1m_1}^{\beta, l_2 m_2} \cdot \fb_{\beta, l_2 m_2}^{(0)}
  + \vb_{\alpha, 1m_1}^{\beta (sol)} \right \},
\end{equation}

\noindent  which differs from analogous system (5.34) for the iteration of
noninteracting particles only by the right-hand side.  This implies that system
(5.110) is consistent only if

\begin{equation}
  \sum\limits_{m_1 = -1}^1 \eb_{m_1} \cdot \sum\limits_{\beta \neq \alpha}
  \left \{ \sum\limits_{m_2 = -l_2}^{l_2}
  \Tb_{\alpha, 1m_1}^{\beta, l_2 m_2} \cdot \fb_{\beta, l_2 m_2}^{(0)}
  + \vb_{\alpha, 1m_1}^{\beta (sol)} \right \} = 0.
\end{equation}

This condition is satisfied because the following relations are true:

\begin{eqnarray}
 \sum\limits_{m_1 = -1}^1 \eb_{m_1} \cdot \Tb_{\alpha, 1m_1}^{\beta,l_2 m_2}
  &=& 0, \\
 \sum\limits_{m_1 = -1}^1 \eb_{m_1} \cdot \vb_{\alpha, 1m_1}^{\beta (sol)}
  &=& 0 \, .
\end{eqnarray}

Therefore, system (5.110) has an infinite number of solutions that differ only
in the potential component.  To determine the unique solution, it is necessary
to introduce an additional equation linearly independent of the equations of
this system.  To derive this equation, we use relations (5.47)--(5.50) and get

\begin{equation}
 Y_\alpha^{(1)} \equiv \sum\limits_{m = -1}^1 \eb_m \cdot \fb_{\alpha,1m}^{(1)}
  = 4 \pi \sqrt{3} \, a_\alpha^2 \sum\limits_{\beta \neq \alpha} \left \{
  p_\beta^{(V)ind}(\Rb_\alpha) + p_\beta^{(S,0)ind}(\Rb_\alpha) \right \},
  \quad  \alpha = 1,2,\ldots, N,
\end{equation}

\noindent  where  $p_\beta^{(S,0)ind}(\Rb_\alpha)$  is defined by relation
(5.50) with  $\fb_{\beta,l_2 m_2} \rightarrow \fb_{\beta,l_2 m_2}^{(0)}$.

Unlike Eq.~(5.48), Eq.~(114) contains the nonzero right-hand side. Therefore,
the interaction between the particles leads to the appearance of the potential
components  $\fb_\alpha^{(p)}(\ab)$  of the induced surface forces densities
$\fb_\alpha(\ab)$.

We solve the system of equation (5.110), (5.114) in a similar way as in the
case of noninteracting particles.  The final result has the form

\begin{equation}
 \fb_{\alpha,1m}^{(1)} = \fb_{\alpha,1m}^{(t,1)} + \fb_{\alpha,1m}^{(r,1)}
  + \fb_{\alpha,1m}^{(ext,1)} + \fb_{\alpha,1m}^{(p,1)},
\end{equation}

\noindent  where

\begin{eqnarray}
 \fb_{\alpha,1m}^{(\epsilon,1)} &=& -\frac{2}{3} \, \xi_\alpha \left \{
  4\bb_{\alpha,m}^{(\epsilon,1)}+\left( \bb_{\alpha,m}^{(\epsilon,1)}\right)^T
  \right \},  \qquad  \qquad \epsilon = t, r, ext,  \\
 \fb_{\alpha,1m}^{(p,1)} &=& \eb_m^* \frac{Y_\alpha^{(1)}}{3},
\end{eqnarray}

\begin{eqnarray}
 \bb_{\alpha,m_1}^{(t,1)} &=& \sum\limits_{\beta \neq \alpha}
  \sum\limits_{l_2 m_2} \Tb_{\alpha,1m_1}^{\beta,l_2 m_2}
  \cdot \fb_{\beta,l_2 m_2}^{(t,0)}
  = \sum\limits_{\beta \neq \alpha} \xi_\beta \Tb_{\alpha,1m_1}^{\beta,00}
  \cdot \Ub_\beta^{(t)},  \\
 \bb_{\alpha,m_1}^{(r,1)} &=& \sum\limits_{\beta \neq \alpha}
  \sum\limits_{l_2 m_2} \Tb_{\alpha,1m_1}^{\beta,l_2 m_2}
  \cdot \fb_{\beta,l_2 m_2}^{(r,0)} \nonumber  \\
  &=& \frac{2}{3\eta} \sum\limits_{\beta \neq \alpha} \xi_\beta
  \sigma_{\alpha\beta} \sigma_{\beta\alpha}^2 \sum\limits_{m = -2}^2
  \left( \Omb_\beta \times \Wb_{1m_1,2m}^{00} \right)
  Y_{2m}(\Theta_{\alpha\beta},\Phi_{\alpha\beta}),  \\
 \bb_{\alpha,m_1}^{(ext,1)} &=& \sum\limits_{\beta \neq \alpha} \left \{
  \sum\limits_{l_2 m_2} \Tb_{\alpha,1m_1}^{\beta,l_2 m_2} \cdot
  \fb_{\beta,l_2 m_2}^{(ext,0)} + \vb_{\alpha,1m_1}^{\beta(sol)} \right \},
\end{eqnarray}

\noindent  and the quantities  $\left( \bb_{\alpha,m}^{(\epsilon,1)}\right)^T$
are determined by relations (5.56), (5.57), and (5.118)--(5.120).

In the first iteration, we obtain that  $\fb_{\alpha,1m}^{(r,1)} \sim \sigma^3$,
$\fb_{\alpha,1m}^{(t,1)}$  has terms proportional to  $\sigma^2$  and
$\sigma^4$,  and  $\fb_{\alpha,1m}^{(ext,1)}$  contains terms
$\sim 1/R_{\alpha\beta}^n$, where $n = 2,3,\ldots$.

For  $l_1 \geq 2$,  the solution of the system of equations (5.104) can be
represented in the form (5.32) with  $n = 1$,  where

\begin{eqnarray}
 \fb_{\alpha,l_1 m_1}^{(t,1)} &=& - \sum\limits_{\beta \neq \alpha} \xi_\beta
  \sum\limits_{m_2 = -l_1}^{l_1} \tilde{\Tb}_{\alpha,l_1 m_1}^{\alpha,l_1 m_2}
  \cdot \Tb_{\alpha,l_1 m_2}^{\beta,00} \cdot \Ub_\beta^{(t)},  \\
 \fb_{\alpha,l_1 m_1}^{(r,1)} &=& - \frac{\sqrt{3}}{2}
  \sum\limits_{\beta \neq \alpha} \frac{\xi_\beta^R}{a_\beta}
  \sum\limits_{m_3 = -l_1}^{l_1} \sum\limits_{m_2 = -1}^1
  \tilde{\Tb}_{\alpha,l_1 m_1}^{\alpha,l_1 m_3} \cdot
  \Tb_{\alpha,l_1 m_3}^{\beta,1m_2} \cdot
  \left( \Omb_\beta \times \eb_{m_2}^* \right)  \nonumber  \\
  &=& -\sqrt{\pi} (2l_1 + 1) \sum\limits_{\beta \neq \alpha}
  \frac{\xi_\beta^R}{a_\beta} \, \sigma_{\alpha\beta}^{l_1 + 1} \sigma_{\beta\alpha}
  \sum\limits_{m_2 = -l_1}^{l_1} \sum\limits_{m = -(l_1 + 1)}^{l_1 + 1}
  \tilde{\Kb}_{l_1 m_1,00}^{l_1 m_2} \nonumber  \\
  & \cdot & \left( \Omb_\beta \times \Wb_{l_1 m_2,l_1 + 1,m}^{00} \right) \,
  Y_{l_1 + 1,m}(\Theta_{\alpha\beta},\Phi_{\alpha,\beta}),  \\
 \fb_{\alpha,l_1 m_1}^{(ext,1)} &=& \sum\limits_{\beta \neq \alpha}
  \sum\limits_{m_3 = -l_1}^{l_1} \tilde{\Tb}_{\alpha,l_1 m_1}^{\alpha,l_1 m_3}
  \cdot \left \{ \Tb_{\alpha,l_1 m_3}^{\beta,00} \cdot \left \{ \xi_\beta
  \overline{\vb^{(0)sol}(\rb)}^{\, S_\alpha} - \tilde{\Fb}_\beta^{(sol)ext}
  \right \} + \frac{2}{3} \, \xi_\beta \sum\limits_{m_2 = -1}^1
  \Tb_{\alpha,l_1 m_3}^{\beta,1m_2} \right. \nonumber \\
  & \cdot& \left. \left \{ 4 \bb_{\beta,m_2}^{(ext,1)} +
  \left ( \bb_{\beta,m_2}^{(ext,1)} \right)^T \right \}
  - \sum\limits_{l_2 = 2}^\infty \sum\limits_{m_2 = -l_2}^{l_2}
  \Tb_{\alpha,l_1 m_3}^{\beta,l_2 m_2} \cdot \fb_{\beta,l_2 m_2}^{(ext,0)}
  - \vb_{\alpha,l_1 m_3}^{\beta(sol)} \right \}.
\end{eqnarray}

Thus, the interaction of particles leads to the appearance of all harmonics for
the components of the induced surface force density connected with the
translational motion of particles and their rotation.

Since the potential component  $\fb_\alpha^{(p,1)}(\ab_\alpha)$ of the induced
surface force density  $\fb_\alpha^{(1)}(\ab_\alpha)$  has only harmonics with
$l = 1$  defined by (5.117), we obtain the following results for the
corresponding fluid velocity and pressure ($r_\alpha \ge a_\alpha$) induced by
these potential forces:

\begin{eqnarray}
 \vb_\beta^{(S,p,1)ind}(\rb) &=& 0,  \qquad  \qquad
  \beta = 1,2,\ldots,N,  \\
 p_\beta^{(S,p,1)ind}(\rb) &=& \delta_{\beta,\alpha} \, \delta_{r_\alpha, a_\alpha}
  \, p_\alpha^{(S,p,1)ind}(\Rb_\alpha + \ab_\alpha + 0),  \quad  \beta = 1,2,\ldots,N,
\end{eqnarray}

\noindent  where

\begin{equation}
 p_\alpha^{(S,p,1)ind}(\Rb_\alpha + \ab_\alpha + 0)
  = -\frac{Y_\alpha^{(1)}}{4\pi \sqrt{3} \, a_\alpha^2}.
\end{equation}

Thus, the potential component  $\fb_\beta^{(p,1)}(\ab_\beta)$,  where  $\beta =
1,2,\ldots N$, of the induced surface force density makes no contribution to
the fluid velocity, while the fluid pressure caused by this potential force is
equal to zero everywhere with the exception of the surface of the sphere where
this force is distributed.

In view of (5.117), the potential component $\fb_\alpha^{(p,1)}(\ab_\alpha)$ is
represented in the form

\begin{equation}
 \fb_\alpha^{(p,1)} = \frac{\ab_\alpha}{a_\alpha}
  \frac{Y_\alpha^{(1)}}{4 \pi \sqrt{3}},
\end{equation}

\noindent  and, hence, has only the radial constant component.  In this case,
it is easy to see that the potential component of the induced surface force has
no effect on the forces and torques exerted by the fluid on the particles
immersed in it.

These conclusions concerning the influence of the potential component of the
induced surface force densities on the fluid velocity and pressure as well as
on the forces and torques exerted by the fluid on particles also remain valid
for any {\it n\/}th iteration if the potential component
$\fb_\alpha^{(p,n)}(\ab_\alpha)$ of the induced surface force density
$\fb_\alpha^{(n)}(\ab_\alpha)$ corresponding to this iteration has only
harmonics with $l = 1$  defined by a relation similar to (5.117).

By virtue of relations (5.105), (5.106), (5.115), (5.116), (5.118), (5.119),
(5.121), and (5.122), in this approximation, the main contribution to the fluid
velocity and pressure in the vicinity of particle $\alpha$  due to the
interaction between the particles caused by the motion of particles and their
rotation is given by the quantities $\vb_\alpha ^{(S,t,1)ind}(\rb)$,
$\vb_\alpha ^{(S,r,1)ind}(\rb)$, $p_\alpha ^{(S,t,1)ind}(\rb)$,  and $p_\alpha
^{(S,r,1)ind}(\rb)$,  moreover, $\vb_\alpha ^{(S,t,1)ind}(\rb)$  and
$\vb_\alpha ^{(S,r,1)ind}(\rb)$  are of the same order in the parameter
$\sigma$  that the quantities  $\vb_\beta ^{(S,t,0)ind}(\rb)$ and  $\vb_\beta
^{(S,r,0)ind}(\rb)$,  where  $\beta \neq \alpha$,  defined, respectively, by
relations (5.78) and (5.79) in the approximation of noninteracting particles.
Near the surface of particle $\alpha$,  for $r_\alpha \ll R_{\alpha\beta}$, the
main terms for the fluid velocity and pressure due to the hydrodynamic
interaction between the particles can be written as follows:

\begin{eqnarray}
 \sum\limits_{\beta \neq \alpha} \vb_\beta ^{(S,t,0)ind}(\Rb_\alpha + \rb_\alpha)
  + \vb_\alpha ^{(S,t,1)ind}(\Rb_\alpha + \rb_\alpha) &\approx&
  \frac{9}{8} \left ( 1 - \frac {a_\alpha}{r_\alpha} \right )
  \left (\Ib - \nb_\alpha \nb_\alpha \right ) \cdot
  \sum\limits_{\beta \neq \alpha} \sigma_{\beta \alpha}
  \left (\Ib \right. \nonumber  \\
  & + &  \left.  \nb_{\alpha\beta} \nb_{\alpha\beta} \right )
  \cdot \Ub_\beta^{(t)}, \\
 \sum\limits_{\beta \neq \alpha} \vb_\beta ^{(S,r,0)ind}(\Rb_\alpha + \rb_\alpha)
  + \vb_\alpha ^{(S,r,1)ind}(\Rb_\alpha + \rb_\alpha) &\approx&
  -\frac{3}{2} \left ( 1 - \frac {a_\alpha}{r_\alpha} \right )
   \left (\Ib - 3 \nb_\alpha \nb_\alpha \right ) \cdot
  \sum\limits_{\beta \neq \alpha} a_\beta  \nonumber  \\
  & \times &\sigma_{\beta \alpha}^2
  \left ( \nb_{\alpha\beta} \times \Omb_\beta \right ), \\
 \sum\limits_{\beta \neq \alpha} p_\beta ^{(S,t,0)ind}(\Rb_\alpha + \rb_\alpha)
  + p_\alpha ^{(S,t,1)ind}(\Rb_\alpha + \rb_\alpha) &\approx&
  p_\alpha ^{(S,t,1)ind}(\Rb_\alpha + \rb_\alpha)
  \approx -\left ( 1 - \delta_{r_\alpha, a_\alpha} \,\frac{1}{2} \right)
  \nb_\alpha \nonumber \\
  & \cdot & \frac{9 \eta}{8 r_\alpha} \,
  \sum\limits_{\beta \neq \alpha} \sigma_{\beta \alpha}
  \left (\Ib  + \nb_{\alpha\beta} \nb_{\alpha\beta} \right )
  \cdot \Ub_\beta^{(t)},
 \end{eqnarray}

\begin{equation}
 p_\alpha ^{(S,r,1)ind}(\Rb_\alpha + \rb_\alpha) =
  \left ( 1 - \delta_{r_\alpha, a_\alpha} \,\frac{1}{2} \right)
  \frac{3 \eta}{2 r_\alpha} \, \nb_\alpha \cdot
  \sum\limits_{\beta \neq \alpha} a_\beta \sigma_{\beta \alpha}^2
  \left ( \nb_{\alpha\beta} \times \Omb_\beta \right ).
\end{equation}

According to (5.128)--(5.131), the main terms of the fluid velocity and
pressure caused by the hydrodynamic interaction between the particles due to
their translational motion and rotation are proportional to the first and
second powers of the parameter $\sigma$,  respectively.

Substituting (5.105) and (5.106) into (5.22) and (5.116), (5.118), and (5.119)
into (5.23) and taking relation (5.97) into account, we obtain the following
relations for the forces and torques exerted by the fluid on sphere  $\alpha$
for this iteration corresponding to the translational motion of particles and
their rotation:

\begin{eqnarray}
 \Fb_\alpha^{(t,1)} &=& \xi_\alpha \sum\limits_{\beta \neq \alpha} \xi_\beta
  \Tb_M(\Rb_{\alpha\beta}) \cdot \Ub_\beta^{(t)},  \\
 \Fb_\alpha^{(r,1)} &=& -\sum\limits_{\beta \neq \alpha} \xi_\beta a_\beta
  \sigma_{\alpha\beta} \sigma_{\beta\alpha}
  \left( \nb_{\alpha\beta} \times \Omb_\beta \right),  \\
 \Tb_\alpha^{(t,1)} &=& -\xi_\alpha a_\alpha \sum\limits_{\beta \neq \alpha}
  \sigma_{\alpha\beta} \sigma_{\beta\alpha}
  \left ( \nb_{\alpha\beta} \times \Ub_\beta^{(t)} \right),  \\
 \Tb_\alpha^{(r,1)} &=& -\frac{\xi_\alpha^R}{2} \sum\limits_{\beta \neq \alpha}
  \sigma_{\beta\alpha}^3
  \left( \Ib - 3 \nb_{\alpha\beta}\nb_{\alpha\beta} \right) \cdot \Omb_\beta.
\end{eqnarray}

Relation (5.132) agrees with the second term in relation (44) in
\cite{ref.Yosh} corresponding to the first iteration.  It is worth noting that
relation (5.133) for the force exerted by the fluid on a sphere due to rotation
of the rest spheres coincides with the known result for two spheres
\cite{ref.Jones2}, while the corresponding force given by the first term in
relation (49) in \cite{ref.Yosh} is proportional to the inverse tensor
$\tilde{\Kb}_{1m_1,00}^{1m_2}$, which, as was mentioned above, does not exist.
Relations (5.134) and (5.135) also agree with the known relations for the
torques exerted by the fluid on a sphere due to the translational motion and
rotation of the rest spheres \cite{ref.Happel,ref.Jones2}.

According to (5.132) and (5.108), the force  $\Fb_\alpha^{(t,1)}$  contains
terms proportional to the first and third powers of the dimensionless parameter
$\sigma$.  Restricting ourselves only to the terms proportional to  $\sigma$,
we obtain

\begin{equation}
 \Fb_\alpha^{(t,1)} \approx \frac{3}{4} \xi_\alpha
  \sum\limits_{\beta \neq \alpha} \sigma_{\beta\alpha}
  \left( \Ib + \nb_{\alpha\beta}\nb_{\alpha\beta} \right) \cdot \Ub_\beta^{(t)}.
\end{equation}

The terms proportional to  $\sigma^3$  in (5.132) should be retained only if
the problem is solved within the framework of at least the third iteration.

Often, it is necessary to determine the velocities of particles in a fluid in
given external force fields (in particular, the sedimentation velocity of
particles in a fluid due to the action of the gravity force) and to investigate
the influence of hydrodynamic interactions between the particles on their
motion.  Let $\Ub_\alpha$  and  $\Omb_\alpha$  be, respectively, the velocity
of translational motion and the angular velocity of a single sphere  $\alpha$
in a fluid caused by the action of a certain external force field.  In the case
of several spheres, these velocities change due to hydrodynamic interactions
between the spheres.  Using relations (5.133)--(5.136), in this approximation,
we can represent the changed velocities  $\tilde{\Ub}_\alpha$ and
$\tilde{\Omb}_\alpha$  as follows:

\begin{eqnarray}
 \tilde{\Ub}_\alpha &=& \Ub_\alpha + \sum\limits_{\beta \neq \alpha}
  \Ub_{\alpha\beta}^{(1)},  \\
 \tilde{\Omb}_\alpha &=& \Omb_\alpha + \sum\limits_{\beta \neq \alpha}
  \Omb_{\alpha\beta}^{(1)},
\end{eqnarray}

\noindent  where  $\Ub_{\alpha\beta}^{(1)}$  and  $\Omb_{\alpha\beta}^{(1)}$
are, respectively, the changes in the velocity of translational motion and the
angular velocity of sphere  $\alpha$  due to the motion and rotation of sphere
$\beta$.  Up to the main terms in powers of the parameter  $\sigma$,  these
quantities have the form

\begin{eqnarray}
 \Ub_{\alpha\beta}^{(1)} &=& \sigma_{\beta\alpha} \left \{ \frac{3}{4}
  \left( \Ib + \nb_{\alpha\beta}\nb_{\alpha\beta} \right ) \cdot \Ub_\beta^{(t)}
  - \sigma_{\beta\alpha} a_\beta
  \left( \nb_{\alpha\beta} \times \Omb_\beta \right ) \right \},  \\
 \Omb_{\alpha\beta}^{(1)} &=& -\frac{3}{2} \sigma_{\beta\alpha} \left \{
  \frac{1}{2R_{\alpha\beta}}
  \left( \nb_{\alpha\beta} \times \Ub_\beta^{(t)} \right )
  + \sigma_{\beta\alpha}^2
  \left( \frac{1}{3} \Ib - \nb_{\alpha\beta}\nb_{\alpha\beta} \right )
  \cdot \Omb_\beta \right \}.
\end{eqnarray}

The first term in relation (5.139) agrees with the main term in the relation
given in \cite{ref.Happel} for the problem of sedimentation of two particles
with constant velocities without rotation under the action of the gravity
force. The first term in relation (5.140) coincides with the well-known result
in \cite{ref.Happel} for the angular velocity of sphere  $\alpha$,  whose
rotation is induced by the motion of sphere  $\beta$  moving in the fluid with
the relative translational velocity  $\Ub_\beta^{(t)}$.


\subsection{$n = 2$.  The Second Iteration}  \label{n = 2}

Using Eqs.~(5.14), (5.49), and (5.50), we obtain the following systems of
equations for the {\it n\/}th iteration, where  $n \geq 2$:

\begin{eqnarray}
 \sum\limits_{m_2 = -l_1}^{l_1}\Tb_{\alpha,l_1 m_1}^{\alpha,l_1 m_2} \cdot
  \fb_{\alpha,l_1 m_2}^{(n)} &=& -\sum\limits_{\beta \neq \alpha}
  \sum\limits_{l_2 m_2} \Tb_{\alpha,l_1 m_1}^{\beta,l_2 m_2} \cdot
  \fb_{\beta,l_2 m_2}^{(n-1)},  \\
 Y_\alpha^{(n)} \equiv \sum\limits_{m = -1}^1 \eb_m \cdot \fb_{\alpha,1m}^{(n)}
  &=& 4\pi \sqrt{3} \, a_\alpha^2
  \sum\limits_{\beta \neq \alpha} p_\beta^{(S,n-1)ind}(\Rb_\alpha),
\end{eqnarray}

\noindent  where  $p_\beta^{(S,n-1)ind}(\Rb_\alpha)$  is defined by relation
(5.50) with  $\fb_{\beta,lm}^{(n-1)}$  substituted for  $\fb_{\beta,lm}$.

Putting  $n = 2$  in (5.141) and (5.142), we obtain the required equations for
the second iteration.  Further, putting  $l_1 = 0$  and using relations
(5.105)--(5.107) obtained for the first iteration, we obtain the following
solution:

\begin{eqnarray}
 \fb_{\alpha,00}^{(t,2)} &=& \xi_\alpha \sum\limits_{\beta \neq \alpha} \left\{
  \xi_\beta \Tb_M(\Rb_{\alpha\beta}) \cdot \sum\limits_{\gamma \neq \beta}
  \xi_\gamma \Tb_M(\Rb_{\beta\gamma}) \cdot
  \Ub_\gamma^{(t)} + \frac{2}{3} \, \xi_\beta \sum\limits_{m_2 = -1}^1
  \Tb_{\alpha,00}^{\beta,1m_2}  \right. \nonumber  \\
  &\cdot & \left. \left \{ 4 \bb_{\beta,m_2}^{(t,1)} +
  \left ( \bb_{\beta,m_2}^{(t,1)} \right)^T \right \}
  - \sum\limits_{l_2 = 2}^\infty \sum\limits_{m_2 = -l_2}^{l_2}
  \Tb_{\alpha,00}^{\beta,l_2 m_2} \cdot \fb_{\beta,l_2 m_2}^{(t,1)} \right \},\\
 \fb_{\alpha,00}^{(r,2)} &=& \xi_\alpha \sum\limits_{\beta \neq \alpha}
  \left\{ \Tb_M(\Rb_{\alpha\beta}) \cdot
  \sum\limits_{\gamma \neq \beta} \xi_\gamma a_\gamma \sigma_{\beta\gamma}
  \sigma_{\gamma\beta} \left( \Omb_\gamma \times \nb_{\beta\gamma} \right)
  + \frac{2}{3} \, \xi_\beta \sum\limits_{m_2 = -1}^1
  \Tb_{\alpha,00}^{\beta,1m_2} \right. \nonumber  \\
  &\cdot& \left. \left \{ 4 \bb_{\beta,m_2}^{(r,1)} +
  \left ( \bb_{\beta,m_2}^{(r,1)} \right)^T \right \}
  - \sum\limits_{l_2 = 2}^\infty \sum\limits_{m_2 = -l_2}^{l_2}
  \Tb_{\alpha,00}^{\beta,l_2 m_2} \cdot \fb_{\beta,l_2 m_2}^{(r,1)} \right \},\\
 \fb_{\alpha,00}^{(ext,2)} &=& -\xi_\alpha \sum\limits_{\beta \neq \alpha}
  \sum\limits_{l_2 m_2}
  \Tb_{\alpha,00}^{\beta,l_2 m_2} \cdot \fb_{\beta,l_2 m_2}^{(ext,1)}.
\end{eqnarray}

According to (5.143) and (5.144), the quantities  $\fb_{\alpha,00}^{(t,2)}$ and
$\fb_{\alpha,00}^{(r,2)}$  are infinite power series in the parameter $\sigma$
starting from the second and third powers, respectively, which essentially
differ them from the corresponding relations (5.105) and (5.106) obtained for
the first iteration.  The quantity  $\fb_{\alpha,00}^{(t,2)}$ contains only
even powers of  $\sigma$.  We note that the representation for a harmonic of
the induced surface force density as an infinite power series in the parameter
$\sigma$ is also valid both for other  $l$ ($l = 3, 4,\ldots$)  and for higher
iterations ($n = 3, 4,\ldots$).

For  $l_1 = 1$,  the system of equations (5.141) has the form

\begin{equation}
 \sum\limits_{m_2 = -1}^1 \Kb_{1m_1,00}^{1m_2} \cdot
  \fb_{\alpha, 1m_2}^{(2)} = - \frac{\xi_\alpha}{\sqrt{\pi}}
  \sum\limits_{\beta \neq \alpha} \sum\limits_{m_2 = -l_2}^{l_2}
  \Tb_{\alpha, 1m_1}^{\beta, l_2 m_2} \cdot \fb_{\beta, l_2 m_2}^{(1)},
\end{equation}

\noindent  which is similar to system (5.110).  Since the determinant of this
system is equal to zero and, by virtue of relation (5.112), this system is
consistent, it has an infinite number of solutions that differ only in the
potential component.  To obtain the unique solution, we add to this system
Eq.~(5.142) with  $n = 2$.  Analogously to the first iteration, we represent
the solution in the form

\begin{equation}
 \fb_{\alpha,1m_1}^{(2)} = \fb_{\alpha,1m_1}^{(t,2)} + \fb_{\alpha,1m_1}^{(r,2)}
  + \fb_{\alpha,1m_1}^{(ext,2)} + \fb_{\alpha,1m_1}^{(p,2)},
\end{equation}

\noindent  where

\begin{eqnarray}
 \fb_{\alpha,1m_1}^{(\epsilon,2)} &=& -\frac{2}{3} \, \xi_\alpha \left \{
  4\bb_{\alpha,m}^{(\epsilon,2)}+\left( \bb_{\alpha,m}^{(\epsilon,2)}\right)^T
  \right \},  \qquad  \qquad \epsilon = t, r, ext,  \\
 \fb_{\alpha,1m_1}^{(p,2)} &=& \eb_m^* \frac{Y_\alpha^{(2)}}{3},  \\
 \bb_{\alpha,m_1}^{(\epsilon,2)} &=& \sum\limits_{\beta \neq \alpha}
  \sum\limits_{l_2 m_2}
  \Tb_{\alpha,1 m_1}^{\beta,l_2 m_2} \cdot \fb_{\beta,l_2 m_2}^{(\epsilon,1)},
  \qquad  \qquad \epsilon = t, r, ext.
\end{eqnarray}

According to (5.148) and (5.150), the quantities  $\fb_{\alpha,1m_1}^{(t,2)}$
and $\fb_{\alpha,1m_1}^{(r,2)}$  are infinite power series in the parameter
$\sigma$  starting from the third and forth powers, respectively.

For  $l_1 \geq 2$,  the solution of system (5.141) can be represented in the
form (5.32) with  $n = 2$,  where

\begin{equation}
 \fb_{\alpha,l_1 m_1}^{(\epsilon,2)} = - \sum\limits_{\beta \neq \alpha}
  \sum\limits_{m_3 = -l_1}^{l_1} \sum\limits_{l_2 m_2}
  \tilde{\Tb}_{\alpha,l_1 m_1}^{\alpha,l_1 m_3} \cdot
  \Tb_{\alpha,l_1 m_3}^{\beta,l_2 m_2} \cdot \fb_{\beta,l_2
  m_2}^{(\epsilon,1)},  \qquad \epsilon = t, r, ext.
\end{equation}

The quantities  $\fb_{\alpha,l_ 1m_1}^{(t,2)}$  and $\fb_{\alpha,l_1
m_1}^{(r,2)}$  are infinite power series in the parameter  $\sigma$  starting
from powers of ($2 + l_1$) and ($3 + l_1$), respectively.

Analogously, we can determine the harmonics of the induced surface force
density (and, hence, the velocity and pressure fields of the fluid) for higher
iterations.

In the investigation of analogous problems, the main question is connected with
the substantiation of the reduction of the infinite system of equations in
unknown quantities \cite{ref.Clercx1,ref.Clercx2} (for the considered approach,
the unknown harmonics  $\fb_{\alpha,lm}$).  Within the framework of this
approach, we can reformulate this question as follows: How many iterations must
be carried out to obtain the velocity and pressure fields of the fluid, the
forces and torques exerted by the fluid on particles immersed in it up to
$\sigma^p$,  where  $p$  is a certain positive integer?  Using the results
obtained above, we can show that for any iteration  $n \geq 1$,  the main terms
of the quantities  $\fb_{\alpha,lm}^{(t,n)}$,  $\fb_{\alpha,lm}^{(ext,n)}$,  and
$\fb_{\alpha,lm}^{(r,n)}$  are proportional, respectively, to  $\sigma^{l + n}$
and  $\sigma^{l + n + 1}$,  moreover, starting from  $n = 2$  [for
$\fb_{\alpha,lm}^{(ext,n)}$,  starting from  $n = 0$], these quantities contain
infinite sums of higher powers of  $\sigma$.  For this reason, to obtain the
induced velocity and pressure fields of the fluid up to  $\sigma^p$,  it is
necessary:

\noindent  (i) to carry out  $p$  iterations,

\noindent  (ii) for each  $s$th iteration, where  $s \leq p$,  to retain only
harmonics with  $l = 0, 1,\ldots,p-s$,

\noindent  (iii) in each retained harmonic, to retain all terms up to terms
proportional to  $\sigma^p$  inclusively.

\noindent  Further simplification of the results depends on the point of
observation.

Substituting (5.143) and (5.144) into (5.22) and (5.147), (5.148), and (5.150)
into (5.23) and taking relation (5.97) into account, after certain
transformations, we obtain the following relations for the forces and torques
exerted by the fluid on particles due to their translational motion and
rotation corresponding to the second iteration:

\begin{eqnarray}
 \Fb_\alpha^{(t,2)} &=& -\xi_\alpha \sum\limits_{\beta \neq \alpha} \Biggl \{
  \xi_\beta \Tb_M(\Rb_{\alpha\beta}) \cdot \sum\limits_{\gamma \neq \beta}
  \xi_\gamma \Tb_M(\Rb_{\beta\gamma}) \cdot \Ub_\gamma^{(t)}
  + \frac{2}{3} \, \xi_\beta \sum\limits_{m_2 = -1}^1
  \Tb_{\alpha,00}^{\beta,1m_2}  \nonumber  \\
  &\cdot & \left \{ 4 \bb_{\beta,m_2}^{(t,1)} +
  \left ( \bb_{\beta,m_2}^{(t,1)} \right)^T \right \}
  - \sum\limits_{l_2 = 2}^\infty \sum\limits_{m_2 = -l_2}^{l_2}
  \Tb_{\alpha,00}^{\beta,l_2 m_2} \cdot \fb_{\beta,l_2 m_2}^{(t,1)} \Biggr \},\\
 \Fb_\alpha^{(r,2)} &=& \xi_\alpha \sum\limits_{\beta \neq \alpha}
  \Biggl \{\Tb_M(\Rb_{\alpha\beta}) \cdot
  \sum\limits_{\gamma \neq \beta} \xi_\gamma a_\gamma \sigma_{\beta\gamma}
  \sigma_{\gamma\beta} \left( \nb_{\beta\gamma} \times \Omb_\gamma \right)
  - \frac{2}{3} \, \xi_\beta \sum\limits_{m_2 = -1}^1
  \Tb_{\alpha,00}^{\beta,1m_2}  \nonumber  \\
  &\cdot&  \left \{ 4 \bb_{\beta,m_2}^{(r,1)} +
  \left ( \bb_{\beta,m_2}^{(r,1)} \right)^T \right \}
  + \sum\limits_{l_2 = 2}^\infty \sum\limits_{m_2 = -l_2}^{l_2}
  \Tb_{\alpha,00}^{\beta,l_2 m_2} \cdot \fb_{\beta,l_2 m_2}^{(r,1)} \Biggr \},\\
 \Tb_\alpha^{(t,2)} &=& a_\alpha \sum\limits_{\beta \neq \alpha} \xi_\beta
  \sigma_{\alpha\beta}^2 \Biggl \{ \sum\limits_{\gamma \neq \beta}
  \xi_\gamma \Biggl ( \nb_{\alpha\beta} \times \Bigl (\Tb_M(\Rb_{\beta\gamma})
  \cdot \Ub_\gamma^{(t)} \Bigr ) \Biggr ) + \frac{2}{3} \, \sigma_{\beta\alpha}
  \sum\limits_{m_2 = -1}^1 \sum\limits_{m = -2}^2
  \Biggl ( \Wb_{00,2m}^{1m_2}  \nonumber  \\
  & \times &  \Bigl ( 4 \bb_{\beta,m_2}^{(t,1)} +
  \left ( \bb_{\beta,m_2}^{(t,1)} \right )^T \Bigr ) \Biggr ) \,
  Y_{2m}(\Theta_{\alpha\beta},\Phi_{\alpha\beta})
  - \frac{4\pi}{\xi_\beta} \sum\limits_{l_2 = 2}^\infty
  \sum\limits_{m_2 = -l_2}^{l_2} \sum\limits_{m = -(l_2 + 1)}^{l_2 + 1}
  \sigma_{\beta\alpha}^{l_2}  \nonumber \\
  & \times &   \left( \Wb_{00,l_2 + 1,m}^{l_2 m_2} \times
  \fb_{\beta,l_2 m_2}^{(t,1)} \right)
  Y_{l_2 + 1,m}(\Theta_{\alpha\beta},\Phi_{\alpha\beta}) \Biggr \},  \\
 \Tb_\alpha^{(r,2)} &=& -a_\alpha \sum\limits_{\beta \neq \alpha}
  \sigma_{\alpha\beta}^2 \Biggl \{ \xi_\alpha a_\alpha \sigma_{\alpha\beta}
  \sigma_{\beta\alpha} \left(\Ib - \nb_{\alpha\beta}\nb_{\alpha\beta} \right)
  \cdot \Omb_\alpha - \Theta(N - 3) \sum\limits_{\gamma \neq \alpha \beta}
  \xi_\gamma a_\gamma \sigma_{\beta\gamma}\sigma_{\gamma\beta}
  \nonumber  \\
  & \times & \Bigl ( \nb_{\alpha\beta} \times \left( \nb_{\gamma\beta}
 \times \Omb_\gamma \right) \Bigr )
  - \frac{2}{3} \, \xi_\beta \sigma_{\beta\alpha} \sum\limits_{m_2 = -1}^1
  \sum\limits_{m = -2}^2 \Biggl ( \Wb_{00,2m}^{1m_2} \times
  \Bigl ( 4 \bb_{\beta,m_2}^{(r,1)} +
  \left( \bb_{\beta,m_2}^{(r,1)} \right)^T \Bigr ) \Biggr) \nonumber  \\
  & \times &  Y_{2m}(\Theta_{\alpha\beta},\Phi_{\alpha\beta})
  + 4\pi \sum\limits_{l_2 = 2}^\infty \sigma_{\beta\alpha}^{l_2}
  \sum\limits_{m_2 = -l_2}^{l_2}
  \sum\limits_{m = -(l_2 + 1)}^{l_2 + 1}
  \left( \Wb_{00,l_2 + 1,m}^{l_2 m_2} \times \fb_{\beta,l_2 m_2}^{(r,1)} \right)
  \nonumber  \\
  & \times &
  Y_{l_2 + 1,m}(\Theta_{\alpha\beta},\Phi_{\alpha\beta}) \Biggr \}.
\end{eqnarray}

If we restrict our consideration to the second iteration, then only the main
terms in powers of  $\sigma$  should be taken into account in relations
(5.152)--(5.154).  As a result, we get

\begin{eqnarray}
 \Fb_\alpha^{(t,2)} & \approx & -\xi_\alpha \frac{9}{16}
  \sum\limits_{\gamma \neq \alpha} \sigma_{\gamma\alpha} \Biggl \{
  \sigma_{\alpha\gamma} \left( \Ib + 3 \nb_{\alpha\gamma}\nb_{\alpha\gamma}
  \right) \cdot \Ub_\alpha^{(t)} \nonumber \\
  &+& \Theta(N - 3)
  \left(\Ib + \nb_{\alpha\gamma}\nb_{\alpha\gamma}\right) \cdot
  \sum\limits_{\beta \neq \gamma,\alpha} \sigma_{\beta\gamma}
  \left(\Ib + \nb_{\beta\gamma}\nb_{\beta\gamma}\right)
  \cdot \Ub_\beta^{(t)} \Biggr \}, \\
 \Fb_\alpha^{(r,2)} & \approx & -\frac{3}{4} \sum\limits_{\gamma \neq \alpha}
  \sigma_{\alpha\gamma} \Biggl \{\xi_\alpha a_\alpha \sigma_{\alpha\gamma}
  \sigma_{\gamma\alpha} \left( \nb_{\alpha\gamma} \times \Omb_\alpha \right)
  \nonumber \\
  &+& \Theta(N - 3) \left(\Ib + \nb_{\alpha\gamma}\nb_{\alpha\gamma} \right)
  \cdot \sum\limits_{\beta \neq \gamma,\alpha} \xi_\beta a_\beta
  \sigma_{\beta\gamma} \sigma_{\gamma\beta}
  \left( \nb_{\beta\gamma} \times \Omb_\beta \right) \Biggr \},  \\
 \Tb_\alpha^{(t,2)} & \approx & \xi_\alpha a_\alpha \frac{3}{4}
  \sum\limits_{\gamma \neq \alpha} \sigma_{\alpha\gamma} \sigma_{\gamma\alpha}
  \Biggl \{ \sigma_{\alpha\gamma}
  \left(\nb_{\alpha\gamma} \times \Ub_\alpha^{(t)} \right) \nonumber \\
  &+& \Theta(N - 3) \sum\limits_{\beta \neq \gamma,\alpha} \sigma_{\beta\gamma}
  \Biggl ( \nb_{\alpha\gamma} \times \Bigl ( \left(\Ib + \nb_{\beta\gamma}
  \nb_{\beta\gamma} \right) \cdot \Ub_\beta^{(t)} \Bigr ) \Biggr ) \Biggr \},  \\
 \Tb_\alpha^{(r,2)} & \approx & -a_\alpha
  \sum\limits_{\gamma \neq \alpha} \sigma_{\alpha\gamma}^2 \Biggl \{
  \xi_\alpha a_\alpha \sigma_{\alpha\gamma}\sigma_{\gamma\alpha}
  \left(\Ib - \nb_{\alpha\gamma}\nb_{\alpha\gamma}\right) \cdot \Omb_\alpha
  \nonumber \\
  &-& \Theta(N - 3) \sum\limits_{\beta \neq \gamma,\alpha} \xi_\beta a_\beta
  \sigma_{\beta\gamma}\sigma_{\gamma\beta}
  \Bigl ( \nb_{\alpha\gamma} \times \left (\nb_{\beta\gamma} \times \Omb_\beta
  \right ) \Bigr )\Biggr \}.
\end{eqnarray}

The first terms in (5.156)--(5.159) correspond to the self-interaction of
particle  $\alpha$  due to the action of the fluid induced by this particle and
reflected from the rest particles.  The second terms in these relations, which
are nonequal to zero only for  $N \geq 3$,  correspond to three-particle
interaction.  For fixed  $\alpha$  and  $\beta$,  these terms describe the
contribution to the force and torque acting on particle  $\alpha$  by the fluid
induced by particle  $\beta$  and scattered by all rest particles (except for
$\alpha$).  For a system of two spheres, relations (5.156)--(5.159) agree with
the well-known results given in \cite{ref.Happel,ref.Jones2}

Taking relations (5.132)--(5.135) and (5.156)--(5.159) into account, we can
represent the velocity of translational motion  $\tilde{\Ub}_\alpha$  and
angular velocity  $\tilde{\Omb}_\alpha$  of particle  $\alpha$  in a fluid in
given force fields in terms of the velocities  $\Ub_\beta$  and the angular
velocities  $\Omb_\beta$  of noninteracting particles immersed in the fluid in
these fields as follows:

\begin{eqnarray}
 \tilde{\Ub}_\alpha &=& \Ub_\alpha + \sum\limits_{\beta = 1}^N
  \Ub_{\alpha\beta}, \\
 \tilde{\Omb}_\alpha &=& \Omb_\alpha + \sum\limits_{\beta = 1}^N
  \Omb_{\alpha\beta}.
\end{eqnarray}

\noindent  For  $\beta \neq \alpha$

\begin{eqnarray}
 \Ub_{\alpha\beta} &=& \Ub_{\alpha\beta}^{(1)} - \Theta(N - 3)\,
  \frac{3}{4} \sum\limits_{\gamma \neq \alpha, \beta}
  \sigma_{\beta\gamma} \Biggl \{ \frac{3}{4} \sigma_{\gamma\alpha}
  \left( \Ib + \nb_{\alpha\gamma}\nb_{\alpha\gamma} \right) \cdot
  \left( \Ib + \nb_{\beta\gamma}\nb_{\beta\gamma} \right) \cdot \Ub_\beta^{(t)}
  \nonumber  \\
  &+& a_\beta \sigma_{\beta\gamma} \sigma_{\gamma\beta}
  \left( \Ib + \nb_{\alpha\gamma}\nb_{\alpha\gamma} \right) \cdot
  \left( \nb_{\beta\gamma} \times \Omb_\beta \right) \Biggr \}, \\
 \Omb_{\alpha\beta} &=& \Omb_{\alpha\beta}^{(1)} + \Theta(N - 3)\,
  \frac{3}{4a_\alpha}
  \sum\limits_{\gamma \neq \alpha, \beta} \sigma_{\alpha\gamma}
  \sigma_{\beta\gamma} \sigma_{\gamma\alpha} \Biggl \{a_\beta \sigma_{\beta\gamma}
  \Bigl ( \nb_{\alpha\gamma} \times \left (\nb_{\beta\gamma} \times \Omb_\beta
  \right ) \Bigr )  \nonumber \\
  &+& \frac{3}{4} \left \{
  \left(\nb_{\alpha\gamma} \times \Ub_\beta^{(t)} \right)
  + \left(\nb_{\beta\gamma} \cdot \Ub_\beta^{(t)} \right)
  \left(\nb_{\alpha\gamma} \times \nb_{\beta\gamma} \right) \right \} \Biggr \}
\end{eqnarray}

\noindent  are, respectively, the velocity of the translational motion of
particle  $\alpha$  and its angular velocity  induced due to motion and
rotation of particle  $\beta$.  The first terms in (5.162) and (5.163)
characterizing two-particle interaction are determined by relations (5.139) and
(5.140), respectively, describing the corresponding changes in the velocities
of particles within the framework of the first iteration. The second terms in
relations (5.162) and (5.163) characterize the changes in the translational
velocity of particles and their angular velocities due to three-particle
interaction.  The power of the parameter  $\sigma$  contained in these terms is
greater than that in the first terms by one.  The second iteration not only
changes the quantities  $\Ub_{\alpha\beta}$  and $\Omb_{\alpha\beta}$  but also
leads to the appearance of new terms $\Ub_{\alpha\alpha}$  and
$\Omb_{\alpha\alpha}$  equal to

\begin{eqnarray}
 \Ub_{\alpha\alpha} &=& -\frac{3}{4} \sum\limits_{\gamma \neq \alpha}
  \sigma_{\alpha\gamma} \sigma_{\gamma\alpha} \left \{ \frac{3}{4}
  \left( \Ib + 3 \nb_{\alpha\gamma}\nb_{\alpha\gamma} \right)
  \cdot \Ub_\alpha^{(t)} +  a_\alpha \sigma_{\alpha\gamma}
  \left( \nb_{\alpha\gamma} \times \Omb_\alpha \right) \right \}, \\
 \Omb_{\alpha\alpha} &=& \frac{3}{4a_\alpha}
  \sum\limits_{\gamma \neq \alpha} \sigma_{\alpha\gamma}^2 \sigma_{\gamma\alpha}
  \left \{ a_\alpha \sigma_{\alpha\gamma}
  \Bigl ( \nb_{\alpha\gamma} \times \left (\nb_{\alpha\gamma} \times
  \Omb_\alpha  \right ) \Bigr )
  + \frac{3}{4} \left( \nb_{\alpha\gamma} \times \Ub_\alpha^{(t)} \right) \right\},
  \end{eqnarray}

\noindent  which are absent for the first iteration and characterize the
changes in the velocity of translational motion  $\Ub_\alpha$  and the angular
velocity $\Omb_\alpha$  of particle  $\alpha$  due to its motion in the
presence of other (not necessary moving or rotating) particles.  The second
term in (5.165) agrees with the expression given in \cite{ref.Happel} for the
angular velocity of a sphere that can freely rotate in the fluid due to its
motion with constant velocity  $\Ub_\alpha$  in the presence of other spheres.
Note that relations (5.164) and (5.165) also follows from the terms in
relations (5.162) and (5.163) proportional to $\Theta(N - 3)$  by formally
replacing  $\Theta(N - 3)$ by 1 and putting $\beta = \alpha$.


\subsection{Friction and Mobility Tensors}  \label{Mobilities}

Within the framework of the second iteration, using relations (5.22), (5.23),
(5.92), (5.93), (5.98), (5.99), (5.132)--(5.135), and (5.156)--(5.159), we
represent the forces  $\Fb_\alpha^{(t)}$  and  $\Fb_\alpha^{(r)}$  and the
torques $\Tb_\alpha^{(t)}$  and  $\Tb_\alpha^{(r)}$   exerted by the fluid on
particle $\alpha$  as follows:

\begin{eqnarray}
 \Fb_\alpha^{(t)} &=& \sum_{k = 0}^2 \Fb_\alpha^{(t,k)} = -\sum_{\beta = 1}^N
  \xib_{\alpha\beta}^{TT} \cdot \Ub_\beta^{(t)},  \\
 \Fb_\alpha^{(r)} &=& \sum_{k = 0}^2 \Fb_\alpha^{(r,k)} = -\sum_{\beta = 1}^N
  \xib_{\alpha\beta}^{TR} \cdot \Omb_\beta,  \\
 \Tb_\alpha^{(t)} &=& \sum_{k = 0}^2 \Tb_\alpha^{(t,k)} = -\sum_{\beta = 1}^N
  \xib_{\alpha\beta}^{RT} \cdot \Ub_\beta^{(t)},  \\
 \Tb_\alpha^{(r)} &=& \sum_{k = 0}^2 \Tb_\alpha^{(r,k)} = -\sum_{\beta = 1}^N
  \xib_{\alpha\beta}^{RR} \cdot \Omb_\beta,
\end{eqnarray}

\noindent  where  $\xib_{\alpha\beta}^{TT}$  and  $\xib_{\alpha\beta}^{RR}$
are, respectively, the translational and rotational friction tensors and
$\xib_{\alpha\beta}^{TR}$  and  $\xib_{\alpha\beta}^{RT}$  are the friction
tensors that couple translational motion of particles and their rotation. These
quantities are infinite power series in the dimensionless parameter $\sigma$.
Up to the main terms corresponding to the second iteration, these quantities
have the form

\begin{eqnarray}
 \xib_{\alpha\beta}^{TT} &=& \xi_\alpha \left(\delta_{\alpha\beta} \Ib
  + \lambdab_{\alpha\beta}^{TT} \right),  \\
 \xib_{\alpha\beta}^{RR} &=& \xi_\alpha^R \left(\delta_{\alpha\beta} \Ib
  + \lambdab_{\alpha\beta}^{RR} \right),  \\
 \xib_{\alpha\alpha}^{TR} &=& -\xi_\alpha a_\alpha \frac{3}{4}
  \sum\limits_{\gamma \neq \alpha} \sigma_{\alpha\gamma}^2 \sigma_{\gamma\alpha}
  \left(\eb \cdot \nb_{\alpha\gamma} \right),  \\
  \xib_{\alpha\alpha}^{RT} &=& -\xib_{\alpha\alpha}^{TR}
  = \left( \xib_{\alpha\alpha}^{TR} \right)^T,
\end{eqnarray}

\begin{eqnarray}
 \xib_{\alpha\beta}^{TR} &=& -\xi_\beta a_\beta \Biggl \{ \sigma_{\alpha\beta}
  \sigma_{\beta\alpha} \left( \eb \cdot \nb_{\alpha\beta} \right)\nonumber  \\
  &+& \Theta(N - 3) \, \frac{3}{4} \sum\limits_{\gamma \neq \alpha, \beta}
  \sigma_{\alpha\gamma} \sigma_{\gamma\beta} \sigma_{\beta\gamma}
  \Bigl \{ \left(\eb \cdot \nb_{\beta\gamma} \right)
  - \nb_{\alpha\gamma} \left( \nb_{\alpha\gamma} \times \nb_{\beta\gamma}
  \right) \Bigr \} \Biggr \},
  \qquad \beta \neq \alpha,  \\
  \xib_{\alpha\beta}^{RT} &=& -a_\alpha \xi_\alpha \Biggl \{ \sigma_{\alpha\beta}
  \sigma_{\beta\alpha} \left( \eb \cdot \nb_{\alpha\beta} \right) \nonumber  \\
  &-& \Theta(N - 3) \, \frac{3}{4} \sum\limits_{\gamma \neq \alpha, \beta}
  \sigma_{\alpha\gamma} \sigma_{\gamma\alpha} \sigma_{\beta\gamma}
  \Bigl \{ \left(\eb \cdot \nb_{\alpha\gamma} \right)
  + \left( \nb_{\beta\gamma} \times \nb_{\alpha\gamma} \right)
  \nb_{\beta\gamma} \Bigr \} \Biggr \},  \qquad \beta \neq \alpha,
\end{eqnarray}

\noindent  where  $\eb$  is the absolutely antisymmetric unit tensor of the
third rank,  $(\eb \cdot \nb) = e_{ijk} n_k$,

\begin{eqnarray}
 \lambdab_{\alpha\alpha}^{TT} &=& \frac{9}{16} \sum\limits_{\gamma \neq \alpha}
  \sigma_{\alpha\gamma} \sigma_{\gamma\alpha} \left( \Ib + 3 \nb_{\alpha\gamma}
  \nb_{\alpha\gamma} \right),  \\
 \lambdab_{\alpha\alpha}^{RR} &=& \frac{3}{4} \sum\limits_{\gamma \neq \alpha}
  \sigma_{\alpha\gamma}^3 \sigma_{\gamma\alpha} \left( \Ib - \nb_{\alpha\gamma}
  \nb_{\alpha\gamma} \right),
\end{eqnarray}

\begin{eqnarray}
 \lambdab_{\alpha\beta}^{TT} &=& -\frac{3}{4} \Biggl \{ \sigma_{\beta\alpha}
  \left(\Ib + \nb_{\alpha\beta} \nb_{\alpha\beta} \right) \nonumber  \\
  &-& \Theta(N - 3) \, \frac{3}{4} \sum\limits_{\gamma \neq \alpha, \beta}
  \sigma_{\gamma\alpha} \sigma_{\beta\gamma}
  \left( \Ib + \nb_{\alpha\gamma} \nb_{\alpha\gamma} \right)
  \cdot \left( \Ib + \nb_{\beta\gamma} \nb_{\beta\gamma} \right) \Biggr \},  \\
 \lambdab_{\alpha\beta}^{RR} &=& \frac{1}{2} \Biggl \{ \sigma_{\beta\alpha}^3
  \left(\Ib - 3 \nb_{\alpha\beta} \nb_{\alpha\beta} \right)  \nonumber  \\
  &+&\Theta(N - 3) \, \frac{3}{2} \sum\limits_{\gamma \neq \alpha, \beta}
  \left( \frac{a_\beta}{a_\alpha} \right)^2
  \sigma_{\beta\gamma} \sigma_{\gamma\beta} \sigma_{\alpha\gamma}^2
  \Bigl \{ \left( \nb_{\alpha\gamma} \cdot \nb_{\beta\gamma} \right) \Ib
  - \nb_{\beta\gamma} \nb_{\alpha\gamma} \Bigr \} \Biggr \}.
\end{eqnarray}

Taking the explicit form of relations (5.170)--(5.179) into account, we can
show that the friction tensors satisfy the Onsager symmetry relations
\cite{ref.Happel}

\begin{equation}
 \left(\xib_{\beta\alpha}^{TT} \right)^T = \xib_{\alpha\beta}^{TT}, \quad
 \left(\xib_{\beta\alpha}^{RR} \right)^T = \xib_{\alpha\beta}^{RR}, \quad
 \left(\xib_{\beta\alpha}^{RT} \right)^T = \xib_{\alpha\beta}^{TR}, \quad
 \alpha, \beta = 1, 2, \ldots, N,
\end{equation}

\noindent   where   $\left(\xib_{\beta\alpha}^{TR} \right)^T$  means the
transposition of the matrix  $\xib_{\beta\alpha}^{TR}$  with respect to the
space variables  $i,j = x,y,z$.

Relations for the translational friction tensors  $\xib_{\alpha\beta}^{TT}$,
where  $\beta = 1, 2,\ldots, N$,  coincide with the corresponding well-known
relations given in \cite{ref.Mazur} retaining in them terms up to the second
order in  $\sigma$  inclusively.

According to (5.170)--(5.179), the friction tensors of translational
($\xib_{\alpha\beta}^{TT}$) and rotational ($\xib_{\alpha\beta}^{RR}$) motions
of particles are determined up to  $\sigma^2$  and  $\sigma^4$,  respectively,
and the tensors  $\xib_{\alpha\beta}^{TR}$  and  $\xib_{\alpha\beta}^{RT}$  are
determined up to  $\sigma^3$.  In the particular case of two particles, the
friction tensors defined by (5.170)--(5.175) with  $N = 2$  calculated up to
the above-mentioned orders in  $\sigma$  agree with the results given in
\cite{ref.Jones2}.

In view of (5.166)--(5.169), the force and the torque exerted by the fluid on
particle  $\alpha$  due to the translational motion and rotation of all
particles are defined as follows:

\begin{eqnarray}
 \Fbc_\alpha &=& \Fb_\alpha^{(t)} + \Fb_\alpha^{(r)}
  = -\sum\limits_{\beta = 1}^N \left \{ \xib_{\alpha\beta}^{TT} \cdot
  \Ub_\beta^{(t)} + \xib_{\alpha\beta}^{TR} \cdot \Omb_\beta \right \},  \\
 \Tbc_\alpha &=& \Tb_\alpha^{(t)} + \Tb_\alpha^{(r)}
  = -\sum\limits_{\beta = 1}^N \left \{ \xib_{\alpha\beta}^{RT} \cdot
  \Ub_\beta^{(t)} + \xib_{\alpha\beta}^{RR} \cdot \Omb_\beta \right \}.
\end{eqnarray}

Solving the system of equations (5.181), (5.182) for the quantities
$\Ub_\beta^{(t)}$  and  $\Omb_\beta$,  we obtain

\begin{eqnarray}
 \Ub_\alpha^{(t)} &=& = -\sum\limits_{\beta = 1}^N \left \{
  \mub_{\alpha\beta}^{TT} \cdot \Fbc_\beta + \mub_{\alpha\beta}^{TR} \cdot
  \Tbc_\beta \right \},  \\
 \Omb_\alpha &=& = -\sum\limits_{\beta = 1}^N \left \{ \mub_{\alpha\beta}^{RT}
  \cdot \Fbc_\beta + \mub_{\alpha\beta}^{RR} \cdot \Tbc_\beta \right \},
\end{eqnarray}

\noindent  where the translational ($\mub_{\alpha\beta}^{TT}$) and rotational
($\mub_{\alpha\beta}^{RR}$) mobility tensors as well as the tensors
$\mub_{\alpha\beta}^{TR}$  and  $\mub_{\alpha\beta}^{RT}$  that couple
translational and rotational motions of particles determined up to the same
orders in the parameter  $\sigma$  that the corresponding friction tensors have
the form

\begin{eqnarray}
 \mub_{\alpha\beta}^{TT} &=& \frac{1}{\xi_\alpha} \left \{ \delta_{\alpha\beta}
  \Ib + \left( 1 - \delta_{\alpha\beta} \right) \frac{3}{4} \sigma_{\alpha\beta}
  \left( \Ib + \nb_{\alpha\beta} \nb_{\alpha\beta} \right) \right \},  \\
 \mub_{\alpha\beta}^{RR} &=& \frac{1}{\xi_\alpha^R} \left \{ \delta_{\alpha\beta}
  \Ib + \left( 1 - \delta_{\alpha\beta} \right) \frac{\sigma_{\alpha\beta}^3}{2}
  \left(3 \nb_{\alpha\beta} \nb_{\alpha\beta} - \Ib \right) \right \},  \\
 \mub_{\alpha\beta}^{TR} &=& \left( 1 - \delta_{\alpha\beta} \right)
  \frac{1}{8\pi \eta R_{\alpha\beta}^2}
  \left( \eb \cdot \nb_{\alpha\beta} \right),  \\
 \mub_{\alpha\beta}^{RT} &=& \mub_{\alpha\beta}^{TR}.
\end{eqnarray}

It is easy to see that the mobility tensors defined by relations
(5.185)--(5.188) satisfy the Onsager symmetry relations \cite{ref.Happel}

\begin{equation}
 \left(\mub_{\beta\alpha}^{TT} \right)^T = \mub_{\alpha\beta}^{TT}, \quad
 \left(\mub_{\beta\alpha}^{RR} \right)^T = \mub_{\alpha\beta}^{RR}, \quad
 \left(\mub_{\beta\alpha}^{RT} \right)^T = \mub_{\alpha\beta}^{TR}, \quad
 \alpha, \beta = 1, 2, \ldots, N.
\end{equation}

Within the framework of the considered approximation, for which
$\mub_{\alpha\beta}^{TT}$,  $\mub_{\alpha\beta}^{RR}$,
$\mub_{\alpha\beta}^{RT}$,  and  $\mub_{\alpha\beta}^{TR}$  are determined up
to terms of  $\sigma^2$,  $\sigma^4$,  $\sigma^3$,  and  $\sigma^3$,
respectively, relations (5.185)--(5.188) for the mobility tensors coincide with
known results given in \cite{ref.MazurBed}.  Unlike the friction tensors
calculated in this approximation, the mobility tensors are determined only by
two-particle interactions and  $\mub_{\alpha\alpha}^{TR} =
\mub_{\alpha\alpha}^{RT} = 0$.  In order to take into account three-particle
interactions in the mobility tensors, it is necessary to carry out calculations
for higher iterations.


\section{Conclusions}  \label{Conclusions}

In the present paper, we have proposed the procedure for the determination of
the time-dependent velocity and pressure fields of an unbounded incompressible
viscous fluid in an external force field induced by an arbitrary number of
spheres moving and rotating in it as well as the forces and torques exerted by
the fluid on the particles.  The corresponding quantities are expressed in
terms of the induced surface force densities.  We showed that the relations for
the required harmonics of the induced surface force densities given in
\cite{ref.Yosh} for the stationary case are expressed in terms on nonexistent
inverse tensors.  We analyzed in detail the reasons for zero of the determinant
of the corresponding system of equations.  We formulated the consistent system
of algebraic equations in the harmonics of these induced surface force
densities.  In the stationary case, we obtained relations for the harmonics up
to the second approximation inclusively. The obtained general results for the
fluid velocity and pressure and the friction and mobility tensors corresponding
to the second approximation agree with the well-known results obtained by other
methods in various particular cases.

The proposed procedure can be used for the investigation of hydrodynamic
interactions of particles in the fluid for higher-order approximations in a
similar way as it is realized in the present paper.  At the same time, the
relations for the fluid velocity and pressure and the forces and torques
exerted by the fluid on the particles expressed in the present paper in terms
of induced surface forces and the proposed procedure for the determination of
these forces can be regarded as a basis for the study of hydrodynamic
interactions between particles in the nonstationary case.  Certain results
concerning the time-dependent hydrodynamic interactions of particles in a
nonstationary fluid will be given in subsequent papers.

\acknowledgments

The author express the deep gratitude to Prof. I.\,P.~Yakimenko for drawing his
attention to this problem and for useful discussions and advices.


\appendix
\section*{}

To determine the explicit form of the quantities
$F_{l_1l_2,l}(r_\alpha,r_\beta,R_{\alpha\beta},\omega)$,
$C_{l_1l_2,l}(r_\alpha,r_\beta,R_{\alpha\beta})$,  and
$P_{l_1l_2,l}(r_\alpha,a_\beta,R_{\alpha\beta},\omega)$  defined by relations
(3.36), (3.40), and (3.42), it is necessary to take integrals of the product of
three spherical Bessel functions.  For this purpose, first, we consider the
integral

\begin{equation}
 \int\limits_0^\infty \!\!dx \,\frac{x^{\alpha-1}}{x^2+\kappa^2}
 J_\nu(ax)J_\mu(bx)J_\gamma(cx),
\end{equation}

\noindent  where  $\alpha$  is real,  $\nu$,  $\mu$,  and  $\gamma$  are real
quantities nonequal to negative integers,  $\Real \kappa > 0$,  and  $a$, $b$,
and $c$  are real positive quantities such that  $c \geq a + b$.  The case
where the order of a Bessel function is a negative integer is reduced to the
above-considered integral using the known relation  $J_{-n}(x) =  (-1)^n
J_n(x)$ valid for integer  $n$.  To take this integral, we use the Hankel
method \cite{ref.Watson}.  To this end, we consider the following auxiliary
integral:

\begin{equation}
 \int\limits_0^\infty \!\!dx \,\frac{x^{\alpha-1}}{x^2+\kappa^2}
  J_\nu(ax)J_\mu(bx) \left[H_\gamma^{(1)}(cx) + (-1)^p H_\gamma^{(1)}(-cx)
  \right],
\end{equation}

\noindent  where  $H_\gamma^{(1)}(x)$  is the Bessel function of the third kind
and  $p$  is a certain quantity.  For  $p = 1 - \alpha -\nu - \mu$, integral
(A2) can be reduced to the form

\begin{equation}
 \int\limits_{-\infty}^\infty \!\!dx \,\frac{x^{\alpha-1}}{x^2+\kappa^2}
  J_\nu(ax)J_\mu(bx)H_\gamma^{(1)}(cx).
\end{equation}

We consider the integral

\begin{equation}
 \int\limits_C \!\!dz \,\frac{z^{\alpha-1}}{z^2+\kappa^2}
 J_\nu(az)J_\mu(bz)H_\gamma^{(1)}(cz).
\end{equation}

\noindent  in the plane of the complex variable  $z$  along the closed contour
$C$  consisting of the large  $C_R$  and small  $C_r$  semicircles of radii $R$
and  $r$,  respectively, centered at the origin of coordinates and lying above
the real axis, and the segments  $[-R, -r]$  and  $[r, R]$  along the real axis
as  $R \rightarrow \infty$  and  $r \rightarrow 0$.  Taking into account that
the functions  $J_\nu(z)$  and  $H_\nu^{(1)}(z)$  are analytic functions of $z$
in the entire complex plane of  $z$  with the cut  $(-\infty, 0]$,  we obtain
that integral (A4) is determined by the simple pole of the integrand at the
point  $z = i\kappa$  provided that the integrals along the contours $C_R$  and
$C_r$  are finite.  Since  $R \rightarrow \infty$,  for  $c > a + b$,  by the
Jordan lemma, the integral along the contour  $C_R$  is equal to zero if
$\alpha < 4\frac{1}{2}$.  In the case where  $c = a + b$,  the integral along
the contour  $C_R$  is equal to zero for  $\alpha < 3\frac{1}{2}$  and finite
for  $\alpha = 3\frac{1}{2}$.  The condition of finiteness of the integral
along the contour  $C_r$  as  $r \rightarrow 0$  imposes the following
constraint on the quantity  $\alpha$  for  $\gamma \neq 0$:

\begin{equation}
 \alpha \geq \gamma - \nu - \mu.
\end{equation}

\noindent  Furthermore, the integral along the infinitely small semicircle
$C_r$  is nonzero only in the case of equality (A5).  In the particular case
$\gamma = 0$,  inequality (A5) is strict.

Passing in (A4) to the limit as  $R \rightarrow \infty$  and  $r \rightarrow
0$,  we can represent integral (A2) in the form

\begin{eqnarray}
 \int\limits_0^\infty \!\!dx \,\frac{x^{\alpha-1}}{x^2+\kappa^2}
  && \! \! \! \! \! \! J_\nu(ax) J_\mu(bx) \left[ H_\gamma^{(1)}(cx) \right.
   + \left. (-1)^p H_\gamma^{(1)}(-cx) \right]
  = 2(i\kappa)^{\alpha-2}\, i^{\nu+\mu-\gamma}
  I_\nu(a\kappa)I_\mu(b\kappa)K_\gamma(c\kappa)
  \nonumber \\
  &+& \delta_{\alpha,\gamma-\nu-\mu} \, (1-\delta_{\gamma,0})
  \frac{2^\alpha \Gamma(\gamma)}{\Gamma(\nu+1)\Gamma(\mu+1)}
  \frac{a^\nu b^\mu}{\kappa^2 c^\gamma}
  - \delta_{c,a+b}\, \delta_{\alpha,\frac{7}{2}} \,
  \frac{i^{1 + \nu + \mu - \gamma}}{2\sqrt{\pi abc}}(1 + i),  \\
   &&p = 1 - \alpha -\nu -\mu,  \qquad  \alpha \geq \gamma - \mu -\nu,
  \nonumber \\
  &&\alpha < \alpha_c \quad \mbox{if} \quad c > a + b \quad \mbox{or} \quad
  \alpha \leq \alpha_c \quad \mbox{if} \quad c = a + b, \nonumber
\end{eqnarray}

\noindent  where

\begin{equation}
  \alpha_c = \left \{
  \begin{array}{rrl}
   4\frac{1}{2},     \quad  &  \mbox{if}  &  c > a + b,  \\
   3\frac{1}{2},     \quad  &  \mbox{if}  &  c = a + b  \\
  \end{array}  \right.
\end{equation}

\noindent  and  $I_\nu(x)$  and  $K_\nu(x)$  are the modified Bessel functions
of the first and second kinds, respectively.

To determine the required integral (A1), we note that in (A2)

\begin{equation}
 \left[H_\gamma^{(1)}(cx) + (-1)^p H_\gamma^{(1)}(-cx) \right]
 = \left[1 + (-1)^{\gamma - \nu - \mu - \alpha}\right] J_\nu(cx)
 + \left[1 + (-1)^{\gamma - \nu - \mu - \alpha + 1}\right] Y_\nu(cx),
\end{equation}

\noindent  where  $Y_\nu(x)$  is the Bessel function of the second kind.  In
the particular case where  $\alpha = \gamma - \nu - \mu + 2k$,  where  $k = 0,
1, 2,\ldots$,  condition (A5) is true and we obtain the following result for
the required integral (A1):

\begin{eqnarray}
 \int\limits_0^\infty \!\!dx \,\frac{x^{\alpha-1}}{x^2+\kappa^2}
  J_\nu(ax)J_\mu(bx)J_\gamma(cx)
  &=& (i\kappa)^{\alpha-2}i^{\nu+\mu-\gamma}I_\nu(a\kappa)I_\mu(b\kappa)K_\gamma(c\kappa)
  + \delta_{\alpha,\gamma-\nu-\mu} \, (1-\delta_{\gamma,0}) \nonumber \\
  &\times &\frac{2^{\alpha-1} \Gamma(\gamma)}{\Gamma(\nu+1)\Gamma(\mu+1)}
  \frac{a^\nu b^\mu}{\kappa^2 c^\gamma}
  + \delta_{c,a+b}\, \delta_{\alpha,\frac{7}{2}}
  \, \frac{(-1)^k}{2\sqrt{2\pi abc}},  \\
  && \alpha = \gamma -\nu -\mu + 2k,  \quad  k = 0, 1, 2,\ldots, \nonumber \\
  &&\alpha < \alpha_c \quad \mbox{if} \quad c > a + b \quad \mbox{or} \quad
  \alpha \leq \alpha_c \quad \mbox{if} \quad c = a + b. \nonumber
\end{eqnarray}

In the particular case where  $\gamma = \nu$,  $c > a + b$,  and  $\alpha = 2k
- \mu < 9/2$,  where  $k = 1,2,\ldots$,  relation (A9) agrees with the known
result given in \cite[p.~232]{ref.Prudnikov}.

Setting in (A9)  $\nu = l_1 + \frac{1}{2}$,  $\mu = l_2 + \frac{1}{2}$,  and
$\gamma = l + \frac{1}{2}$,  where  $l_1$,  $l_2$,  and  $l$  are nonnegative
integers, we get

\begin{eqnarray}
 \int\limits_0^\infty \!\!dx \,\frac{x^\alpha}{x^2+\kappa^2}
  \, j_{l_1}(ax)j_{l_2}(bx)j_l(cx)
  &=& (-1)^{k-1}\kappa^{\alpha-1}\tilde{j}_{l_1}(a\kappa)\tilde{j}_{l_2}(b\kappa)
  \tilde{h}_l(c\kappa) + \delta_{\alpha,l- l_1 - l_2} \,
  \pi^{\frac{3}{2}}2^{\alpha-3}\nonumber \\
  &\times & \frac{\Gamma\left(l + \frac{1}{2}\right)}{\Gamma\left(l_1 + \frac{3}{2}\right)
  \Gamma\left(l_2 + \frac{3}{2}\right)} \frac{a^{l_1} b^{l_2}}{\kappa^2 c^{l+1}}
  + \delta_{c,a+b}\, \delta_{\alpha,4} \, \frac{(-1)^k \pi}{8abc},  \nonumber \\
  && \alpha = l - l_1 - l_2 + 2k,  \quad  k = 0, 1, 2,\ldots, \\
  && \alpha < \alpha_3 \quad \mbox{if} \quad c > a + b \quad \mbox{or} \quad
  \alpha \leq \alpha_3 \quad \mbox{if} \quad c = a + b, \nonumber
\end{eqnarray}

\noindent  where  $\tilde{j}_l(x) = \sqrt{\pi/(2x)} I_{l +\frac{1}{2}}(x)$ and
$\tilde{h}_l(x) = \sqrt{\pi/(2x)} K_{l +\frac{1}{2}}(x)$ are the modified
spherical Bessel functions of the first and third kind, respectively,
\cite{ref.Abram} and

\begin{equation}
  \alpha_3 = \left \{
  \begin{array}{rrl}
   5,     \quad  &  \mbox{if}  &  c > a + b,  \\
   4,     \quad  &  \mbox{if}  &  c = a + b.  \\
  \end{array}  \right.
\end{equation}

Passing in (A10) to the limit as  $\kappa \to 0$,  we get

\begin{eqnarray}
 \int\limits_0^\infty \!\!dx \, x^{\alpha-2} \, j_{l_1}(ax)j_{l_2}(bx)j_l(cx)
  &=& \pi^{\frac{3}{2}}2^{l-(l_1 + l_2 + 3)}
  \frac{\Gamma\left(l + \frac{1}{2}\right)}{\Gamma\left(l_1 + \frac{3}{2}\right)
  \Gamma\left(l_2 + \frac{3}{2}\right)} \Biggl \{\delta_{k,1}
  - \delta_{k,0}\, \frac{1}{2}\left(\frac{a^2}{2l_1 + 3} \right.\nonumber \\
  &+& \left. \frac{b^2}{2l_2 + 3}
  - \frac{c^2}{2l^2 -1}\right)\Biggr\} \frac{a^{l_1} b^{l_2}}{c^{l+1}}
  + \delta_{c,a+b}\, \delta_{\alpha,4} \, \frac{(-1)^k \pi}{8abc}, \\
  && c \geq a + b, \quad a, b > 0,
  \quad \alpha = l - l_1 - l_2 + 2k,  \, \,  k = 0, 1,2,\ldots, \nonumber \\
  && \alpha < \alpha_3 \quad \mbox{if} \quad c > a + b \quad \mbox{or} \quad
  \alpha \leq \alpha_3 \quad \mbox{if} \quad c = a + b, \nonumber
\end{eqnarray}

\noindent  except for the special case where  $\alpha, l_1, l_2, l = 0$.

For  $c > a + b$,  relation (A12) agrees with the result given in
\cite[p.~239]{ref.Prudnikov}.

Note that we can apply the method used above for the determination of integral
(A1) to the integral

\begin{equation}
 \int\limits_0^\infty \!\!dx \, x^{\alpha-2} \, j_{l_1}(ax)j_{l_2}(bx)j_l(cx).
\end{equation}

In a similar way as for integral (A1), we get

\begin{eqnarray}
 \int\limits_0^\infty \!\!dx \, x^{\alpha-2} \, j_{l_1}(ax)j_{l_2}(bx)j_l(cx)
  &=& \delta_{k,1} \, \pi^{\frac{3}{2}}2^{\alpha - 5}
  \frac{\Gamma\left(l + \frac{1}{2}\right)}{\Gamma\left(l_1 + \frac{3}{2}\right)
  \Gamma\left(l_2 + \frac{3}{2}\right)} \frac{a^{l_1} b^{l_2}}{c^{l+1}}
  + \delta_{c,a+b}\, \delta_{\alpha,4} \, \frac{(-1)^k \pi}{8abc}, \nonumber \\
  && c \geq a + b, \quad a, b > 0,
  \qquad \alpha = l - l_1 - l_2 + 2k,  \quad  k = 1, 2,\ldots, \nonumber \\
  && \alpha < \alpha_3 \quad \mbox{if} \quad c > a + b \quad \mbox{or} \quad
  \alpha \leq \alpha_3 \quad \mbox{if} \quad c = a + b.
\end{eqnarray}

Relation (A14) coincides with (A12) for  $k \geq 1$.  At the same time, we see
that the direct application of this method to (A13) does not enable us to
determine integral (A13) for  $\alpha = l - l_1 - l_2$  ($k = 0$) because
relation (A14) is true for  $k > 0$.

Analogously, we can determine the integral of the product of two spherical
Bessel functions.  The final result is presented as follows:

\begin{eqnarray}
 \int\limits_0^\infty \!\!dx \,\frac{x^\alpha}{x^2+\kappa^2}
  \, j_{l_1}(bx)j_{l_2}(cx)
  &=& (-1)^{k - 1} \kappa^{\alpha - 1} \tilde{j}_{l_2}(c\kappa)\tilde{h}_{l_1}(b\kappa)
  \nonumber \\
  &+& \delta_{\alpha,l_1 -l_2} \, \pi 2^{\alpha-2}
  \frac{\Gamma\left(l_1 +\frac{1}{2}\right)}{\Gamma\left(l_2 +\frac{3}{2}\right)}
  \frac{c^{l_2}}{\kappa^2 b^{l_1 +1}}
  + \delta_{b,c}\, \delta_{\alpha,3} \, \frac{(-1)^k \pi}{4b^2},  \nonumber \\
  && b \geq c > 0,  \qquad
  \alpha = l_1 - l_2 + 2k,  \quad  k = 0, 1, 2,\ldots, \nonumber \\
  && \alpha < \alpha_2 \quad \mbox{if} \quad b > c \quad \mbox{or} \quad
  \alpha \leq \alpha_2 \quad \mbox{if} \quad b = c,
\end{eqnarray}

\noindent  where

\begin{equation}
  \alpha_2 = \left \{
  \begin{array}{rrl}
   4,     \quad  &  \mbox{if}  &  b >  c,  \\
   3,     \quad  &  \mbox{if}  &  b =  c.  \\
  \end{array}  \right.
\end{equation}

For  $b > c$  and  $k > 0$,  integral (A15) agrees with the result given in
\cite[p.~213]{ref.Prudnikov}.

Passing in (A15) to the limit as  $\kappa \rightarrow 0$,  we obtain the known
Weber--Schafheitlin integral

\begin{eqnarray}
 \int\limits_0^\infty \!\!dx \, x^{\alpha-2} \, j_{l_1}(bx)j_{l_2}(cx)
  &=& \pi 2^{\l_1 - l_2 - 2} \frac{c^{l_2}}{b^{l_1 +1}}
  \frac{\Gamma\left(l_1 +\frac{1}{2}\right)}{\Gamma\left(l_2 +\frac{3}{2}\right)}
  \left \{ \delta_{k,1} + \delta_{k,0} \, \frac{1}{2}
  \left(\frac{b^2}{2l_1 -1} - \frac{c^2}{2l_2 + 3}\right) \right\} \nonumber  \\
  &+& \delta_{b,c}\, \delta_{\alpha,3} \, \frac{(-1)^k \pi}{4b^2},  \\
  && b \geq c > 0,  \qquad
  \alpha = l_1 - l_2 + 2k,  \quad  k = 0, 1, 2,\ldots, \nonumber \\
  && \alpha < \alpha_2 \quad \mbox{if} \quad b > c \quad \mbox{or} \quad
  \alpha \leq \alpha_2 \quad \mbox{if} \quad b = c, \nonumber
\end{eqnarray}

\noindent  except for the special case  $\alpha, l_1, l_2 = 0$.

Relation (A17) agrees with the results given in \cite[pp.~239 and
209]{ref.Prudnikov} obtained for $b > c$ and $b = c$, $\alpha = 3$,
respectively.

Note that except for the particular case where  $b = c$  and  $\alpha < 3$,  it
is simpler to determine the integrals of the product of two spherical Bessel
functions (A15) and (A17) on the basis of relations (A10) and (A12) for the
integrals of the product of three spherical Bessel functions found for $a,b,c
> 0$.  For this purpose, it is necessary to pass to the limit  $a = 0$ (or $b =
0$) in (A10) and (A12) defined for  $\alpha < \alpha_3$  and to take into
account that  $\tilde{j}_l(0) = \delta_{l,0}$  and the fact that for  $a = 0$
(or  $b = 0$), the condition of the finiteness of integral (A4) along the
contour $C_r$  leads to the decrease in the maximum possible value of  $\alpha$
by one, i.e.,  $\alpha_3 \to \alpha_2$.  After the corresponding renaming, we
obtain, respectively, (A15) and (A17) defined for  $\alpha < \alpha_2$.

In the particular case  $b = c$,  it is convenient to represent integrals (A15)
and (A17) as follows:

\begin{eqnarray}
 \int\limits_0^\infty \!\!dx \,\frac{x^\alpha}{x^2+\kappa^2}
  \, j_{l_1}(bx)j_{l_2}(cx)
  &=& (-1)^{k - 1} \kappa^{\alpha - 1} \tilde{j}_{l_{max}}(b\kappa)
  \tilde{h}_{l_{min}}(b\kappa) + \delta_{k,0} \, \frac{\pi}{2(b\kappa)^2}
  \left(\frac{2}{b}\right)^{\alpha - 1}  \nonumber \\
  & \times & \frac{\Gamma\left(l_{min} +\frac{1}{2}\right)}
  {\Gamma\left(l_{max} +\frac{3}{2}\right)}
  + \delta_{\alpha,3} \, \frac{(-1)^k \pi}{4b^2}, \nonumber  \\
  && \alpha = l_{min} - l_{max} + 2k,  \quad  k = 0, 1, 2,\ldots,
\end{eqnarray}

\begin{eqnarray}
 \int\limits_0^\infty \!\!dx \,x^{\alpha-2} \, j_{l_1}(bx)j_{l_2}(bx)
  &=& \frac{\pi}{4b} \left( \frac{b}{2} \right)^ {2k - \alpha}
  \frac{\Gamma\left(l_{min} +\frac{1}{2}\right)}
  {\Gamma\left(l_{max} +\frac{3}{2}\right)}
  \left \{ \delta_{k,1} + \delta_{k,0}
  \left(\frac{1}{2l_{min} - 1} - \frac{1}{2l_{max} + 3}\right)
  \frac{b^2}{2}\right \} \nonumber \\
  &+& \delta_{\alpha,3} \, \frac{(-1)^k \pi}{4b^2},
  \qquad \alpha = l_{min} - l_{max} + 2k < 3,  \quad  k = 0, 1, 2,\ldots,
\end{eqnarray}

\noindent  where  $l_{max} = \max(l_1, l_2)$  and  $l_{min} = \min(l_1, l_2)$.

To determine the quantity
$F_{l_1l_1,l}(r_\alpha,r_\beta,R_{\alpha\beta},\omega)$  defined by relation
(3.36) for  $l = l_1 + l_2 - 2p \geq 0$,  where  $p = -1, 0, 1,\ldots,p_{max}$,
we use the above obtained integrals (A10) (for  $r_\beta \neq 0$) and (A15)
(for  $r_\beta = 0$) setting  $\alpha = 2$  in them.  Indeed, in this case, the
representation of  $\alpha$  in the form  $\alpha = l - l_1 - l_2 + 2k$,  where
$k = 0, 1, 2,\ldots$,  corresponds to  $k = 1 + p$.  Since  $0 \geq r_\beta
\geq a_\beta$,  depending on the distance  $r_\alpha$  to the point of
observation, two cases are possible, namely,  $r_\alpha \leq R_{\alpha\beta} -
r_\beta$  and  $r_\alpha \geq R_{\alpha\beta} + r_\beta$  [the domain
$R_{\alpha\beta} + r_\beta > r_\alpha > R_{\alpha\beta} - r_\beta$  should be
considered separately because it does not satisfy the condition  $c \geq a + b$
of integral (A10)].  Depending on the case,  $R_{\alpha\beta}$  or $r_\alpha$
should be taken as the quantity  $c$  in relation (A10).  As a result, we
obtain

\begin{eqnarray}
 F_{l_1 l_2,l}(r_\alpha,r_\beta,R_{\alpha\beta},\omega) &=&
  (-1)^p \frac{2\kappa}{\pi\eta} \Biggl \{ \tilde{j}_{l_1}(x_\alpha)
  \tilde{j}_{l_2}(x_\beta) \tilde{h}_l(y_{\alpha\beta})
  - \delta_{l, l_1 + l_2 + 2}\,
  \frac{\pi^{3/2}}{2}
  \frac{\Gamma\left(l_1 + l_2 + \frac{5}{2} \right)}
  {\Gamma\left(l_1 + \frac{3}{2} \right)\Gamma\left(l_2 + \frac{3}{2} \right)}
  \nonumber \\
  & \times &  \frac{r_\alpha^{l_1} r_\beta^{l_2}}
  {y_{\alpha\beta}^3 R_{\alpha\beta}^{l_1 + l_2}} \Biggr \},
  \quad  l = l_1 + l_2 -2p \geq 0,  \quad  p = -1, 0, 1,\ldots, p_{max},
\end{eqnarray}

\noindent  for  $r_\alpha \leq R_{\alpha\beta} - r_\beta$  and

\begin{eqnarray}
 F_{l_1 l_2,l}(r_\alpha,r_\beta,R_{\alpha\beta},\omega) &=&
  (-1)^{l_{2} - p} \frac{2\kappa}{\pi\eta} \Biggl \{ \tilde{h}_{l_1}(x_\alpha)
  \tilde{j}_{l_2}(x_\beta) \tilde{j}_l(y_{\alpha\beta})
  - \delta_{l, l_1 - l_2 - 2}\,
  \frac{\pi^{3/2}}{2}  \nonumber \\
  & \times &  \frac{\Gamma\left(l_1 + \frac{1}{2} \right)}
  {\Gamma\left(l_2 + \frac{3}{2} \right)
  \Gamma\left(l_1 - l_2 - \frac{1}{2} \right)}
  \frac{r_\beta^{l_2} R_{\alpha\beta}^l}
  {x_\alpha^3 r_\alpha^{l_1 - 2}} \Biggr \},  \nonumber  \\
  && \qquad  l = l_1 + l_2 -2p \geq 0,  \quad  p = -1, 0, 1,\ldots, p_{max},
\end{eqnarray}

\noindent  for  $r_\alpha \geq R_{\alpha\beta} + r_\beta$.  Here,
$x_\alpha = \kappa r_\alpha,$  $x_\beta = \kappa r_\beta,$  and
$y_{\alpha\beta} = \kappa R_{\alpha\beta}$.

Note that in view of the representation (A10) for the integral of the product
of three spherical Bessel functions, relation (A21) for the quantity $F_{l_1
l_2,l}(r_\alpha,r_\beta,R_{\alpha\beta},\omega)$  valid for $r_\alpha \geq
R_{\alpha\beta} + r_\beta$  can also be represented in the form (A20) valid for
$r_\alpha \leq R_{\alpha\beta} - r_\beta$  with the changes $r_\alpha
\leftrightarrow R_{\alpha\beta}$  and  $l_1 \leftrightarrow l$ putting  $l_1 =
l + l_2 - 2s$  (i.e.,  $p = l_2 - s$), where  $s = -1,0,1,\ldots,s_{max}$  and
$s_{max} = \min\Bigl(\left[(l + l_2)/2\right], 1 + \min (l, l_2) \Bigr)$.

Putting  $r_\beta = 0$  and  $r_\alpha = R_{\alpha\beta}$  in (A20) and (A21)
and equating the obtained relations to one another, we obtain the following
relation for the modified spherical Bessel functions:

\begin{equation}
 \tilde{j}_l(x) \tilde{h}_{l+2}(x) - \tilde{j}_{l+2}(x) \tilde{h}_l(x)
 = \frac{\pi}{x^3} \left(l + \frac{3}{2} \right ), \qquad l \geq 0.
\end{equation}

Note that this relation can also be derived by equating relations (A.15) for  $c
= b$  and (A.18) putting  $\alpha = 2$  and  $l_1 = l_2 + 2$  in them.

To determine the quantity  $C_{l_1l_1,l}(r_\alpha,r_\beta,R_{\alpha\beta})$
defined by relation (3.40) for $l = l_1 + l_2 - 2p + 1 \geq 0$,  where  $p = 0,
1,\ldots,\tilde{p}_{max}$,  we use relation (A12) with  $\alpha = 3$  and take
into account that  $r_\beta \leq a_\beta$.  As a result, for  $r_\beta \neq 0$,
we get

\begin{eqnarray}
 C_{l_1 l_2,l}(r_\alpha,r_\beta,R_{\alpha\beta}) &=&
  \delta_{l, l_1 + l_2 + 1} \, \frac{\sqrt{\pi}}{2}
  \frac{\Gamma\left(l_1 + l_2 + \frac{3}{2} \right)}
  {\Gamma\left(l_1 + \frac{3}{2} \right)\Gamma\left(l_2 + \frac{3}{2} \right)}
  \frac{r_\alpha^{l_1} r_\beta^{l_2}}{R_{\alpha\beta}^{l_1 + l_2 +2}}
\end{eqnarray}

\noindent  for  $r_\alpha \leq R_{\alpha\beta} - r_\beta$  and

\begin{eqnarray}
 C_{l_1 l_2,l}(r_\alpha,r_\beta,R_{\alpha\beta}) &=&
  \frac{\sqrt{\pi}}{2} \frac{\Gamma\left(l_1 + \frac{1}{2} \right)}
  {\Gamma\left(l_2 + \frac{3}{2} \right)}
  \frac{r_\beta^{l_2} R_{\alpha\beta}^{l_1 - l_2 - 1}}
  {r_\alpha^{l_1 + 1}} \left \{ \delta_{l, l_1 - l_2 - 1}\,
  \frac{1}{\Gamma\left(l_1 - l_2 + \frac{1}{2} \right)}
  + \delta_{l, l_1 - l_2 - 3} \right. \nonumber  \\
  & \times & \left. \frac{2}{\Gamma\left(l_1 - l_2 - \frac{3}{2} \right)}
  \left [ \frac{1}{2l_1 - 1} \left (\frac{r_\alpha}{R_{\alpha\beta}} \right )^2
  - \frac{1}{2l_2 + 3}
  \left (\frac{r_\beta}{R_{\alpha\beta}} \right )^2
  - \frac{1}{2l + 3} \right ] \right \}
\end{eqnarray}

\noindent  for  $r_\alpha \geq R_{\alpha\beta} + r_\beta$.

Just as for the representation of the quantity $F_{l_1
l_2,l}(r_\alpha,r_\beta,R_{\alpha\beta},\omega)$  for $r_\alpha \geq
R_{\alpha\beta} + r_\beta$,  we can show the possibility of the representation
of relation (A24) for the quantity $C_{l_1
l_2,l}(r_\alpha,r_\beta,R_{\alpha\beta})$  in the form (A23) valid for
$r_\alpha \leq R_{\alpha\beta} - r_\beta$  with the changes $r_\alpha
\leftrightarrow R_{\alpha\beta}$  and  $l_1 \leftrightarrow l$ putting,  $l_1 =
l + l_2 + 1 - 2s$,  where  $s = 0,1,\ldots,\tilde{s}_{max}$  and
$\tilde{s}_{max} = \min\Bigl(\left[(l + l_2 + 1)/2\right], 1 + \min (l, l_2)
\Bigr)$.

In the particular case  $r_\beta  = 0$,  to determine the quantity
$C_{l_1l_2,l}(r_\alpha,0,R_{\alpha\beta})$,  it is necessary to use relation
(A17).  As a result, we obtain

\begin{equation}
 C_{l_1 l_2,l}(r_\alpha,0,R_{\alpha\beta}) = \delta_{l_2, 0}
  C_{l_1,l}(r_\alpha,R_{\alpha\beta}), \qquad l =l_1 \pm 1 \geq 0,
\end{equation}

\noindent  where

\begin{equation}
 C_{l_1,l}(r_\alpha,R_{\alpha\beta}) = \frac{1}{R_{\alpha\beta}^2}
  \left \{\delta_{l,l_1 + 1}\left(\frac{r_\alpha}{R_{\alpha\beta}}\right)^{l_1}
  + \delta_{r_\alpha,R_{\alpha\beta}} \, \frac{1}{2} \left(\delta_{l,l_1 - 1}
  - \delta_{l,l_1 + 1}\right) \right \}
\end{equation}

\noindent  for  $r_\alpha \leq R_{\alpha\beta}$  and

\begin{equation}
 C_{l_1,l}(r_\alpha,R_{\alpha\beta}) = \frac{1}{r_\alpha^2}
  \left \{ \delta_{l,l_1-1} \left(\frac{R_{\alpha\beta}}{r_\alpha} \right)^{l_1-1}
  + \delta_{r_\alpha,R_{\alpha\beta}} \, \frac{1}{2} \left(\delta_{l,l_1 + 1}
  - \delta_{l,l_1 - 1}\right) \right \}
\end{equation}

\noindent  for  $r_\alpha \geq R_{\alpha\beta}$.  In this case, the quantities
$C_{l_1,l}(r_\alpha,R_{\alpha\beta})$  should be considered only for  $l = l_1
\pm 1 \geq 0$  because for  $l_2 = 0$,  $\tilde{p}_{max} = 0$  if  $l_1 = 0$
or  $\tilde{p}_{max} = 1$  if  $l_1 \geq 1$.

Note that simply passing to the limit in (A23) and (A24) as  $r_\beta \to 0$, we
obtain

\begin{equation}
 C_{l_1 l_2,l}(r_\alpha,0,R_{\alpha\beta}) = \delta_{l_2, 0}
  \tilde{C}_{l_1,l}(r_\alpha,R_{\alpha\beta}), \qquad l =l_1 \pm 1 \geq 0,
\end{equation}

\noindent  where

\begin{equation}
 \tilde{C}_{l_1,l}(r_\alpha,R_{\alpha\beta}) = \delta_{l,l_1 + 1} \,
  \frac{1}{R_{\alpha\beta}^2}
  \left(\frac{r_\alpha}{R_{\alpha\beta}} \right)^{l_1}
\end{equation}

\noindent  for  $r_\alpha \leq R_{\alpha\beta}$  and

\begin{equation}
 \tilde{C}_{l_1,l}(r_\alpha,R_{\alpha\beta}) = \delta_{l,l_1 - 1}
  \frac{1}{r_\alpha^2}
  \left(\frac{R_{\alpha\beta}}{r_\alpha} \right)^{l_1-1}
\end{equation}

\noindent  for  $r_\alpha \geq R_{\alpha\beta}$.

Comparing relations (A29) and (A30) with relations (A26) and (A27), we see that
the function  $\tilde{C}_{l_1,l}(r_\alpha,R_{\alpha\beta})$  corresponds to the
first term of the function  $C_{l_1,l}(r_\alpha,R_{\alpha\beta})$.  Furthermore,
in view of (A29) and (A30), the function
$\tilde{C}_{l_1,l}(r_\alpha,R_{\alpha\beta})$  has a discontinuity at the point
$r_\alpha = R_{\alpha\beta}$,  while, according to (A26) and (A27), the function
$C_{l_1,l}(r_\alpha,R_{\alpha\beta})$  is continuous at this point and equal to

\begin{equation}
 C_{l_1,l}(R_{\alpha\beta},R_{\alpha\beta}) = \frac{1}{2 R_{\alpha\beta}^2}
  \left(\delta_{l,l_1+1} + \delta_{l,l_1-1}\right ),\qquad l = l_1 \pm 1 \geq 0.
\end{equation}

In a similar way, using (A10), we can find the quantity
$P_{l_12,l}(r_\alpha,a_\beta,R_{\alpha\beta},\omega)$  defined by relation
(3.42) with  $l_2 = 2$.  Indeed, according to (3.31) and (3.51), this quantity
should be determined only for  $l = l_1 \pm 1 \geq 0$, which agrees with the
constraint for the quantities  $l_1$,  $l_2$,  $l$, $\alpha$,  and  $k$  in the
integral defined by relation (A10).  As a result, we get

\begin{equation}
 P_{l_1 2,l}(r_\alpha,a_\beta,R_{\alpha\beta},\omega) = (-1)^p
  \frac{2}{\pi \eta} \, \tilde{j}_{l_1}(x_\alpha)
  \tilde{j}_2(b_\beta) \tilde{h}_l(y_{\alpha\beta})
\end{equation}

\noindent  for  $r_\alpha \leq R_{\alpha\beta} - a_\beta$  and

\begin{equation}
 P_{l_1 2,l}(r_\alpha,a_\beta,R_{\alpha\beta},\omega) = (-1)^{p + 1}
  \frac{2}{\pi \eta} \, \tilde{h}_{l_1}(x_\alpha)
  \tilde{j}_2(b_\beta) \tilde{j}_l(y_{\alpha\beta})
\end{equation}

\noindent  for  $r_\alpha \geq R_{\alpha\beta} + a_\beta$.  Here,  $b_\beta =
\kappa a_\beta$.

Using the obtained integrals of the product of two spherical Bessel functions
(A15) and (A17), we can determine the quantities
$F_{l_1l_2}(r_\alpha,r_\alpha^{\,\prime},\omega)$,
$C_{l_1l_2}(r_\alpha,r_\alpha^{\,\prime})$,  and
$P_{1,2}(r_\alpha,a_\alpha,\omega)$,  defined, respectively, by relations
(3.64), (3.66), and (3.67), for  $r_\alpha \geq a_\alpha$  and $0 \leq
r_\alpha^{\,\prime} \leq a_\alpha$,  taking into account that
$F_{l_1,l_2}(r_\alpha,r_\alpha^{\,\prime},\omega)$  and
$C_{l_1l_2}(r_\alpha,r_\alpha^{\,\prime})$  should be determined only for $l_2
= l_1 + 2p \ge 0$,  where  $p = 0, \pm 1$,  and  $l_ 2 = l_1 \pm 1 \geq 0$,
respectively.  As a result, we get

\begin{eqnarray}
 F_{l_1 l_2}(r_\alpha,r_\alpha^{\,\prime},\omega) &=&
  (-1)^p \, \frac{2\kappa}{\pi \eta} \left \{
  \tilde{h}_{l_1}(x_\alpha) \tilde{j}_{l_2}(x_\alpha^{\,\prime})
  - \delta_{l_2, l_1 -2} \left (l_1 - \frac{1}{2} \right) \frac{\pi}{x_\alpha^3}
  \left( \frac{r_\alpha^{\,\prime}}{r_\alpha} \right)^{l_1 -2} \right \},
  \nonumber \\
  && \qquad \qquad \qquad \qquad l_2 = l_1 + 2p \geq 0,  \quad p = 0, \pm 1, \\
 C_{l_1 l_2}(r_\alpha,r_\alpha^{\,\prime}) &=&
  \delta_{l_2, l_1 - 1} \frac{1}{r_\alpha^2}
  \left( \frac{r_\alpha^{\,\prime}}{r_\alpha} \right)^{l_1 - 1}
  + \delta_{r_\alpha, a} \, \delta_{r_\alpha^{\,\prime}, a} \, \frac{1}{2 a_\alpha^2}
  \left( \delta_{l_2, l_1 + 1} - \delta_{l_2, l_1 - 1} \right), \nonumber \\
  && \qquad  \qquad \qquad \qquad l_2 = l_1 \pm 1 \geq 0,
\end{eqnarray}

\begin{equation}
 P_{1,2}(r_\alpha,a_\alpha,\omega) =
  \frac{2}{\pi \eta} \, \tilde{h}_1(x_\alpha) \tilde{j}_2(b_\alpha).
\end{equation}

In the important particular case where  $r_\alpha = a_\alpha$  and
$r_\alpha^{\,\prime} = a_\alpha$,  using relation (A22), we reduce relation
(A34) to the form

\begin{equation}
 F_{l_1 l_2}(a_\alpha,a_\alpha,\omega) =
  (-1)^p \, \frac{2\kappa}{\pi \eta} \,
  \tilde{j}_{l_{max}}(b_\alpha) \tilde{h}_{l_{min}}(b_\alpha),
  \quad  l_2 = l_1 + 2p \geq 0,  \quad  p = 0, \pm 1.
\end{equation}

We also need the explicit form for the quantities
$C_{l_1l_2,l}(r_\alpha,a_\beta,R_{\alpha\beta})$  and
$C_{l_1l_2,l}(r_\alpha,a_\alpha)$  for  $r_\alpha = 0$.  Setting  $r_\alpha = 0$
in (3.40) and using (A17), we obtain

\begin{equation}
 C_{l_1 l_2,l}(0,a_\beta,R_{\alpha\beta}) = \delta_{l_1,0} \,
  \delta_{l,l_2 + 1}\,
  \frac{\sigma_{\beta\alpha}^{l_2}}{R_{\alpha\beta}^2},
  \qquad l = l_2 \pm 1 \geq 0.
\end{equation}

Note that this result can be also obtained from the general relation (A23) for
$C_{l_1 l_2,l}(r_\alpha,r_\beta,R_{\alpha\beta})$  found for  $r_\alpha > 0$  if
we set  $r_\beta = a_\beta$  in it and pass to the limit as  $r_\alpha
\rightarrow 0$.

To determine the quantity  $C_{l_1 l_2}(r_\alpha,a_\alpha)$  defined by
relation (3.66), where  $l_2 = l_1 \pm 1 \geq 0$,  for  $r_\alpha = 0$,  we,
first, find the quantity  $C_{l_1 l_2}(r_\alpha,a_\alpha)$  for  $r_\alpha <
a_\alpha$.  To this end, using relation (A17), we get

\begin{equation}
 C_{l_1 l_2}(r_\alpha,a_\alpha) = \delta_{l_2,l_1 + 1}\, \frac{1}{a_\alpha^2}
 \,
  \left(\frac{r_\alpha}{a_\alpha} \right)^{l_1},
  \qquad  l_2 = l_1 \pm 1 \geq 0.
\end{equation}

Passing in (A39) to the limit as  $r_\alpha \rightarrow 0$,  we obtain

\begin{equation}
 \lim\limits_{r_\alpha \to 0} C_{l_1 l_2}(r_\alpha,a_\alpha)
  =  \delta_{l_1,0}\, \delta_{l_2,1} \, \frac{1}{a_\alpha^2}.
\end{equation}

At the same time,according to relation (3.66) for the quantity $C_{l_1
l_2}(r_\alpha,r_\alpha^\prime)$,  we obtain the following relation for
$r_\alpha = 0$,  $r_\alpha^\prime = a_\alpha$,  and $l_2 = l_1 \pm 1 \ge 0$:

\begin{equation}
 C_{l_1 l_2}(0,a_\alpha) =  \delta_{l_1,0} \, \delta_{l_2,1}\,
  \frac{1}{a_\alpha^2}\left(1 -
  \frac{2}{\pi} \left. \sin x \right |_{x \, = \, \infty} \right )
\end{equation}

\noindent  that has not any definite limit.



\newpage

\begin{thebibliography}{99}

\bibitem{ref.Happel}
 J.~Happel and H.~Brenner, {\it Low Reynolds Number Hydrodynamics\/}
 (Prentice-Hall, London, 1965).

\bibitem{ref.Lamb}
 H.~Lamb, {\it Hydrodynamics\/} (Dover, New York, 1945).

\bibitem{ref.Milne}
 L.\,M.~Milne-Thomson, {\it Theoretical Hydrodynamics\/}
 (Macmillan, London, 1960).

\bibitem{ref.Batch}
 G.\,K.~Batchelor, {\it An Introduction to Fluid Dynamics\/}
 (Cambridge, 1970).

\bibitem{ref.Landau}
 L.\,D.~Landau and E.\,M.~Lifshitz, {\it Fluid Mechanics\/}
 (Pergamon, New York, 1978).

\bibitem{ref.Loitsyan}
 L.\,G.~Loitsyanskii, {\it Mechanics of Fluid and Gas\/}
 (Nauka, Moscow, 1987).

\bibitem{ref.Stimson}
 M.~Stimson and G.\,B.~Jeffery, Proc. Roy. Soc. London, {\bf A111}, 110 (1926).

\bibitem{ref.Maude}
 A.\,D.~Maude, {\it End effects in a falling-sphere viscometer,\/}
 Brit. J. Apl. Phys., {\bf 12}, No.~6, 293--295 (1961).

\bibitem{ref.Wakiya}
 S.~Wakiya, {\it Slow motions of a viscous fluid around two spheres,\/}
 J. Phys. Soc. Japan, {\bf 22}, No.~4, 1101--1109 (1966).

\bibitem{ref.Davis}
 M.\,H.~Davis, {\it The slow translation and rotation of two unequal spheres in a
 viscous fluid,\/} Chem. Eng. Sci., {\bf 24}, No.~12, 1769--1776 (1969).

\bibitem{ref.Jones1}
 R.\,B.~Jones, {\it Hydrodynamic interaction of two permeable spheres.  I:
 The method of reflections,\/}
 Physica, {\bf 92A}, No.~3--4, 545--556 (1978).

\bibitem{ref.Jones2}
 R.\,B.~Jones, {\it Hydrodynamic interaction of two permeable spheres.  II:
 Velocity field and friction constants,\/}
 Physica, {\bf 92A}, No.~3--4, 557--570 (1978).

\bibitem{ref.Jones3}
 R.\,B.~Jones, {\it Hydrodynamic interaction of two permeable spheres.  III:
 Mobility tensors,\/} Physica, {\bf 92A}, No.~3--4, 571--583 (1978).

\newpage

\bibitem{ref.Feld1}
 B.\,U.~Felderhof, {\it Force density induced on a sphere in linear
 hydrodynamics.  II.  Moving sphere, mixed boundary conditions,\/}
 Physica, {\bf 84A}, No.~3, 569--576 (1976).

\bibitem{ref.Feld2}
 B.\,U.~Felderhof, {\it Hydrodynamic interaction between two spheres,\/}
 Physica, {\bf 89A}, No.~2, 373--384 (1977).

\bibitem{ref.Schmitz1}
 R.~Schmitz and B.\,U.~Felderhof, {\it Creeping flow about a sphere,\/}
 Physica, {\bf 92A}, No.~3--4, 423--437 (1978).

\bibitem{ref.Schmitz2}
 R.~Schmitz and B.\,U.~Felderhof, {\it Creeping flow about a spherical
 particle,\/} Physica, {\bf 113A}, 90--102 (1982).

\bibitem{ref.Schmitz3}
 R.~Schmitz and B.\,U.~Felderhof, {\it Friction matrix for two spherical
 particles with hydrodynamic interaction,\/} Physica, {\bf 113A},
 103--116 (1982).

\bibitem{ref.MazurBed}
 P.~Mazur and D.~Bedeaux, {\it A generalization of Fax\'{e}n theorem to
 nonsteady motion of a sphere through an incompressible fluid in arbitrary
 flow,\/} Physica, {\bf 76},  No.~2, 235--246 (1974).

\bibitem{ref.Mazur}
 P.~Mazur, {\it On the motion and Brownian motion of n spheres in a viscous
 fluid,\/} Physica, {\bf 110A}, 128--146 (1982).

\bibitem{ref.MazurSaarl}
 P.~Mazur and W.~van~Saarloos, {\it Many-sphere hydrodynamic interactions and
 mobilities in a suspension,\/} Physica, {\bf 115A}, 21--57 (1982).

\bibitem{ref.Saarl}
 W.~van~Saarloos and P.~Mazur, {\it Many-sphere hydrodynamic interactions.  II.
 Mobilities at finite frequencies,\/}
 Physica, {\bf 120A}, 77--102 (1983).

\bibitem{ref.Freed1}
 K.\,F.~Freed and M.~Muthukumar, {\it On the Stokes problem for a suspension of
 spheres at finite concentrations,\/}
 J. Chem. Phys., {\bf 68}, No. 5, 2088--2096 (1978).

\bibitem{ref.Freed2}
 K.\,F.~Freed and M.~Muthukumar, {\it Dynamics and hydrodynamics of
 translational-rotational Brownian particles at finite concentrations,\/}
 J. Chem. Phys., {\bf 69}, No. 6, 2657--2671 (1978).

\bibitem{ref.Pien1}
 I.~Pie\'{n}kowska, {\it An unsteady Faxen's relation for the force including
 interaction effects,\/} Arch. Mech, {\bf 34}, No.~3, 297--306 (1982).

\bibitem{ref.Pien2}
 I.~Pie\'{n}kowska, {\it Unsteady friction and mobility relations for Stokes flow,\/}
 Arch. Mech, {\bf 36}, No.~5--6, 746--769 (1984).

\bibitem{ref.Clercx1}
 H.\,J.\,H.~Clercx and P.\,P.\,J.\,M.~Schram, {\it Quasistatic hydrodynamic
 interaction in suspensions,\/} Physica A, {\bf 174}, 293--324 (1991).

\bibitem{ref.Clercx2}
 H.\,J.\,H.~Clercx and P.\,P.\,J.\,M.~Schram, {\it Retarded hydrodynamic
 interaction in suspensions,\/} Physica A, {\bf 174}, 325--354 (1991).

\bibitem{ref.Hofman}
 J.\,M.\,A.~Hofman, H.\,J.\,H.~Clercx, and P.\,P.\,J.\,M.~Schram, {\it Effective
 viscosity of dense colloidal crystals,\/} Phys. Rev. E, {\bf 62}, No.~6,
 8212--8333 (2000).

\bibitem{ref.Yosh}
 T.~Yoshizaki and H.~Yamakawa, {\it Validity of the superposition approximation
 in an application of the modified Oseen tensor to rigid polymers,\/}
 J. Chem. Phys., {\bf 73},  No.~1, 578--582 (1980).

\bibitem{ref.Nikiforov}
 A\,F.~Nikiforov and V.\,B.~Uvarov, {\it Special Functions of Mathematical
 Physics\/} (Nauka, Moscow, 1978).

\bibitem{ref.Varshalovich}
 D.\,A~Varshalovich, A\,N.~Moskalev, and V.\,K.~Khersonskii,
 {\it Quantum Theory of Angular Momentum\/} (Nauka, Leningrad, 1975).

 \bibitem{ref.Davydov}
  A\,S.~Davydov, {\it Quantum Mechanics\/} (Macmillan, London, 1960).

\bibitem{ref.Abram} M.~Abramowitz and I.\, A~Stegun (eds.), {\it Handbook of
 Mathematical Functions with Formulas, Graphs and Mathematical Tables\/},
 (National Bureau of Standards, Appl. Math., Ser. 55, 1964).

\bibitem{ref.Watson}
 G.\,N.~Watson, {\it A Treatise of the Theory of Bessel Functions\/}
 (Cambridge University, Cambridge, 1945).

\bibitem{ref.Prudnikov}
 A\,P.~Prudnikov, Yu.\,A~Brychkov, and O.\,I.~Marichev, {\it Integrals and
 Series.  Special Functions\/} (Nauka, Moscow, 1983).

\end{thebibliography}
\end{document}